\documentclass{iopjournal}

\usepackage{ragged2e}
\usepackage{times,graphicx,amssymb,amsmath,amsfonts,bbm,psfrag,xcolor}
\usepackage{booktabs}
\usepackage{diagbox}
\usepackage{graphicx}
\usepackage{amssymb}
\usepackage{amsmath}
\usepackage{amsfonts}
\usepackage{bbm}
\usepackage{psfrag}
\usepackage{xcolor}
\usepackage{epstopdf}
\usepackage{graphicx}
\usepackage{subcaption}
\usepackage{times}
\usepackage{commath}
\usepackage{thmtools}
\usepackage{thm-restate}
\usepackage{times}
\usepackage{graphicx}
\usepackage{natbib}
\usepackage{dirtytalk}
\usepackage{empheq}
\usepackage{arydshln}
\usepackage{ulem}

\definecolor{ddarkbrown}{rgb}{0.5,0.2,0.05} \definecolor{bbluegray}{rgb}{0.05,0,0.5}

\newtheorem{theorem}{Theorem}

\usepackage{mathtools}

\usepackage{algorithm,algcompatible,algpseudocode}

\usepackage{multicol,lipsum}
\algnewcommand{\Inputs}[1]{%
	\State \textbf{Inputs: \:}{#1}
}

\algnewcommand{\Output}[1]{%
	\State \textbf{Output: \:}{#1}
}
\algnewcommand{\Initialize}[1]{%
	\State \textbf{Initialize: \:}{#1}
}

\algnewcommand{\IIf}[1]{\State\algorithmicif\ #1\ \algorithmicthen}
\algnewcommand{\EndIIf}{\unskip\ \algorithmicend\ \algorithmicif}

\let \oldsection \section
\renewcommand{\section}{\vspace{3ex plus 1ex}\oldsection}

\newcommand{\BEAS}{\begin{eqnarray*}}
	\newcommand{\EEAS}{\end{eqnarray*}}
\newcommand{\BEA}{\begin{eqnarray}}
\newcommand{\EEA}{\end{eqnarray}}

\newcommand{\BEQ}{\begin{equation}}
\newcommand{\EEQ}{\end{equation}}
\newcommand{\BIT}{\begin{itemize}}
	\newcommand{\EIT}{\end{itemize}}
\newcommand{\BNUM}{\begin{enumerate}}
	\newcommand{\ENUM}{\end{enumerate}}

 \newcommand{\Ft}{\mathcal{F}_\theta}

	\newcommand{\R}{\mathbb{R}}

\newcommand{\Fth}{\mathcal{F}_\theta}

\newcommand{\Adagger}{A^\dagger}

\newcommand{\BA}{\begin{array}}
	\newcommand{\EA}{\end{array}}

 \numberwithin{dummy}{section}

\numberwithin{mythm}{section}
\numberwithin{mydef}{section}
\numberwithin{myprop}{section}
\numberwithin{mylem}{section}
\numberwithin{mycor}{section}

\newcommand{\xgx}[1]{{\color{purple}{#1}}}

\begin{document}


\title{Equivariant Test-Time Training with Operator Sketching for Imaging Inverse Problems}

\author{Guixian Xu$^1$, Jinglai Li$^1$ and Junqi Tang$^{1,*}$}

\affil{$^1$School of Mathematics, University of Birmingham, Birmingham, UK}

\affil{$^*$Author to whom any correspondence should be addressed.}

\email{j.tang.2@bham.ac.uk}


\begin{abstract}
\justify
Equivariant Imaging (EI) regularization has become the de-facto technique for unsupervised training of deep imaging networks, without any need of ground-truth data. Observing that the EI-based unsupervised training paradigm currently has significant computational redundancy leading to inefficiency in high-dimensional applications, we propose a sketched EI regularization which leverages the randomized sketching techniques for acceleration. We apply our sketched EI regularization to develop an accelerated deep internal learning framework, which can be efficiently applied for test-time network adaptation. Additionally, for network adaptation tasks, we propose a parameter-efficient approach to accelerate both EI and Sketched-EI via optimizing only the normalization layers. Our numerical study on X-ray CT and multicoil magnetic resonance image reconstruction tasks demonstrate that our approach can achieve significant computational acceleration over the standard EI counterpart, especially in test-time training tasks.

\end{abstract}

\section{Introduction}
\justify

Unsupervised training has emerged as a powerful approach for solving imaging inverse problems such as computed tomography (CT) and magnetic resonance imaging (MRI) reconstruction \cite{carioni2023unsupervised, tirer2024deep}. Unlike conventional supervised learning methods that rely on large datasets of paired ground-truth and measurement samples, unsupervised approaches learn directly from the measurement data themselves. In these settings, only the measured signals and the associated imaging model are available, while the underlying clean images remain unknown.

Many unsupervised approaches are based on  the idea of Deep Internal Learning, exemplified by  the Deep Image Prior (DIP) \cite{ulyanov2018deep, tachella2020neural, mataev2019deepred, liu2019image}, where a network learns to reconstruct an image from a single noisy input without external supervision.
Other developments of unsupervised approaches, including the Noise2X family of methods \cite{lehtinen2018noise2noise, batson2019noise2self} and its variant Artifact2Artifact \cite{liu2020rare}, 
train networks from collections of measurements, but also do not  clean ground truth images.
More recently, a particularly promising advance along this research direction is the Equivariant Imaging (EI) framework \cite{chen2021equivariant, chen2022robust, tachella2023sensing}.
Simply put, EI is an unsupervised framework that, via the so-called EI regularization, encourages the network to produce consistent reconstructions under transformations of the input, effectively allowing it to learn information beyond the range space of the measurement operator.

Despite their success,  unsupervised approaches 
generally rely on the assumption that training and testing data share the same distribution, which may not be true in real-world applications. The Test-time training (TTT) and adaptation (TTA) methods address this issue by updating 
trained models directly on unlabeled test data~\cite{sun2020test, wang2020tent}. 
The pretrained model used for TTT can itself be obtained through either supervised or unsupervised learning. 
The core idea is to treat each test input (or small batch) as a self-supervised learning problem. For example, \cite{sun2020test} proposes converting a single unlabeled test sample into a self-supervised task and updating the model parameters on it before making a prediction. Similarly, \cite{wang2020tent} adapts by minimizing the entropy of predictions: it updates the model’s batch-normalization affine parameters online to increase confidence in each test batch. In general, TTT methods use unsupervised objectives (such as predicting known image transformations or enforcing consistency with measurements) at test time and can operate on either individual input or small minibatches. This on-the-fly adaptation allows the model to specialize in the specific test distribution. \cite{liu2021ttt++} analyzes when TTT succeeds or fails under large shifts and introduces feature-alignment and contrastive losses (the “TTT++” method) to stabilize adaptation. \cite{gandelsman2022test} shows that using a masked-autoencoder reconstruction loss as self-supervision yields a robust generalization on visual distribution-shift benchmarks. More recent TTT work adds practical considerations: \cite{niu2022efficient} notes that some test samples (e.g. those with high entropy) can yield noisy gradients, so they propose selecting only reliable samples for adaptation and adding a Fisher-information regularizer to prevent catastrophic forgetting. \cite{yuan2023robust} extends TTT to streaming/dynamic test environments and develops a ``RoTTA'' scheme with robust batch norm estimation and memory-bank sampling for continual adaptation.

In this work, we incorporate the EI unsupervised formulation and the test-time training/adaptation. The EI formulation provides a natural self-supervised objective based on equivariance consistency, which enforces 
the invariance of the reconstructed images under input transformations.
As a result, the model can better exploit the physical model and the inherent symmetries of the imaging process, leading to more stable and effective adaptation than conventional TTT methods that rely solely on statistical or entropy-based objectives. We emphasize that in our method both the  pretraining and the inference-time adaptation use the EI formulation, 
and an advantage of this design is that the offline training provides a natural warm start for the subsequent test-time training.

EI methods, like most other unsupervised approaches for inverse problems, are computationally demanding because each optimization step requires evaluating the forward operator, its adjoint, or a pseudoinverse, all of which are expensive in imaging applications such as CT or MRI. This computational burden is even more critical in the test-time training  setting, where adaptation must be performed at inference time for each individual test sample. To address this issue, we accelerate EI-based learning through operator sketching, a technique from randomized linear algebra that compresses the forward operator using a structured random projection. The resulting sketched operator reduces dimensionality while approximately preserving the action of the original operator, thereby lowering the cost of the forward–adjoint evaluations required by EI regularization. Specifically, we use operator sketching to reduce the computational cost in both the model pretraining and the TTT stages.

We summarize the specific technical contributions of this work as follows:

\begin{itemize}
\item \textbf{EI-based test-time training --}  
    We extend the EI framework to a test-time training (TTT) setting, enabling single-input, task-adapted image reconstruction via inference-time network adaptation. The proposed TTT–EI approach refines a pre-trained model directly on the test measurement using EI regularization, improving reconstruction quality under distribution shift or limited measurements. 

    \item \textbf{Sketched EI regularization --} We propose an efficient variant of the EI regularizer, mitigating the computational inefficiency of the original approach by \cite{chen2021equivariant}. We provide a motivational theoretical analysis on the approximation bound of our sketched EI regularizer, demonstrating that our approach admits a nice mathematical interpretation.

    \item \textbf{Coil-sketched EI for multicoil MRI --}  For a special but important medical imaging application, multicoil MRI, we design a special variant of sketched EI that utilizes the coil-sketching technique \cite{Oscanoa2024Coil}. We numerically observe a striking ``less is more" effect \cite{rudi2015less}, that is, our coil-sketched EI can significantly improve both the reconstruction accuracy and computational complexity at the same time over standard EI due to an implicit regularization effect of this tailored dimensionality reduction.

        \item \textbf{Parameter-efficient network adaptation --} Building on the EI and Sketched EI framework, we propose an even more computationally efficient approach for network adaptation {/Test Time Training}, which takes a pre-trained network and fine-tunes on the given inverse problem at hand. Our new approaches ({TTT-BN-EI and TTT-BN-SkEI}) select only a fraction of the network parameters (which typically are batch-norm layers) in the EI and Sketched EI. Our numerical results demonstrate the remarkable computational efficiency of this strategy in accelerating both the original EI and the sketched EI.

\end{itemize}




\section{Background and preliminaries}

\subsection{Problem setup}
In this work, we consider linear (imaging) inverse problems in the form of:
\begin{equation}
    y \approx Ax,
\end{equation}
where  $x \in \R^d$ is the image to estimate, $A \in \R^{n \times d}$ is the forward or measurement operator, and $y\in \R^n$ is the noisy measurement data. Many learning based methods  seek to learn a reconstruction network (set of network parameters denoted as $\theta$ here)
\begin{equation}
    \Ft(A^\dagger y) \rightarrow x
\end{equation}
which can be used to directly reconstruct image $x$ from a measurement input $y$.
Here we denote $A^\dagger$ as the pseudo-inverse of $A$, or a stable approximation of it (such as the FBP for X-ray CT). 
The network $\Ft$ can be trained in either a supervised or unsupervised manner. 
In the supervised case, the network is trained using paired datasets of measurements and corresponding ground-truth images $\{(y_1, x_1),..., {(y_i, x_i)}..., \}$, usually by minimizing a loss such as $\sum_{i}\| \Ft(A^\dagger y_i) - x_i \|^2$.
In an unsupervised setting, the network is trained without access to ground-truth images (i.e., using the measurements only), typically by enforcing data consistency with the measurement model and additional regularization or self-supervised objectives.

\subsection{Equivariant Imaging Regularization} 
Among unsupervised approaches for training deep imaging networks, the EI
framework proposed in \cite{chen2021equivariant,chen2022robust} was the first to
explicitly address the issue of learning beyond the range space of the forward
operator $A$ by exploiting inherent symmetries of imaging systems.  Let
$\{y_j\}_{j=1}^{N_{{Y}}}$ be a finite collection of
measurements, and let $\{T_{i}\}_{i=1}^{N_T}$ be
a finite set of system-specific transformations (e.g., rotations for CT).  
The EI loss can then be written in the empirical summation form as
\begin{multline}\label{e:eiloss}
    \theta^\star 
    =
    \arg\min_{\theta} 
    \mathcal{L}_{\mathrm{EI}}(\theta)
    :=
    \frac{1}{N_{{Y}} N_{{T}}}
    \sum_{j=1}^{N_{{Y}}}
    \sum_{i=1}^{N_{{T}}}
    \Big(
        \underbrace{\|y_j - A(\mathcal{F}_\theta(A^\dagger y_j))\|_2^2}_{\text{MC loss}}
       \\ +
        \lambda \,
        \underbrace{
        \big\|
            T_{i}\mathcal{F}_\theta(A^\dagger y_j)
            -
            \mathcal{F}_\theta\!\big(A^\dagger A\,T_{i}\mathcal{F}_\theta(A^\dagger y_j)\big)
        \big\|_2^2}_{\text{EI regularization}}
    \Big),
\end{multline}
where the first term is the measurement-consistency loss (MC), while the second term is the EI regularizer which allows the training program to learn in the null space of $A$. 
That is, recall that the space $\mathbb{R}^n$ can be decomposed as:
\begin{equation}
    \mathbb{R}^d = \text{range}(A^\top) \oplus \text{null}(A),
\end{equation}
where $\text{range}(A^\top) = \{ A^\top y, y \in \mathbb{R}^n \}$ and $\text{null}(A) = \{ x \in \mathbb{R}^d, Ax=0 \}$. For reconstruction in range space, the MC loss in~\eqref{e:eiloss} can be used for supervision. In the null space, since the measurements provide no information of $x$, the EI regularization in~\eqref{e:eiloss}  uses the output of $f$ as a noisy version of the GT image and its equivariant property, to supervise the reconstruction.

Once $\theta^\star$ is obtained, the underlying image can be easily reconstructed as $\mathcal{F}_{\theta^\star}(A^\dagger y)$. 
Variants of EI have been developed to enhance the robustness to measurement noise, using an additional G-SURE \cite{chen2022robust} or UNSURE \cite{tachella2024unsure} regularizer alongside EI. Meanwhile, extensions of EI have been proposed to new group actions tailored for different inverse problems \cite{wang2024perspective,wang2024fully,scanvic2023self}. In our work, we 
adopt the vanilla form of the EI regularizer, but the same principle can be easily extended for all these enhanced versions of EI.

\subsection{Operator Sketching and Stochastic Optimization.} 
A key technique that we use to accelerate the EI computation is operator sketching.  
Operator sketching, often linked with stochastic gradient descent techniques, has been widely applied for machine learning \cite{kingma2014adam,johnson2013accelerating,2015_Pilanci_Randomized,pilanci2017newton, tang2017gradient} and more recently in imaging inverse problems \cite{sun2019online,ehrhardt2024guide}. In the context of imaging inverse problems, given a general loss function:
\[
    \mathcal{L}(x) = \Phi(Ax,\, y),
\]
where  \(\Phi\) is a suitable
discrepancy measure, one may construct a collection of sketching operators
\(\{S_i\}_{i=1}^N\) (e.g. projections, compressions, randomized sketches, or
subsampling maps) such that
\[
    \mathcal{L}(x) = \sum_{i=1}^N \mathcal{L}_i(x),
    \qquad 
    \mathcal{L}_i(x) := \Phi(S_i A x,\, S_i y).
\]
This decomposition expresses the global objective as a sum of sketched
sub-objectives, each involving only a partial view of the data,
and naturally admits stochastic or mini-batch optimization schemes by sampling
from the set \(\{\mathcal{L}_i(x)\}_{i=1}^N\).
It is demonstrated in \cite{tang2020practicality} that the success of operator sketching depends on the spectral structure of the forward operator $A$. 
If $A$ has a fast decay in the singular value spectrum, then we can expect an order-of-magnitude acceleration in terms of computational complexity over deterministic methods such as proximal gradient descent or FISTA \cite{beck2009fast}.
Most computationally intensive imaging inverse problems fall into this category; for example, X-ray CT, multicoil MRI, and positron emission tomography all admit efficient applications of operator sketching and stochastic optimization. 

\section{Sketched Equivariant Imaging for Unsupervised Learning and Test-Time Training}
\subsection{Sketched EI} \label{sec:skei}

We now apply the operator sketch technique to accelerate EI–regularized
learning.  Let $\mathbb{S}=\{S_k\}_{k=1}^{N_S}$ be a collection of sketching
operators, where each $S_k$ is a matrix $m\times d$ that generally
satisfies $\mathbb{E}(S_k^{\!\top} S_k) = I$ and $m \ll n$.  
For any $S_k \in \mathbb{S}$ we define the sketched measurement 
$y_{S_k} := S_k y$, the sketched forward operator $A_{S_k} := S_k A$, and
$A_{S_k}^\dagger := (S_k A)^\dagger$.
By inserting the sketched quantities into Eq.~\eqref{e:eiloss}, we obtain the
empirical sketched EI loss
\begin{multline}
    \label{e:skeiloss}
\theta^\star
=
\arg\min_{\theta}
\mathcal{L}_{\mathrm{SkEI}}(\theta)
:=
\frac{1}{N_Y N_T N_S}
\sum_{j=1}^{N_Y}
\sum_{i=1}^{N_T}
\sum_{k=1}^{N_S}
\Big(
    \underbrace{
    \|S_k y_{ j} - A_{S_k}\mathcal{F}_\theta(A^\dagger y_j) \|_2^2
    }_{\text{Sketched MC loss}}
   \\ +
    \lambda\,
    \underbrace{
    \Big\|
        T_{g_i}\mathcal{F}_\theta(A^\dagger y_j)
        -
        \mathcal{F}_\theta\!\big(
            A_{S_k}^\dagger A_{S_k} \,
            T_{g_i}\mathcal{F}_\theta(A^\dagger  y_j)
        \big)
    \Big\|_2^2
    }_{\text{Sketched EI regularization}}
\Big).
\end{multline}
 It is important to know here that, unlike the conventional application of operator sketching, the sketched EI loss in Eq.~\eqref{e:skeiloss}
 is not equal to the original EI loss in Eq.~\eqref{e:eiloss}, because the regularizations in both loss functions are not the same. 
 Nevertheless, we provide the following theorem to demonstrate that the sketched EI regularization is an effective approximation to the original one:
 \begin{theorem}\label{thm:sketch}
   \textbf{(Approximation bound for Sketched EI regularization)} Suppose that the network $\Ft$ is $L$-Lipschitz, while $\|v\|_2 \leq r$, we have:
    \begin{equation}
      \|v - \Ft(A^\dagger A v)\|_2 - Lr\delta \leq  \frac{1}{N_S}
\sum_{k=1}^{N_S}\|v - \Ft(A_S^\dagger A_S v)\|_2 \leq \|v - \Ft(A^\dagger A v)\|_2 + Lr\delta
    \end{equation}
    almost surely, where $\delta$ is a constant only depending on the sketch-size $m$ and the choice of the sketching operator, and we denote here $v := T_g\Ft(A^\dagger y)$.
\end{theorem}
\leftline{\textit{Proof. The proof is provided in Appendix.}}

 The $L$-Lipschitz continuity assumption of the form $\|\Ft(p) - \Ft(q)\|_2 \leq L\|p- q\|_2, \ \ \forall p, q \in \mathcal{X}$ on the reconstruction network $\Ft$ is standard for the theoretical analysis of deep networks in imaging inverse problems. For example, in the convergence analysis of plug-and-play algorithms \cite{ryu2019plug,tan2024provably} and diffusion-based MCMC \cite{cai2024nf}, such types of assumptions have been used in pretrained denoisers or generative image priors based on deep networks for convergence proofs. The above theorem provides an upper bound and a lower bound that sandwich the sketched EI regularization with the original EI regularization, with a deviation $\delta$ that scales approximately as $O(1/\sqrt{m})$. 

In the appendix, we demonstrate that the theoretical approximation accuracy can be significantly improved for approximately low-rank measurement operators, which have a fast decaying spectrum. This observation is consistent with the findings of \cite{tang2020practicality} on suitable imaging applications of stochastic optimization. In the work of \cite{tang2020practicality}, it has been demonstrated both theoretically and numerically that stochastic gradient methods can only be effective for those inverse problems where the measurement operator $A$ has a fast decaying spectrum structure (approximately low-rank). For example, X-ray CT, multicoil MRI, and PET are all very good applications of stochastic gradient methods with minibatch sampling, while they all have a fast decaying spectrum. Our theory here suggests that we should expect similar efficiency for our Sketched-EI regularization.

This theoretical result, although preliminary and motivational, justifies that the proposed sketching scheme provides a good approximation for the original EI regularizer statistically and admits a nice mathematical interpretation.

In optimizing the sketched EI loss \eqref{e:skeiloss}, we employ a stochastic gradient–based procedure in which randomness arises solely from sampling the measurements and the sketching operators. An SGD update for minimizing \eqref{e:skeiloss} follows the structure illustrated in Alg.~\ref{alg:skei_sgd},
and other stochastic gradient based algorithms such as Adam can be developed similarly.

\begin{algorithm}[t]
\caption{Stochastic gradient iteration for optimizing the SkEI loss}
\label{alg:skei_sgd}
\begin{algorithmic}[1]

\State Sample $y_j$ for $j=1,\ldots,N_Y$, $T_i \sim \mathcal{T}$ and $S_k$ for $k=1,\ldots,N_S$.

\State 
Let $\ell_{\mathrm{MC}}
=
\big\|
S_k y_j
-
S_k A\,\mathcal{F}_\theta(A^\dagger y_j)
\big\|_2^2.$

\State 
Let $\ell_{\mathrm{SkEI}}
=
\Big\|
T_i \mathcal{F}_\theta(A^\dagger y_j)
-
\mathcal{F}_\theta\!\big(
(S_k A)^\dagger S_k A\, T_i \mathcal{F}_\theta(A^\dagger y_j)
\big)
\Big\|_2^2$.

\State Update model parameters:
$\theta \leftarrow \theta - \eta\, P \circ\nabla_\theta (\ell_{\mathrm{MC}}+\lambda \ell_{\mathrm{SkEI}})$ with $\eta$ being the stepsize and $P$ the online preconditioner (such as the one defined by Adam optimizer).

\end{algorithmic}
\end{algorithm}

\subsection{Equivariant Test Time Training with Sketching}

In many practical scenarios, measurement at deployment may deviate from the training distribution due to changes in acquisition geometry, noise characteristics, etc. As EI provides a self-supervised signal that depends solely on the forward operator and the measurement (i.e., no ground-truth), it can be naturally leveraged at inference time to adapt the network to the specific test sample. This test-time training (TTT) step allows the model to refine its prediction using the EI regularization computed on the observed measurement alone, thereby improving robustness and reconstruction fidelity on out-of-distribution measurements.

Assume that we have trained an NN model with external measurement data, 
\begin{equation}
    \mathcal{F}_{\theta^\star}(A^{\dagger}y) : \quad y \to x,
\end{equation}
where ${\theta}^{\star}$ denotes the parameters of the pre-trained model. Given a test sample $\tilde{y}$, we may adapt the pre-trained model $\mathcal{F}(\cdot, \tilde{\theta})$ to $\tilde{y}$ by minimizing the following loss:
\begin{multline}\label{e:eilossttt}
    \tilde{\theta}^\star 
    =
    \arg\min_{\theta} 
    \mathcal{L}_{\mathrm{EI}}(\theta)
    :=
    \frac{1}{N_{{T}}}
    \sum_{i=1}^{N_{{T}}}
    \Big(
        \underbrace{\|\tilde{y} - A(\mathcal{F}_\theta(A^\dagger \tilde{y}))\|_2^2}_{\text{MC loss}}
       \\ +
        \lambda \,
        \underbrace{
        \big\|
            T_{i}\mathcal{F}_\theta(A^\dagger \tilde{y})
            -
            \mathcal{F}_\theta\!\big(A^\dagger A\,T_{i}\mathcal{F}_\theta(A^\dagger \tilde{y})\big)
        \big\|_2^2}_{\text{EI regularization}}
    \Big).
\end{multline}
It should be noted here that the EI regularization is defined over the test data, rather than the training set,
so that we ensure that an out-of-distribution test data also satisfies the EI requirement. 
Eq.~\eqref{e:eilossttt} is optimized, and then we reconstruct the image as
$\mathcal{F}_{\tilde{\theta}^\star}\!\big(A^\dagger \tilde{y})$.  

It should be clearly that the EI loss at the TTT stage can be sketched as well. 
In fact, sketching becomes especially important for TTT because the adaptation must be performed rapidly on a single measurement, and sketching greatly reduces the per-iteration cost
and enables fast inference-time implementation.
In this case, the loss function becomes
\begin{multline}\label{e:skeilossttt}
    \tilde{\theta}^\star 
    =
    \arg\min_{\theta} 
    \mathcal{L}_{\mathrm{SkEI}}(\theta)
    :=
    \frac{1}{N_{{T}}}
    \sum_{i=1}^{N_{{T}}}
    \Big(
        \underbrace{
        \big\|
            y_{S_k} - A_{S_k}\,\mathcal{F}_\theta(A^\dagger \tilde{y})
        \big\|_2^2
        }_{\text{Sketched MC loss}}
        \\
        +
        \lambda \,
        \underbrace{
        \big\|
            T_{i}\mathcal{F}_\theta(A^\dagger \tilde{y})
            -
            \mathcal{F}_\theta\!\big(
                A_{S_k}^\dagger A_{S_k}\,
                T_{i}\mathcal{F}_\theta(A^\dagger \tilde{y})
            \big)
        \big\|_2^2
        }_{\text{{Sketched EI regularization}}}
    \Big),
\end{multline}
where $y_{S_k}$, $A_{S_k}$ and $A_{S_k}^\dagger$ are as defined in Section \ref{sec:skei}. We describe the Sketch-EI TTT in Algorithm \ref{alg:sketched-ei}.

\subsection{Parameter-Efficient Network Adaptation (NA) via Optimizing Only the Batch-Norms}

During TTT, one often restricts the optimization to a selected subset of the model parameters  to prevent overfitting and improve computational efficiency.
In this regard, inspired by the works of \cite{frankle2021training,mueller2024normalization}, we propose to update only the BatchNorm (BN) layers 
at the TTT stage. 
Batch Normalization (BN) is a widely used technique for stabilizing and
accelerating deep network training. A BN layer standardizes its input
activations by subtracting the batch mean and dividing by the batch standard
deviation, and then applies a learnable affine transformation to recover the
appropriate scale and offset. Formally, for an input signal $v$, BN computes
\[
\mathrm{BN}_{\theta_1,\theta_2}(v)
=
\theta_1\!\left(\frac{v - \mu}{\sigma}\right) + \theta_2,
\]
where $\mu$ and $\sigma$ are the empirical mean and standard deviation
estimated from the current batch, and $\theta_1$ and $\theta_2$ are learnable
parameters often referred to as the ``scale'' and ``shift''. During
test-time training, batch statistics $(\mu,\sigma)$ are recomputed from the
test input itself, thus capturing the feature distribution of the specific
measurement. By updating only the affine parameters $(\theta_1,\theta_2)$ while
keeping the rest of the network fixed, we provide a lightweight
mechanism to adapt the model at inference time.

  As a result of applying BN-only optimization on EI and SkEI for network adaptation tasks, we derive and name our new approaches BN-NA-EI and Sketched BN-NA-EI, which are remarkably efficient for network adaptation tasks.

\begin{algorithm}[H]
\caption{Sketched Equivariant Test-Time Training (SkEI-TTT)}
\label{alg:sketched-ei}
\begin{algorithmic}[1]
\Inputs{ Pre-trained model parameters $\theta^\star$, test measurement $\tilde{y}$, forward operator $A$, collection of sketching operators $\mathbb{S}$, set of transformations $\mathcal{T}=\{T_i\}_{i=1}^{N_T}$, stepsize $\eta$, regularization parameter $\lambda$.}

\Initialize{$\theta \leftarrow \theta^\star$}

\For{$t = 1, \dots, \text{MaxIter}$}
    \State Sample a sketching operator $S_k \sim \mathbb{S}$, transform $T_i \sim \mathcal{T}$.
    \State Sketched MC loss:
    \[
      \ell_{\mathrm{MC}} \leftarrow \big\| S_k \tilde{y} - S_k A \Fth(\Adagger \tilde{y}) \big\|_2^2
    \]
    
    \State Sketched EI regularization:
    \[
      \ell_{\mathrm{SkEI}} \leftarrow 
\Big\|
T_i \mathcal{F}_\theta(A^\dagger \tilde{y})
-
\mathcal{F}_\theta\!\big(
(S_k A)^\dagger S_k A\, T_i \mathcal{F}_\theta(A^\dagger \tilde{y})
\big)
\Big\|_2^2
    \]
    
  \State \textbf{Optimization step:}
    \State Update parameters via a stochastic gradient method (such as Adam):
    \[
      \theta \leftarrow \theta - \eta\, P\circ \cdot \nabla_\theta \big( \ell_{\mathrm{MC}} + \lambda \,\ell_{\mathrm{SkEI}} \big)
    \]
\EndFor

\State{\textbf{Output:} Adapted reconstruction $\tilde{x} = \Fth(\Adagger \tilde{y})$.}
\end{algorithmic}
\end{algorithm}

\subsection{Coil-Sketched EI for MRI}
Sketching projection $S$ is chosen so that the action of
forward operator $A$ is approximately preserved in a lower dimension,
and commonly used methods to construct it include subsampling, Gaussian/Rademacher, and so on \cite{woodruff2014sketching}.
For applications in multicoil MRI (see Section~\ref{sec: setup} for details), we present in this section
a special sketched EI tailored for multicoil data, utilizing coil sketching.

We denote by $N_c$ the number of receiver coils.  
Let $\mathbf{x} \in \mathbb{C}^{d}$ be the image to be reconstructed and  
$\mathbf{k} \in \mathbb{C}^{n}$ the stacked multi-coil k-space measurements,
where $n = N_c d$.

The multi-coil MRI forward model is
\begin{equation}
    \mathbf{k} = \mathbf{M}\,\mathbf{F}(\mathbf{C}\mathbf{x}), \label{e:mri}
\end{equation}
where $\mathbf{M} \in \mathbb{R}^{n \times n}$ is a diagonal sampling mask.
The coil-sensitivity operator $\mathbf{C}$ and Fourier operator $\mathbf{F}$
are defined as
\begin{equation*}
\begin{gathered}
\mathbf{C}
=
\begin{bmatrix}
\mathrm{diag}(C_1) \\
\vdots \\
\mathrm{diag}(C_{N_c})
\end{bmatrix}
\in \mathbb{C}^{n \times d},
\qquad
\mathbf{F}
=
\mathbf{I}_{N_c} \otimes \tilde{\mathbf{F}},
\end{gathered}
\end{equation*}
where $\tilde{\mathbf{F}} \in \mathbb{C}^{d \times d}$ is the discrete Fourier
transform and $\mathbf{I}_{N_c}$ is the $N_c \times N_c$ identity matrix.

\subsubsection{Classical Subsampling Sketch}

\begin{figure}[htb!]
    \centering
    \includegraphics[width=0.98\linewidth]{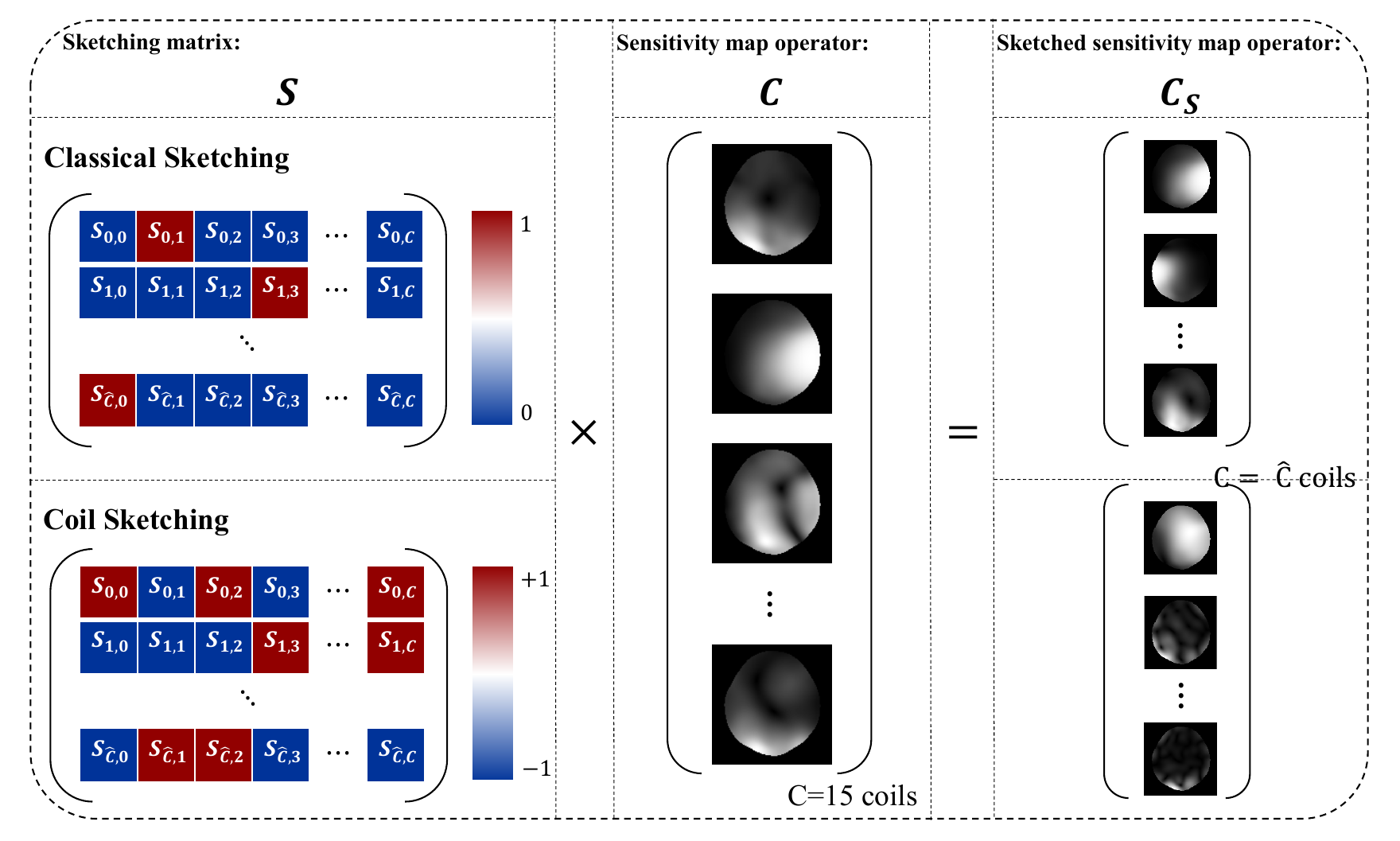}
    \caption{Structured sketching matrix of \cite{Oscanoa2024Coil}. The classical sketching matrix consists of a group binary mask, with each element being an all-ones matrix or an all-zeros matrix. Each row features exactly an all-ones matrix, and the remaining elements are all-zeros matrices. The coil sketching matrix, in contrast, comprises two blocks: one block is a group identity matrix (not shown in the figure), and the other block follows a group Rademacher distribution with probability $p=0.5$ as showed in the figure.}
    \label{fig:sketching theory}
\end{figure}

Classical sketching reduces the number of receiver coils by applying a
randomly generated block-selection matrix
\[
    \mathbf{S} \in \mathbb{R}^{M_c d \,\times\, N_c d},
    \qquad M_c < N_c.
\]
The sketched forward model is then
\begin{equation*}
    \mathbf{k}^s
    = \mathbf{M}_s \, \mathbf{F}_s \big( \mathbf{S} \mathbf{C} \mathbf{x} \big),
\end{equation*}
where $\mathbf{C} \in \mathbb{C}^{N_c d \times d}$ is the coil-sensitivity
operator and $\mathbf{S}$ selects $M_c$ coils out of the total $N_c$ coils.
The matrix $\mathbf{S}$ is a block-binary matrix composed of $d \times d$
blocks:
\[
    \mathbf{S}
    =
    \begin{bmatrix}
        \mathbf{S}_{11} & \cdots & \mathbf{S}_{1N_c} \\
        \vdots          & \ddots & \vdots \\
        \mathbf{S}_{M_c 1} & \cdots & \mathbf{S}_{M_c N_c}
    \end{bmatrix},
\qquad
\mathbf{S}_{ij} \in \{\mathbf{0}, \mathbf{I}_d\},
\]
where $\mathbf{0}$ is the $d \times d$ zero matrix and $\mathbf{I}_d$ is the
$d \times d$ identity matrix. Each row block of $\mathbf{S}$ contains exactly
one identity block, so that $\mathbf{S}$ selects exactly $M_c$ coils from the
$N_c$ available coils.
Figure~\ref{fig:sketching theory} illustrates the concept of classical
sketching. Hence one option for us is to apply SkEI directly on multicoil MRI via sweeping through the subsampling sketches. However we found that there is room for improvement for SkEI in this task, and next we are going to present a tailored variant of SkEI for multicoil MRI, namely the C-SkEI, utilizing coil-sketching \cite{Oscanoa2024Coil} for a refined dimensionality reduction.

\subsubsection{Coil Sketch}

Inspired by the work of~\cite{Oscanoa2024Coil}, we integrate their coil–
sketching algorithm into the Equivariant Imaging framework in order to
achieve a more efficient sketching strategy for multicoil MRI.

We start with the original multicoil $k$-space data
\[
    \mathbf{k} = [\mathbf{k}_1, \mathbf{k}_2, \ldots, \mathbf{k}_{N_c}],
\]
where each coil vector $\mathbf{k}_i \in \mathbb{C}^{d}$ contains the
$k$-space samples acquired by the $i$-th receive coil. Coil compression
is first applied to reduce the number of channels, with 
principal component analysis (PCA) being the predominant choice
\cite{Buehrer2007Array, Zhang2013Coil, HUANG2008133}.
For PCA, each $\mathbf{k}_i$ is first centered:
\[
    \overline{\mathbf{k}}_i
    =
    \mathbf{k}_i
    - \frac{1}{d}\sum_{j=1}^{d} k_i^j,
\]
and the covariance between two channels $\overline{\mathbf{k}}_i$ and
$\overline{\mathbf{k}}_j$ is computed as
\[
    v_{ij}
    =
    \mathrm{cov}(\overline{\mathbf{k}}_i, \overline{\mathbf{k}}_j)
    =
    \frac{\overline{\mathbf{k}}_i^{\mathrm{H}} \, \overline{\mathbf{k}}_j}{d-1},
\]
where $(\cdot)^{\mathrm{H}}$ denotes the Hermitian transpose.
Let $V$ be the resulting $N_c \times N_c$ covariance matrix. 
Now we use $Q_L$ to denote the matrix whose columns are the $L$ dominant eigenvectors
of $V$ ($L \le N_c$). These eigenvectors define the PCA-based coil
compression, producing $L$ virtual coils via
\[
    (\tilde{\mathbf{k}}_1, \ldots, \tilde{\mathbf{k}}_{L})
    =
    (\mathbf{k}_1, \ldots, \mathbf{k}_{N_c}) \, Q_L.
\]

Previous work indicates that many low–energy virtual coils can be removed
with negligible loss in reconstruction quality
\cite{Buehrer2007Array, Zhang2013Coil, HUANG2008133}.  Hence, the coil sketching algorithm constructs a sketching matrix that preserves the information in the high-energy virtual coils while selectively compressing only the low-energy components. This targeted strategy yields a more accurate approximation than approaches that reduce information uniformly across all virtual coils. The corresponding coil-sensitivity maps are estimated using the widely adopted ESPIRiT method~\cite{Uecker2013ESPIRiT}.

Let $R$ denote the number of preserved high–energy virtual coils and
let $M_c - R$ denote the number of sketched low–energy coils, where
$M_c \le L$ is the total number of output coils after sketching.
We begin by defining a coil-domain sketching matrix
\[
    S_c \in \mathbb{R}^{M_c \times L},
    \qquad
    S_c
    =
    \begin{bmatrix}
        \mathbf{I}_R & \mathbf{0} \\
        \mathbf{0}   & S_{\text{low}}
    \end{bmatrix},
\]
where $\mathbf{I}_R$ is the $R \times R$ identity matrix and
$S_{\text{low}} \in \mathbb{R}^{(M_c - R) \times (L - R)}$ (this block mixes only the low-energy virtual coils) is a random matrix
with i.i.d.\ Rademacher entries (values $\pm 1$ with equal probability),
following~\cite{Oscanoa2024Coil}.
To extend the sketching operation to the stacked coil–image space,
we construct the full sketching matrix using the Kronecker product:
\[
    S
    =
    S_c \otimes \mathbf{I}_d
    \in \mathbb{R}^{M_c d \times L d},
\]
where $\mathbf{I}_d$ is the $d \times d$ identity matrix.
The sketched coil–sensitivity operator is then defined as
\[
    \mathbf{C}_{S}
    =
    S \, \mathbf{C}_{L},
\]
with $\mathbf{C}_{L} \in \mathbb{C}^{L d \times d}$ denoting the
coil–sensitivity operator associated with the $L$ PCA virtual coils.
The complete procedure is summarized
in Algorithm~\ref{alg:coil-sketch} and illustrated in
Figure~\ref{fig:sketching theory}.

\begin{algorithm}[t]
    \caption{Sketched-EI for Multi-coil MRI with Coil sketching (C-SkEI)}
    \label{alg:coil-sketch}
    \begin{algorithmic}[1]
        \Inputs{$k$-space measurements $\mathbf{k}=[\mathbf{k}_1, \cdots, \mathbf{k}_C]$; EI regularization parameter $\lambda$}; set of transformations $\mathcal{T}=\{T_i\}_{i=1}^{N_T}$ 
        \Initialize{Set parameters $L$,  $\hat{C}$, $R$ and $S$, neural network $\mathcal{F}_{\theta}$};
        \For{$i=1, \cdots, C$}
            \State{$\overline{\mathbf{k}}_i = \mathbf{k}_i - \frac{\sum_{i=j}^n k_i^j}{n}$}
        \EndFor
        \State{Compute covariance matrix $V$ between $\overline{\mathbf{k}}_i$ and $\overline{\mathbf{k}}_j$};
        \State{Define $Q_L$ as the biggest $L$ eigenvectors of $V$;}
        \State{Compute $\hat{\mathbf{k}}_L = \mathbf{k} \cdot Q_L = [\tilde{\mathbf{k}}_1, \cdots, \tilde{\mathbf{k}}_L]$;}
        \State{Estimate the corresponding sensitivity maps $\mathbf{C}_L$ of the compressed $k$-space data $\hat{\mathbf{k}}_L$ with ESPIRiT;}
        \State{Form sketched sensitivity maps with $\hat{C}$ coils: $\mathbf{C_S}= \tilde{\mathbf{S}} \cdot \mathbf{C}_L$.}
        \State{Obtain sketched forward Fourier operator and inverse Fourier operator: $A_S = M_s \circ F_s \mathbf{C_S}$, $A^\dagger_S = (M_S \circ F_s)^{-1}$, $y_S = \hat{\mathbf{k}}_L$, by default: $z = A^\dagger \tilde{y}$};
\For{$t = 1, \dots, \text{MaxIter}$}
    \State Sample transform $T_i \sim \mathcal{T}$.
    \State Sketched MC loss:
    \[
      \ell_{\mathrm{MC}} \leftarrow \big\| y_S - A_S \Fth(\Adagger \tilde{y}) \big\|_2^2
    \]
    
    \State Sketched EI regularization:
    \[
      \ell_{\mathrm{SkEI}} \leftarrow 
\Big\|
T_i \mathcal{F}_\theta(A^\dagger \tilde{y})
-
\mathcal{F}_\theta\!\big(
A_{S}^\dagger A_S\, T_i \mathcal{F}_\theta(A^\dagger \tilde{y})
\big)
\Big\|_2^2
    \]
    
  \State \textbf{Optimization step:}
    \State Update parameters via a stochastic gradient method (such as Adam):
    \[
      \theta \leftarrow \theta - \eta\, P\circ \cdot \nabla_\theta \big( \ell_{\mathrm{MC}} + \lambda \,\ell_{\mathrm{SkEI}} \big)
    \]
\EndFor
    \end{algorithmic}
\end{algorithm}

Compared to the standard SkEI we proposed in previous subsections, our C-SkEI utilizes a single accurate sketch instead of multiple subsampling sketches which require us to sweep through them across stochastic gradient iterations. As we will observe in experiments, this helps in stabilizing the iterations and further increases the convergence rates numerically.

\clearpage

\section{Numerical Experiments}
In this section, we show the performance of the proposed method with two  examples: sparse-view CT and multicoil MRI reconstruction. 

\subsection{Problem Setup and Implementation}\label{sec: setup}
\subsubsection{Sparse-view CT imaging} Sparse-view CT  refers to a CT imaging setup in which only a small number of projection angles (views) are acquired — typically much fewer than in standard CT scans,
and this renders image reconstruction a highly ill-posed inverse problem.
In this example, the imaging physics of X-ray computed tomography (CT) is modeled by the discrete {Radon} transform,
where 100 scans (angles) are uniformly subsampled to produce the sparse-view sinograms (observations) denoted as $y$.
The forward operator $A$ is defined accordingly.
The filter back projection (FBP) is used to compute a stable approximation of $A^{\dagger}$. Training and test images are taken from the CT100 dataset~\cite{clark2013cancer} and resized to $512 \times 512$ pixels. Measurement data are generated by applying the Radon transform to images in the CT100 dataset to produce 100-scan sinograms.

While in this paper we focus on applying Sketched-EI regularization in TTT, we need to note that it can also be effectively applied in a standard unsupervised training setting. Hence we also implement an additional experiment on standard unsupervised training over a dataset of multiple examples in Section \ref{unsup}.\\

\subsubsection{Accelerated Multi-coil MRI imaging} MRI generates images of biological tissues by sampling the Fourier transform of the object $\mathbf{x}$, a domain called the $k$-space, where $k$ represents the spatial wave number. In multi-coil MRI systems, the signal measured by each individual coil is spatially modulated by that coil's distinct sensitivity profile. The forward operator $A$ for this problem is described in detail in Eq.~\eqref{e:mri}. We utilized multi-coil data from the NYU fastMRI Initiative~\cite{zbontar2018fastmri}, specifically acquisitions using 15 receiver coils. The fully reconstructed ground truth images were resized to a target matrix size of $128 \times 128$ pixels. The model was subsequently trained only on the corresponding complex-valued $k$-space measurements, which were retrospectively undersampled at a $4\times$ acceleration factor. We treat the real and imaginary parts of the complex-valued data as separate channels and, for visualization purposes, only display the magnitude image in all the plots. 

\subsubsection{Implementation details}
To provide a comprehensive comparison, we implemented several methods for these two applications: Deep Image Prior, standard EI, 
and the proposed sketched EI. In both EI based methods, we also demonstrated the effect of TTT. 
In both experiments, we used a U-Net~\cite{ronneberger2015u} to build $\mathcal{F}_{\theta}$ as suggested in~\cite{chen2021equivariant}.
We also use the residual U-Net architecture for all counterpart learning methods to ensure fair comparison. 
To implement the EI regularization in both examples, we choose $\mathcal{G}$ to be the group of rotations that span from 1 to 360 degrees (with $\vert \mathcal{G} \vert = 360$). 
 All of our experiments were performed with a NVIDIA RTX 4060ti GPU, along with \texttt{DeepInv} toolbox\footnote{\url{https://deepinv.github.io/deepinv/}}. All these  methods were implemented in PyTorch and optimized by Adam~\cite{kingma2014adam}, for which we set the learning rate to $5 \times 10^{-4}$. We trained all the methods over 5,000 iterations for CT and 10,000 iterations for MRI. 

It should also be noted that in TTT, we tested two strategies: (1) to adapt all the network parameters in the reconstruction model; (2) to update only the BatchNorm layers in the model. In addition, to ensure a fair comparison, hyper-parameters in all the considered methods are either manually tuned to optimality or automatically selected as described in the references.

\subsection{Numerical Results}
 In this section, we present our experimental results with detailed descriptions that demonstrate the effectiveness of our SkEI scheme.
\subsubsection{Sparse-view CT}\label{sec: exp-ct}
\begin{figure}[htp!]
    \centering

    \begin{subfigure}{0.18\textwidth}
        \centering
        \includegraphics[width=\textwidth]{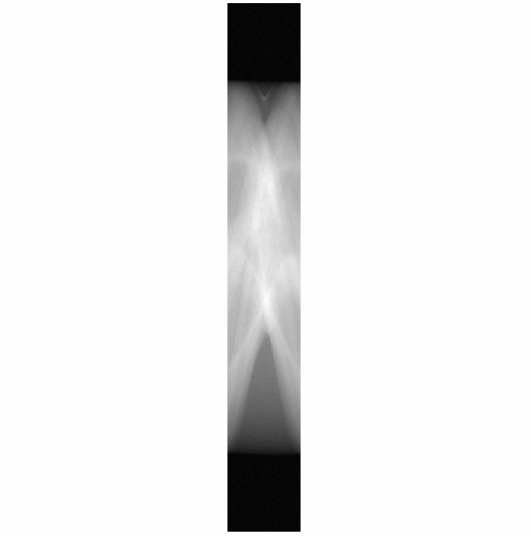} 
        \caption*{$y$}
    \end{subfigure}
    \begin{subfigure}{0.18\textwidth}
        \centering
        \includegraphics[width=\textwidth]{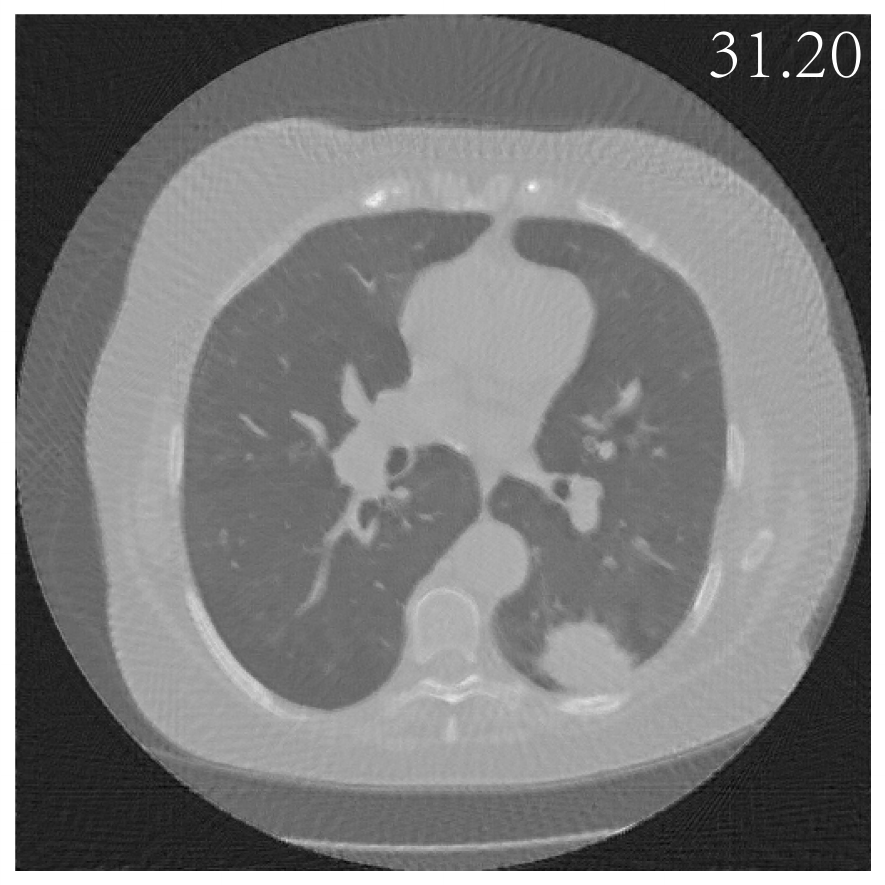} 
        \caption*{DIP-full}
    \end{subfigure}
    \begin{subfigure}{0.18\textwidth}
        \centering
        \includegraphics[width=\textwidth]{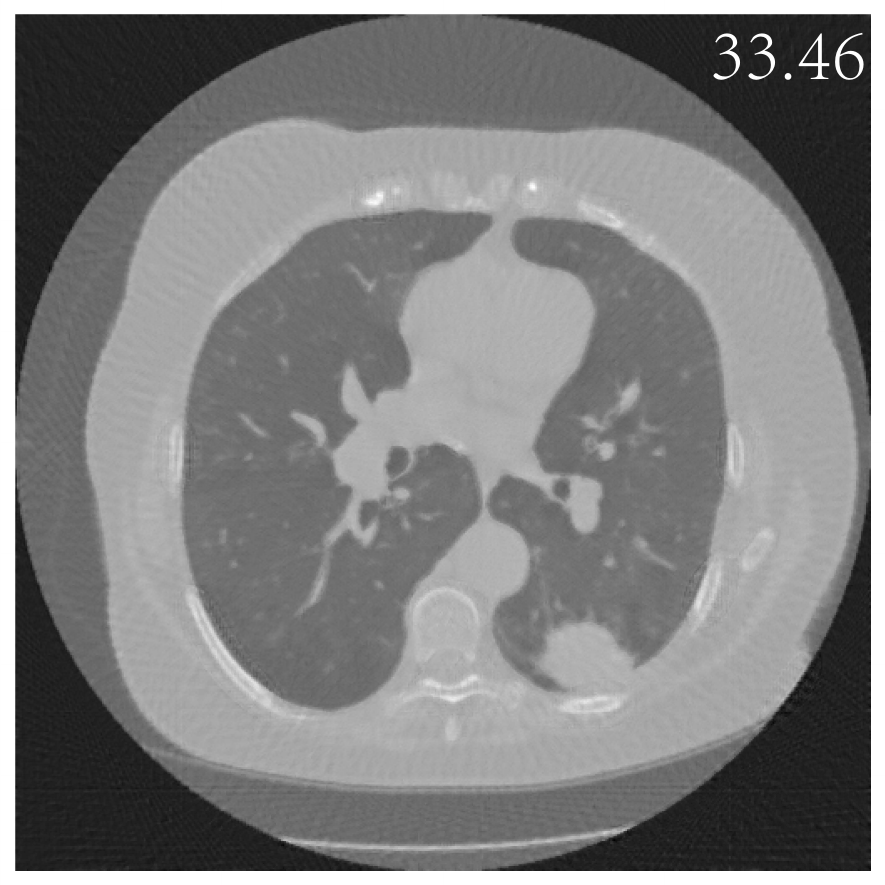} 
        \caption*{EI-full(*)}
    \end{subfigure}
    \begin{subfigure}{0.18\textwidth}
        \centering
        \includegraphics[width=\textwidth]{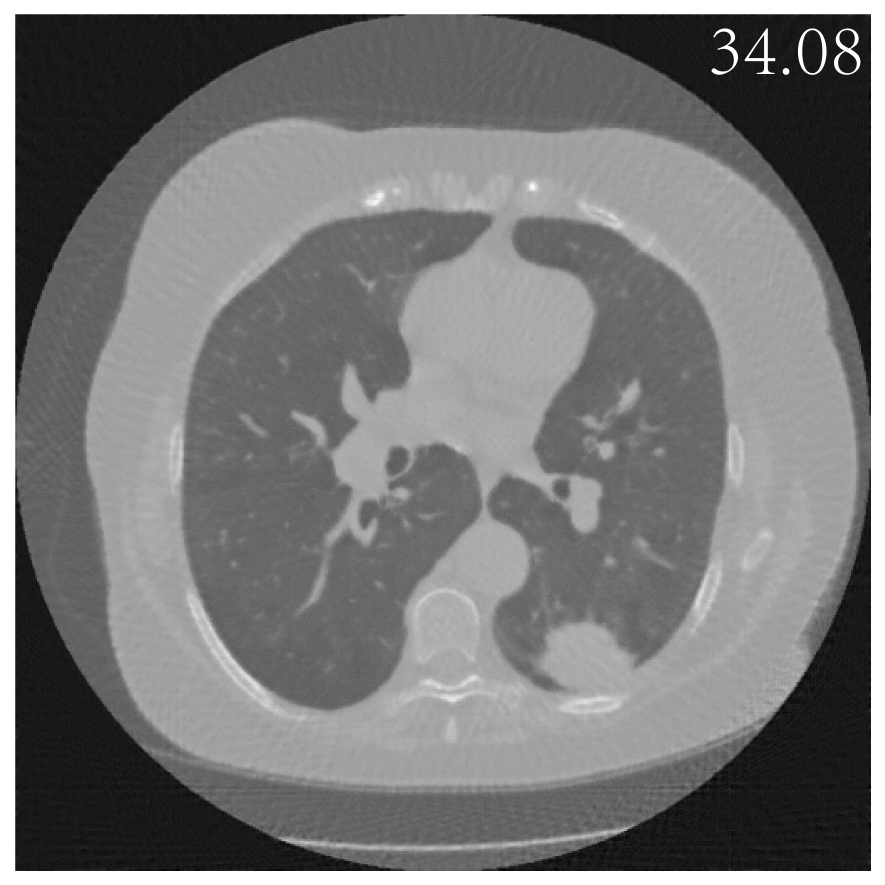} 
        \caption*{SkEI-10\%}
    \end{subfigure}
    \begin{subfigure}{0.18\textwidth}
        \centering
        \includegraphics[width=\textwidth]{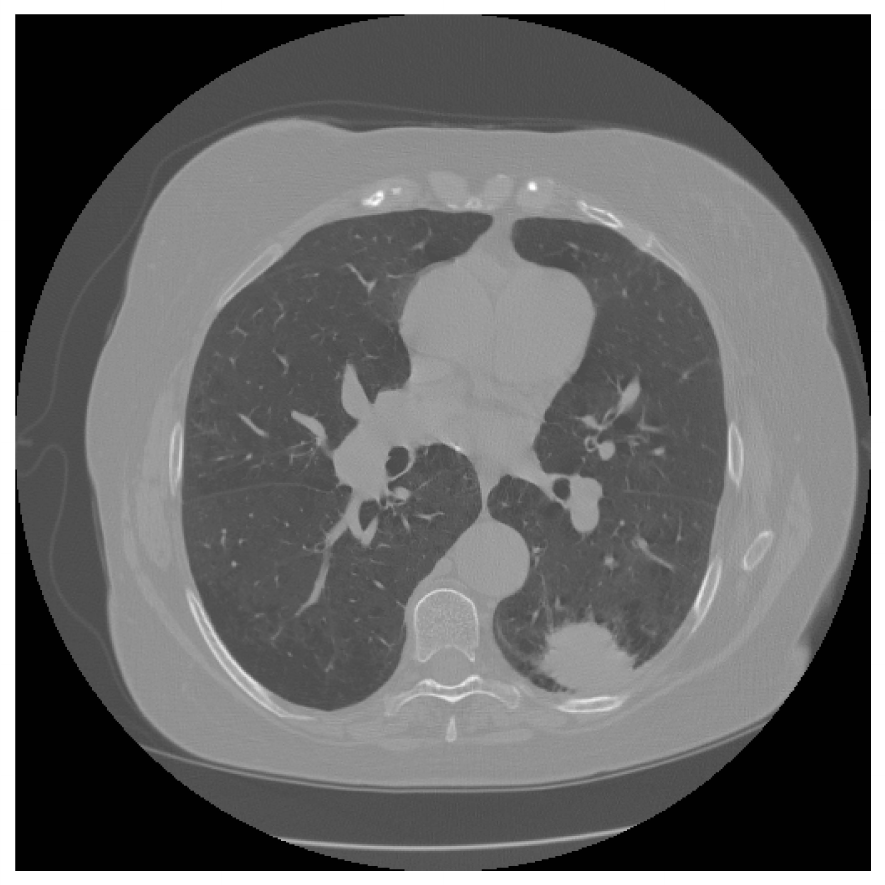} 
        \caption*{$x$ (GT)}
    \end{subfigure}

\caption{CT Images (with corresponding PSNR) reconstructed by DIP, EI and our Sketched EI. DIP-full and EI-full use 100 CT scans, while EI-Sketch-10\% means only 10\% of 100 CT scans used per iteration. (*) denotes the baseline.}
    \label{fig:reconstruction_comparison}
\end{figure}

\begin{figure}[htp!]
    \centering

    \begin{subfigure}{0.15\textwidth}
        \centering
        \includegraphics[width=\textwidth]{figures/3.2/512-CT/index-95/EI-DIP-100Views.pdf} 
        \caption*{EI-\\full (*)}
    \end{subfigure}
    \begin{subfigure}{0.15\textwidth}
        \centering
        \includegraphics[width=\textwidth]{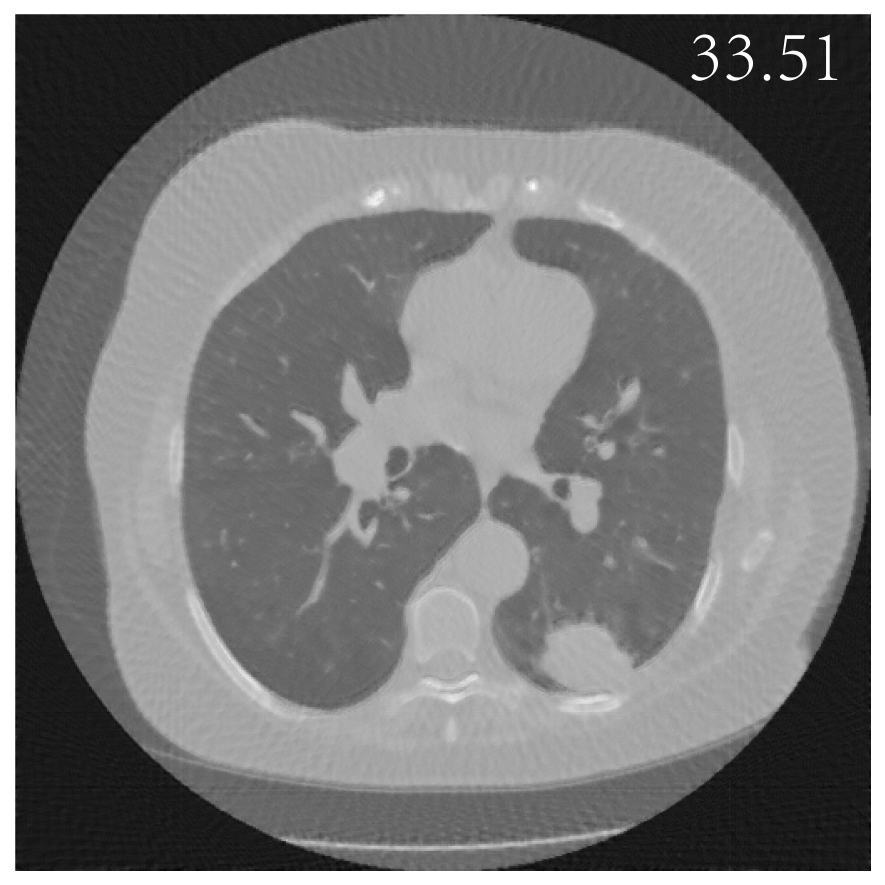} 
        \caption*{\centering SkEI-\\ Sketch-50\%}
    \end{subfigure}
    \begin{subfigure}{0.15\textwidth}
        \centering
        \includegraphics[width=\textwidth]{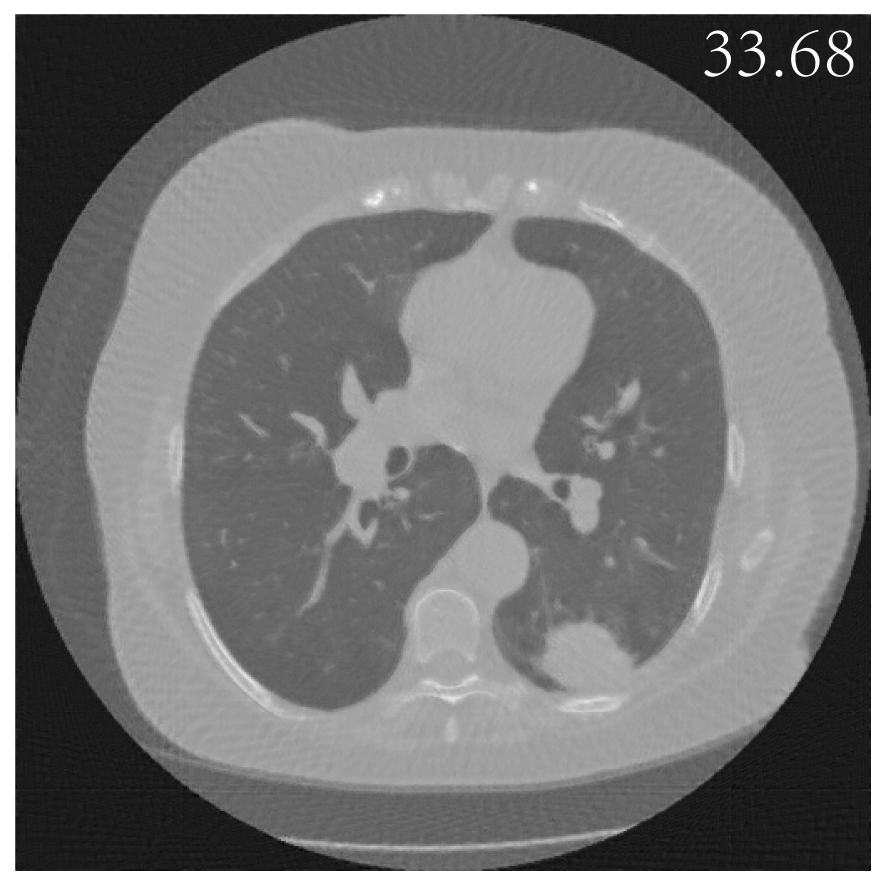} 
        \caption*{\centering SkEI-\\ Sketch-20\%}
    \end{subfigure}
    \begin{subfigure}{0.15\textwidth}
        \centering
        \includegraphics[width=\textwidth]{figures/3.2/512-CT/index-95/EI-DIP-10Views.pdf} 
        \caption*{\centering SkEI-\\ Sketch-10\%}
    \end{subfigure}
    \begin{subfigure}{0.15\textwidth}
        \centering
        \includegraphics[width=\textwidth]{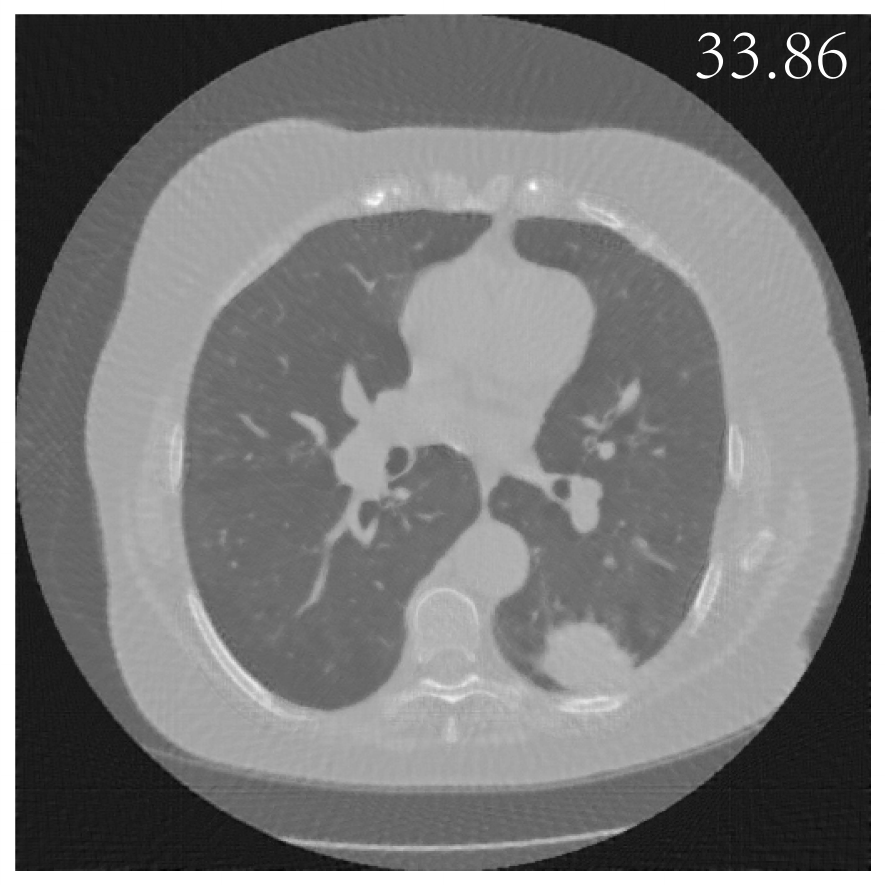} 
        \caption*{\centering SkEI-\\Sketch-5\%}
    \end{subfigure}
    \begin{subfigure}{0.15\textwidth}
        \centering
        \includegraphics[width=\textwidth]{figures/3.2/512-CT/index-95/GTx-512.pdf} 
        \caption*{\centering $x$ (GT) \\ \quad}
    \end{subfigure}

\caption{CT Images (with corresponding PSNR) reconstructed by Sketched EI, with different sketch sizes. EI-full uses 100 CT scans, while 50\%, 20\%, 10\% and 5\% means only 50\%, 20\%, 10\% and 5\% of 100 CT scans used per iteration. (*) denotes the baseline.} 
    \label{fig:sketched}
\end{figure}

\begin{figure}[htp!]
    \centering
    \begin{minipage}{0.48\textwidth}
        \centering
        \includegraphics[width=1.1\textwidth]{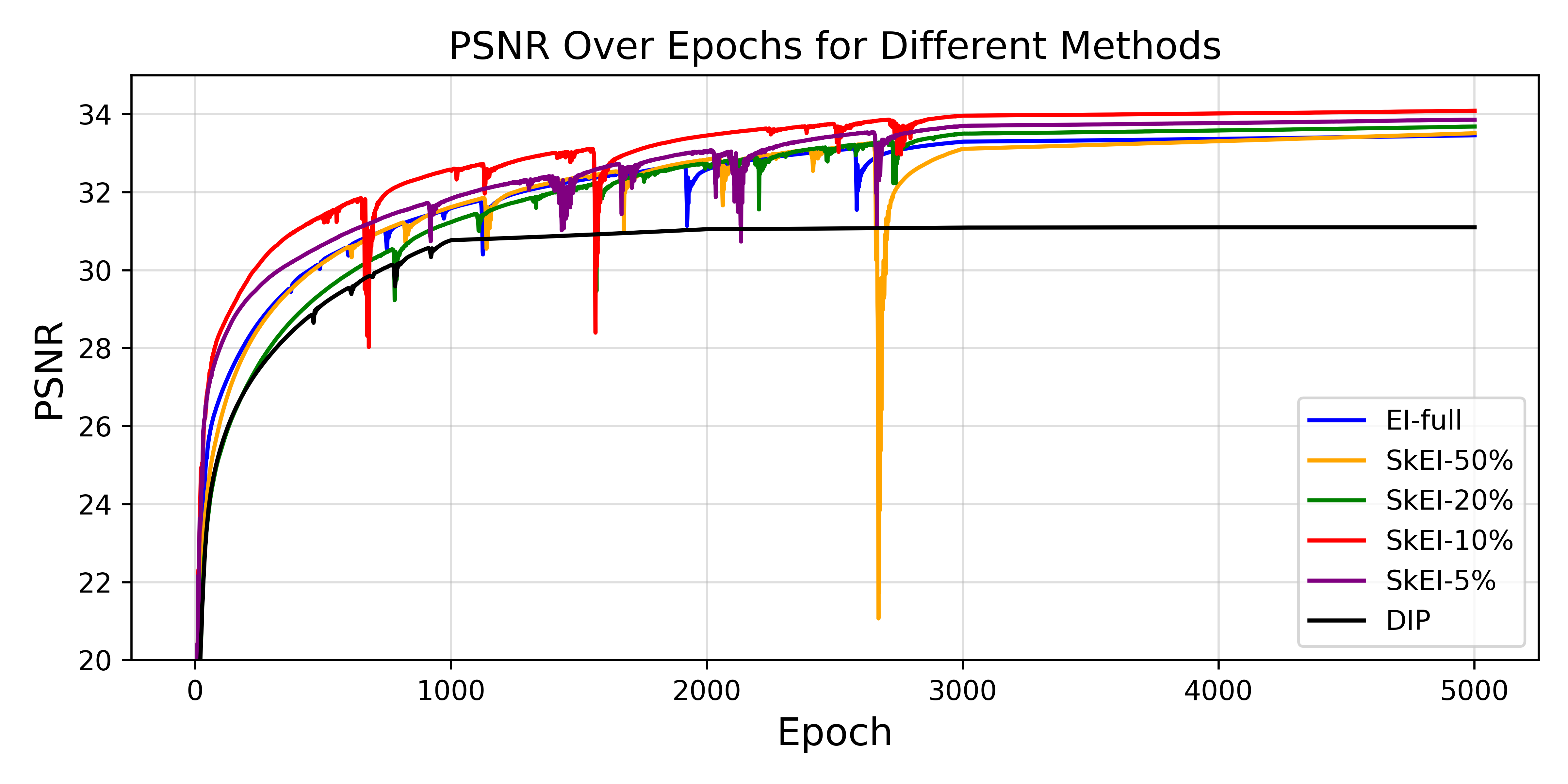} 
        \caption*{(a): Reconstruction accuracy curves for all the compared schemes.}
    \end{minipage}
    \hfill
    \begin{minipage}{0.48\textwidth}
        \centering
        \includegraphics[width=1.1\linewidth]{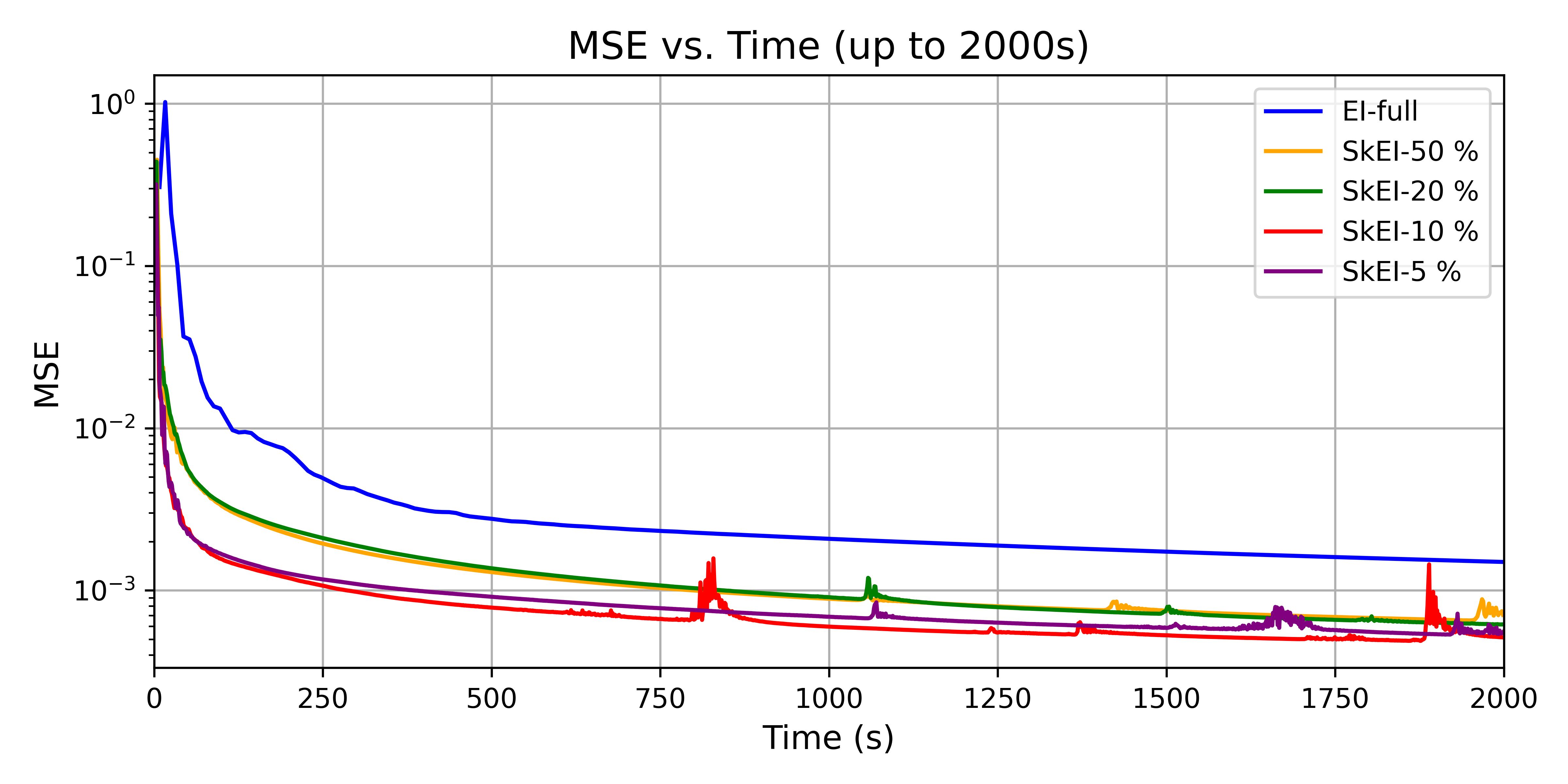}
        \caption*{(b): MSE with time up to 2000 seconds of the proposed EI with different sketched size.}
        \label{fig:mse-v.s.-time}
    \end{minipage}
    
\caption{PSNR and MSE comparisons of the proposed Sketched EI with DIP, EI methods. EI-full uses 100 CT scans, while 50\%, 20\%, 10\% and 5\% means only 50\%, 20\%, 10\% and 5\% of 100 CT scans used per iteration.}
\label{fig:result&time}
\end{figure}

We first evaluate the performance of SkEI using a single sample from the resized \textit{sparse-view CT 100} dataset and then further compare this method with DIP and EI.
For the {vanilla EI} method, we use the architecture suggested in~\cite{chen2021equivariant} to achieve the best performance and build the DIP using the same residual U-Net used in EI. For the sketched EI, we choose the subsampling sketch as our $S$, which splits the measurement operator into $N$ minibatches, $A_{S_1}, A_{S_2}... A_{S_N}$ from interleaved angles. In each iteration, we randomly select one of the minibatch and perform the update. We test on the choices $N=2,5,10,20$ respectively here.
As shown in Figure \ref{fig:reconstruction_comparison}, the baseline DIP frameworks deliver only modest reconstruction fidelity in our study, while the incorporation of the EI regularizer improves reconstruction precision. In particular, the sketch-guided EI model with $N=10$ achieves the highest reconstruction quality of all the evaluated methods. To further study the impacts of the sketching operation, we performed ablation experiments with four different sketch sizes as reported in Figure~\ref{fig:sketched} and Figure~\ref{fig:result&time}(a). We can observe that the results for Sketched EI are slightly better than full EI {(vanilla EI)}, for the number of minibatch splits chosen to be 2, 5, 10, or 20 (corresponds to 50\%, 20\%, 10\% and 5\% of origin, respectively), and reach the best at splits 10. The result would deteriorate if we choose to sketch over aggressively (20 splits in this setting), indicating a phase transition.

 Figure~\ref{fig:result&time}(b) illustrates that all methods achieve a significant decrease in mean square error (MSE) within the first 2000 seconds of training. The Sketched-EI exceeds the full EI (vanilla EI) in the convergence rate, achieving a faster decline in the MSE. Furthermore, among the sketched schemes employing minibatch splits of 2, 5, 10, and 20, the split-10 scheme demonstrates the most pronounced and rapid convergence.

\begin{figure}[htp!]
    \centering

    \begin{subfigure}{0.13\textwidth}
        \centering
        \includegraphics[width=\textwidth]{figures/3.2/512-CT/index-95/GTy-512.pdf} 
        \caption*{{\centering $y \sim$} \\ $\mathcal{N}(y^*, 0.1)$}
    \end{subfigure}
    \begin{subfigure}{0.13\textwidth}
        \centering
        \includegraphics[width=\textwidth]{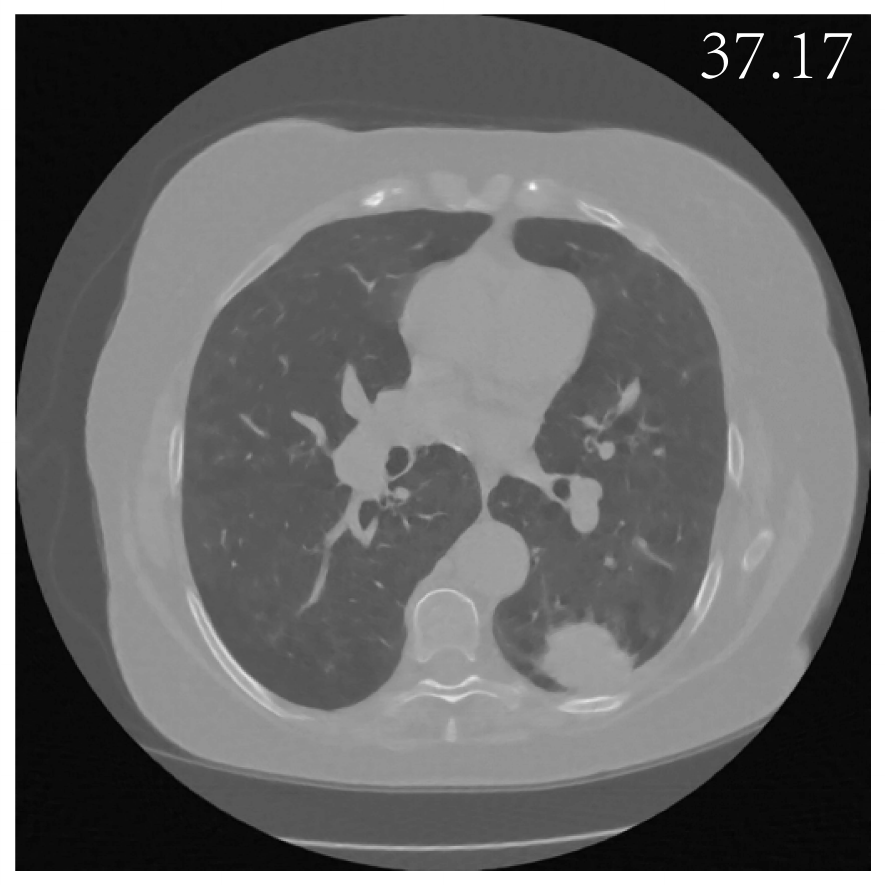} 
        \caption*{{TTT}-EI-\\full(*)}
    \end{subfigure}
    \begin{subfigure}{0.13\textwidth}
        \centering
        \includegraphics[width=\textwidth]{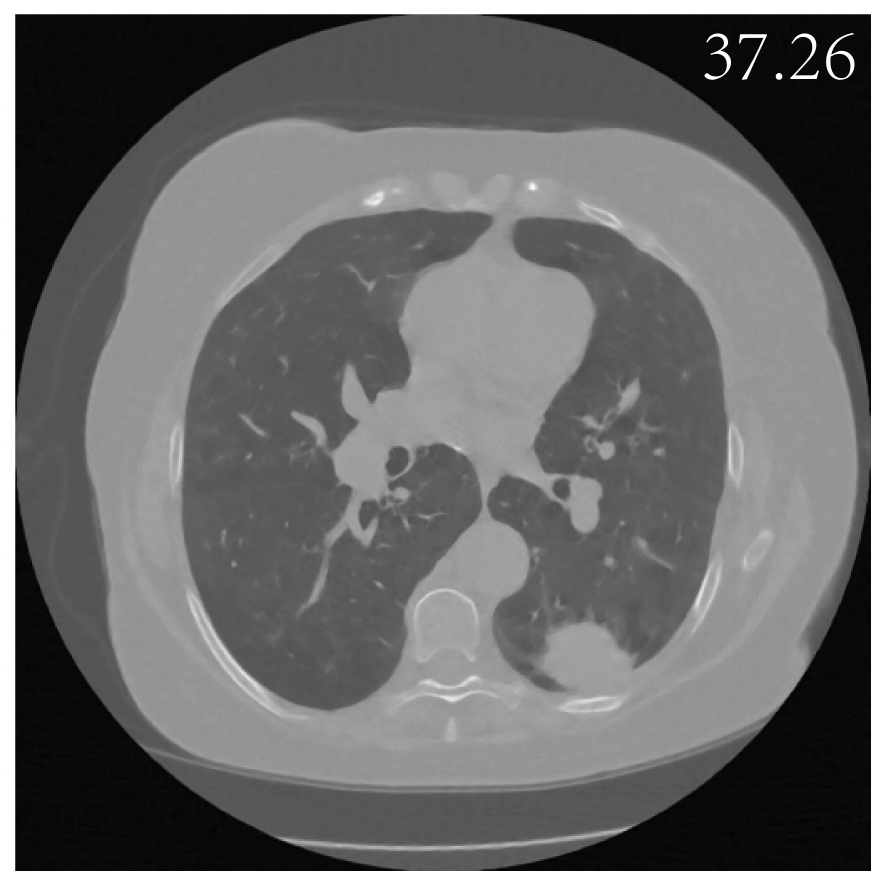} 
        \caption*{\centering{TTT}-\\SkEI-50\%}
    \end{subfigure}
    \begin{subfigure}{0.13\textwidth}
        \centering
        \includegraphics[width=\textwidth]{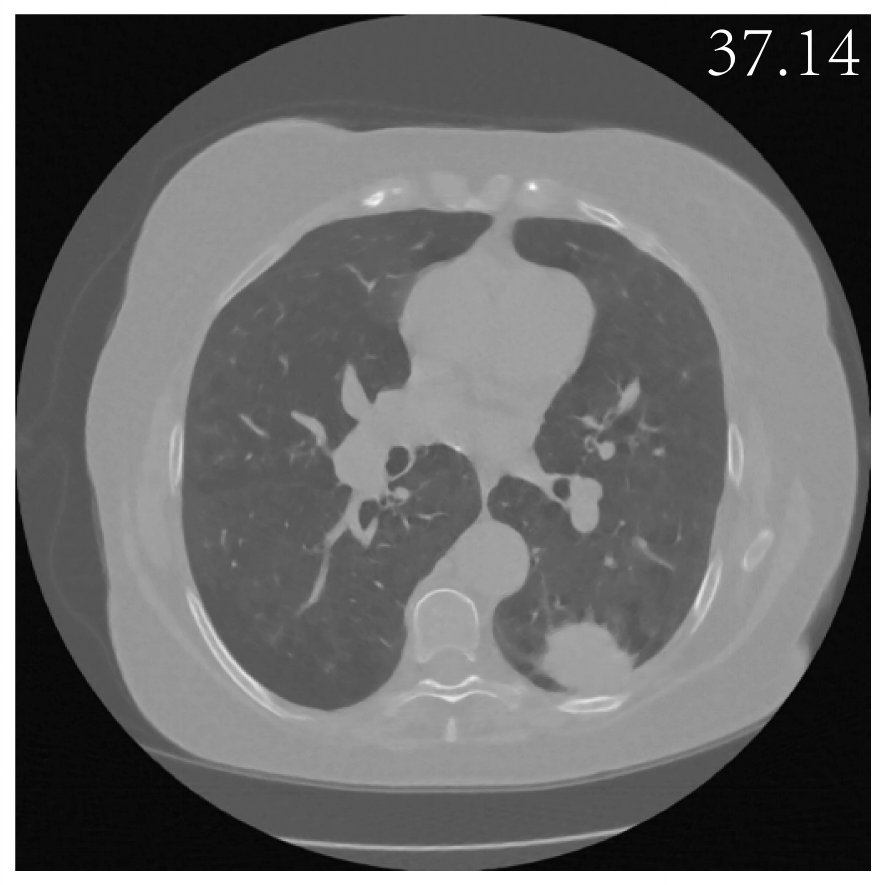} 
        \caption*{\centering{TTT}-\\SkEI-20\%}
    \end{subfigure}
    \begin{subfigure}{0.13\textwidth}
        \centering
        \includegraphics[width=\textwidth]{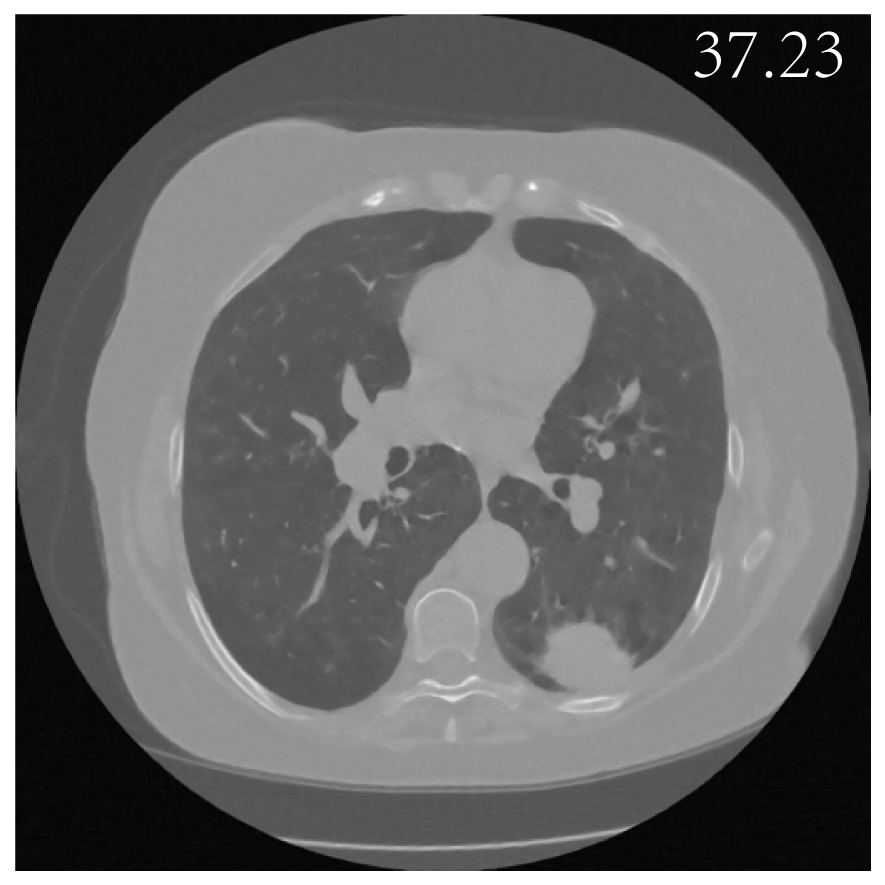} 
        \caption*{\centering{TTT}-\\SkEI-10\%}
    \end{subfigure}
    \begin{subfigure}{0.13\textwidth}
        \centering
        \includegraphics[width=\textwidth]{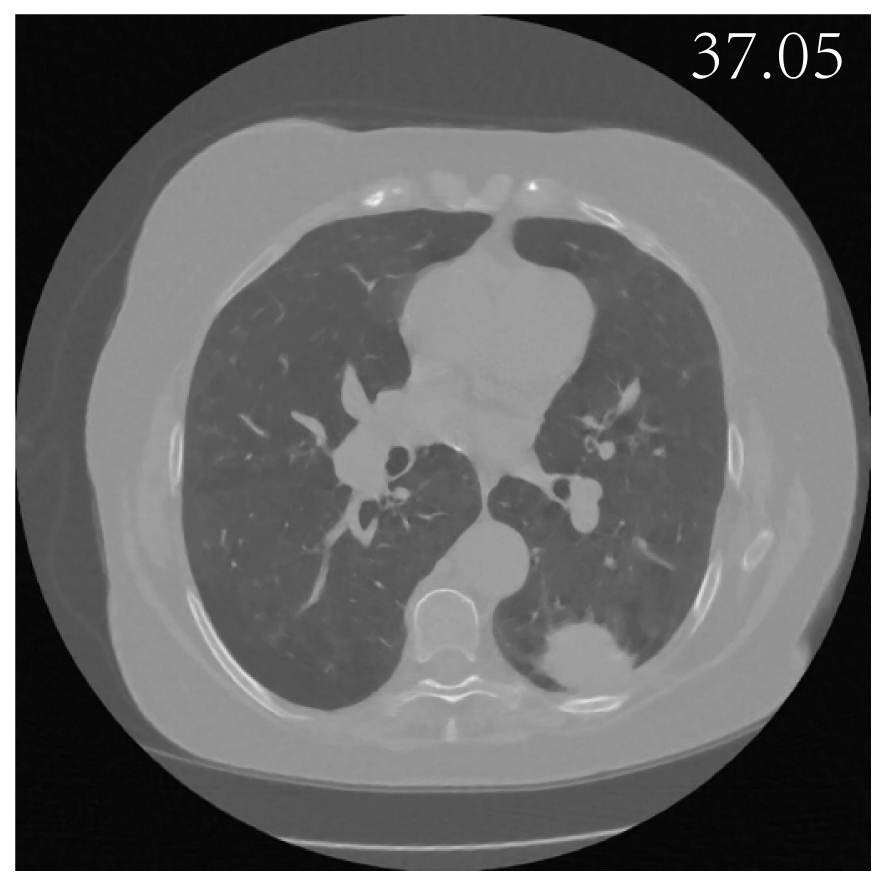} 
        \caption*{\centering{TTT}-\\SkEI-5\%}
    \end{subfigure}
    \begin{subfigure}{0.13\textwidth}
        \centering
        \includegraphics[width=\textwidth]{figures/3.2/512-CT/index-95/GTx-512.pdf} 
        \caption*{$x$ (GT)\\ \quad}
    \end{subfigure}

    \begin{subfigure}{0.13\textwidth}
        \centering
        \includegraphics[width=\textwidth]{figures/3.2/512-CT/index-95/GTy-512.pdf} 
        \caption*{{\centering $y \sim$} \\ $\mathcal{N}(y^*, 0.1)$}
    \end{subfigure}
    \begin{subfigure}{0.13\textwidth}
        \centering
        \includegraphics[width=\textwidth]{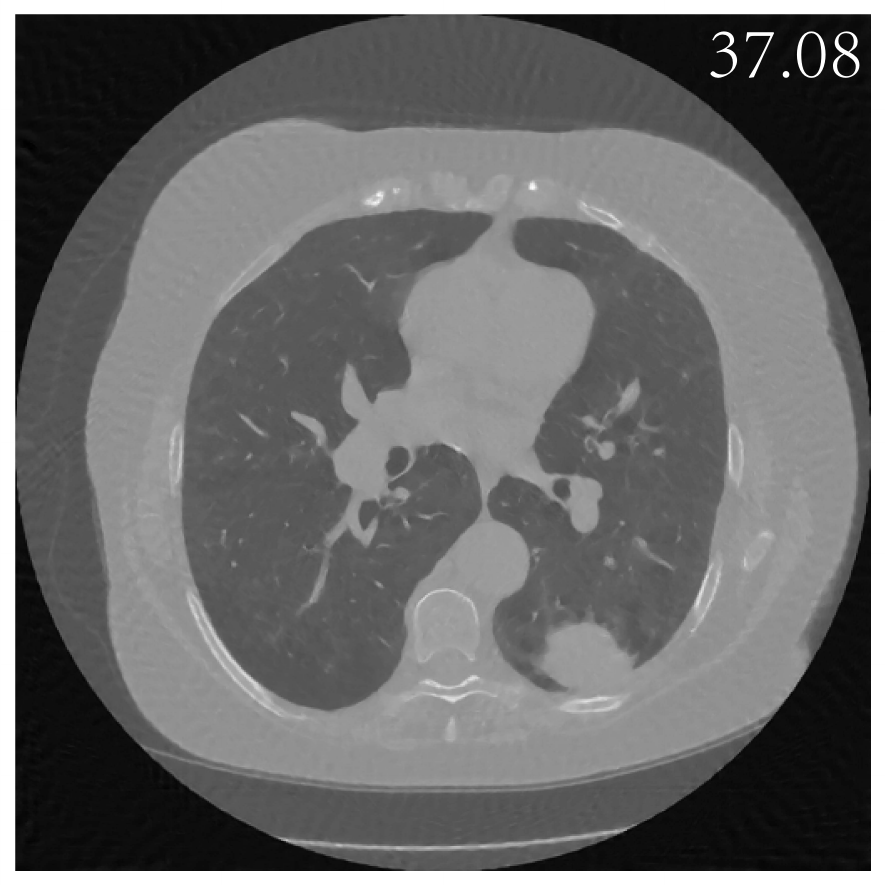} 
        \caption*{\centering {TTT-BN-EI-} \\ full(*)}
    \end{subfigure}
    \begin{subfigure}{0.13\textwidth}
        \centering
        \includegraphics[width=\textwidth]{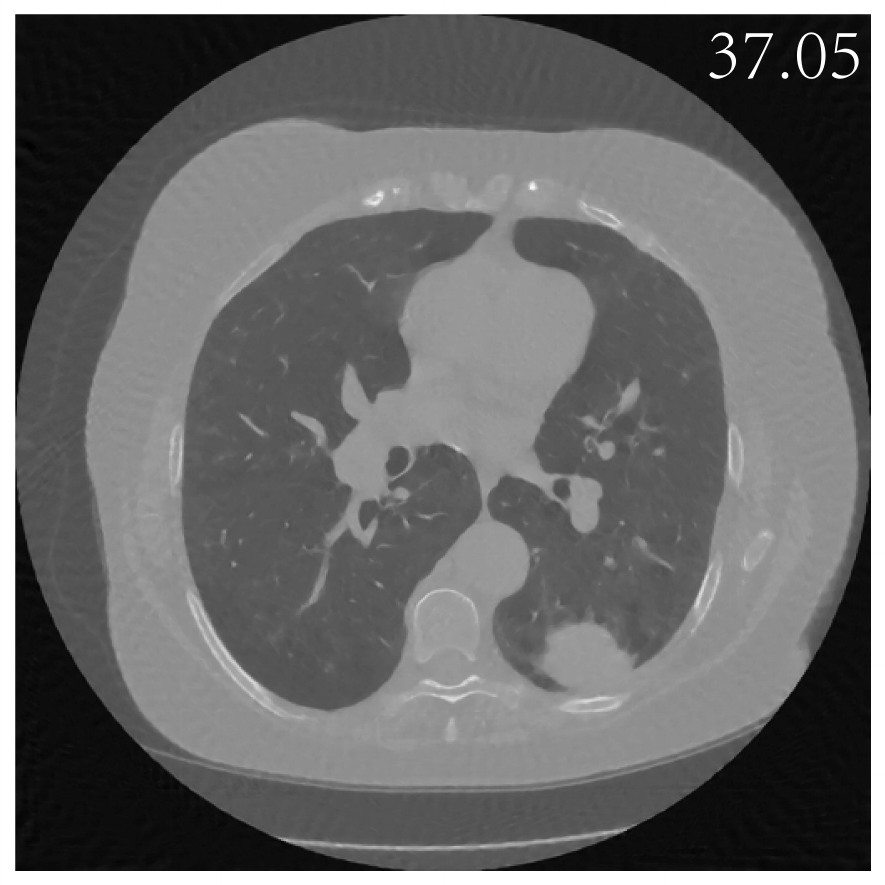} 
        \caption*{\centering {TTT-BN}- \\ SkEI-50\%}
    \end{subfigure}
    \begin{subfigure}{0.13\textwidth}
        \centering
        \includegraphics[width=\textwidth]{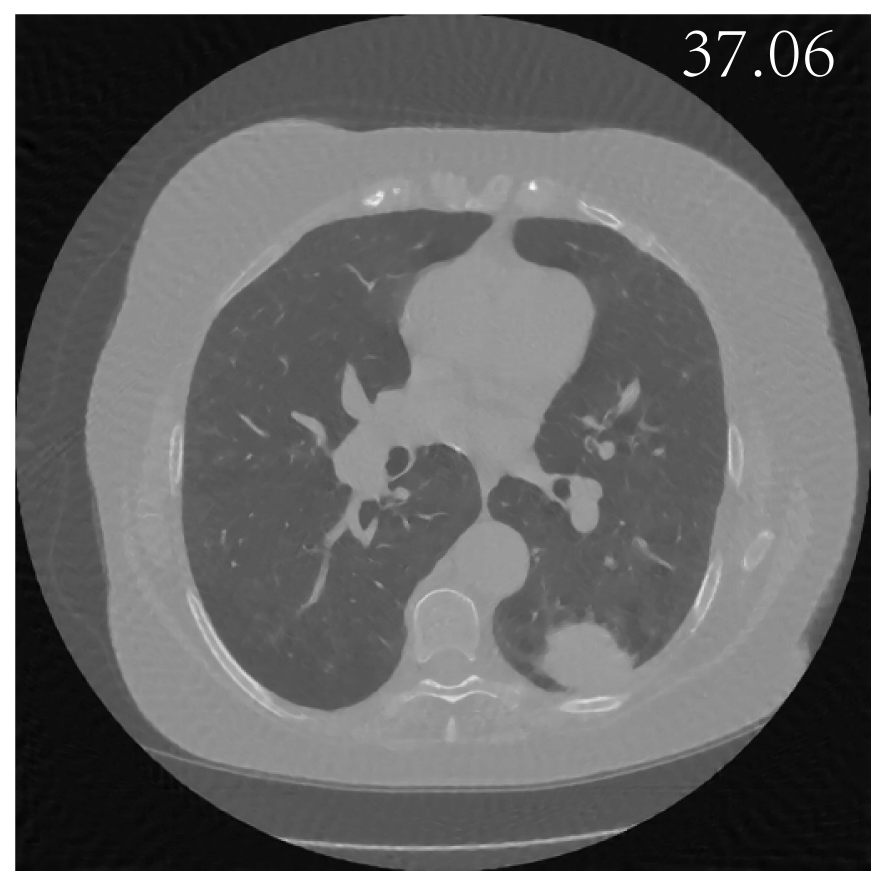} 
        \caption*{\centering {TTT-BN}- \\ SkEI-20\%}
    \end{subfigure}
    \begin{subfigure}{0.13\textwidth}
        \centering
        \includegraphics[width=\textwidth]{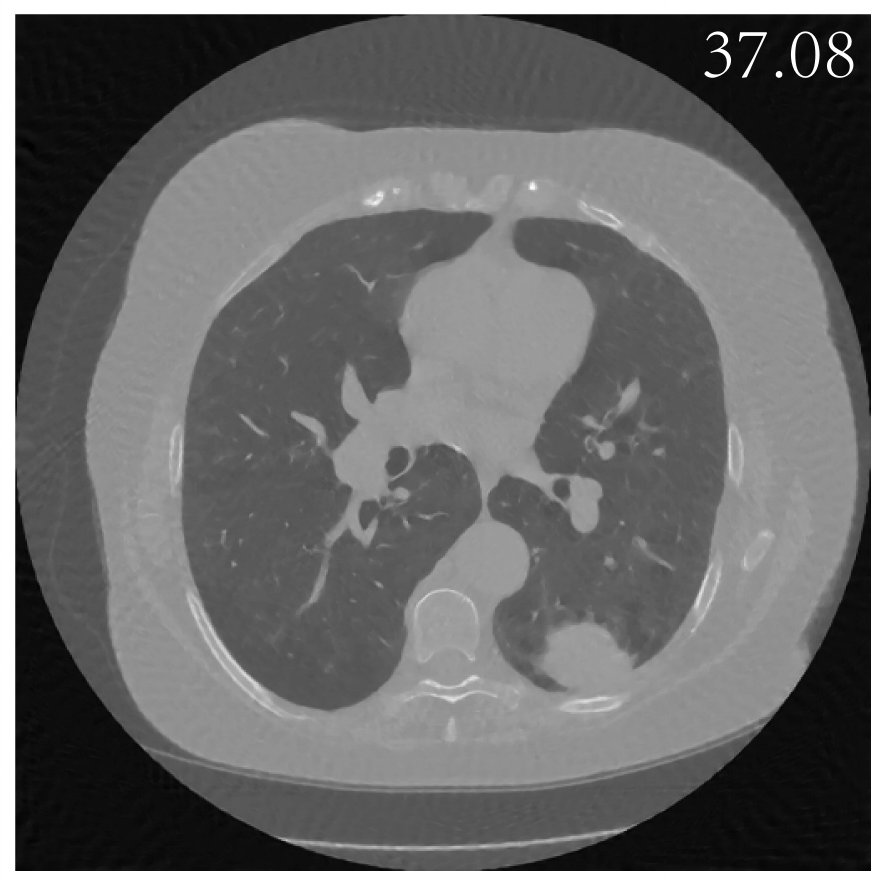} 
        \caption*{\centering {TTT-BN}- \\ SkEI-10\%}
    \end{subfigure}
    \begin{subfigure}{0.13\textwidth}
        \centering
        \includegraphics[width=\textwidth]{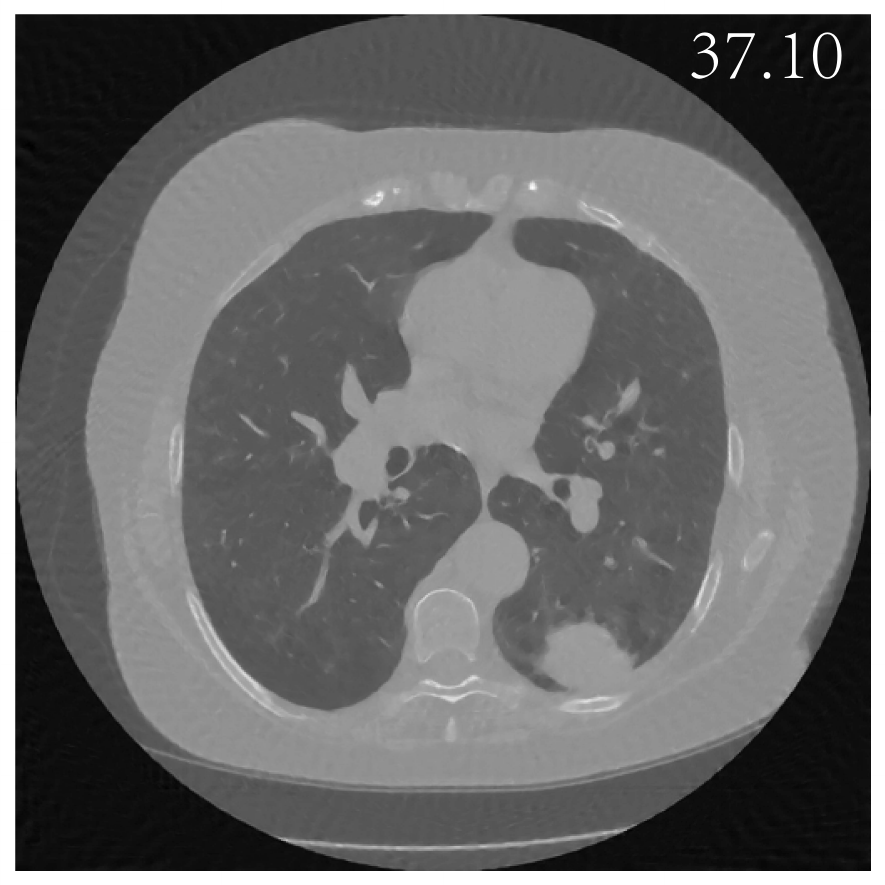} 
        \caption*{\centering {TTT-BN}- \\ SkEI-5\%}
    \end{subfigure}
    \begin{subfigure}{0.13\textwidth}
        \centering
        \includegraphics[width=\textwidth]{figures/3.2/512-CT/index-95/GTx-512.pdf} 
        \caption*{$x$ (GT) \\ \quad}
    \end{subfigure}

\caption{CT image reconstructions by Sketched EI in Network {Test Time Training} with a single noisy measurement. Top row shows reconstructions of fine-tuning entire network (TTT-EI) while bottom row fine-tuning only the BatchNorm (TTT-BN-EI) layers, both with various sketch size. '-full' uses 100 CT scans, while 50\%, 20\%, 10\% and 5\% means only 50\%, 20\%, 10\% and 5\% of 100 CT scans used per iteration. (*) denotes the baseline.}
    \label{fig:NA-EI-DIP}
\end{figure}

We further investigate the application of the Sketched-EI method to the Test-Time training task~\cite{darestani2022test}, which involves adjusting a pretrained model to accommodate variations in new data or tasks. 
Specifically, we applied the model pre-trained~\footnote{\url{https://drive.google.com/drive/folders/1Io0quD-RvoVNkCmE36aQYpoouEAEP5pF}} using the first 90 samples' measurements of \textit{CT 100} dataset to reconstruct an unseen CT image (index 95 sample in \textit{CT 100} dataset) using only noisy measurement with a known Gaussian noise level of 0.1 (domain shift), with the experimental results presented in the top row of Figure~\ref{fig:NA-EI-DIP}.
The visualized results of the Test Time Training (TTT-EI) demonstrate that the sketched EI method achieves a comparable reconstruction performance compared to the standard EI method without sketching, while reaching superior reconstructions compared to their counterparts, which were training from scratch.

\begin{figure}[htp!]
    \centering
    \begin{minipage}{0.48\textwidth}
        \centering
        \includegraphics[width=1.0\textwidth]{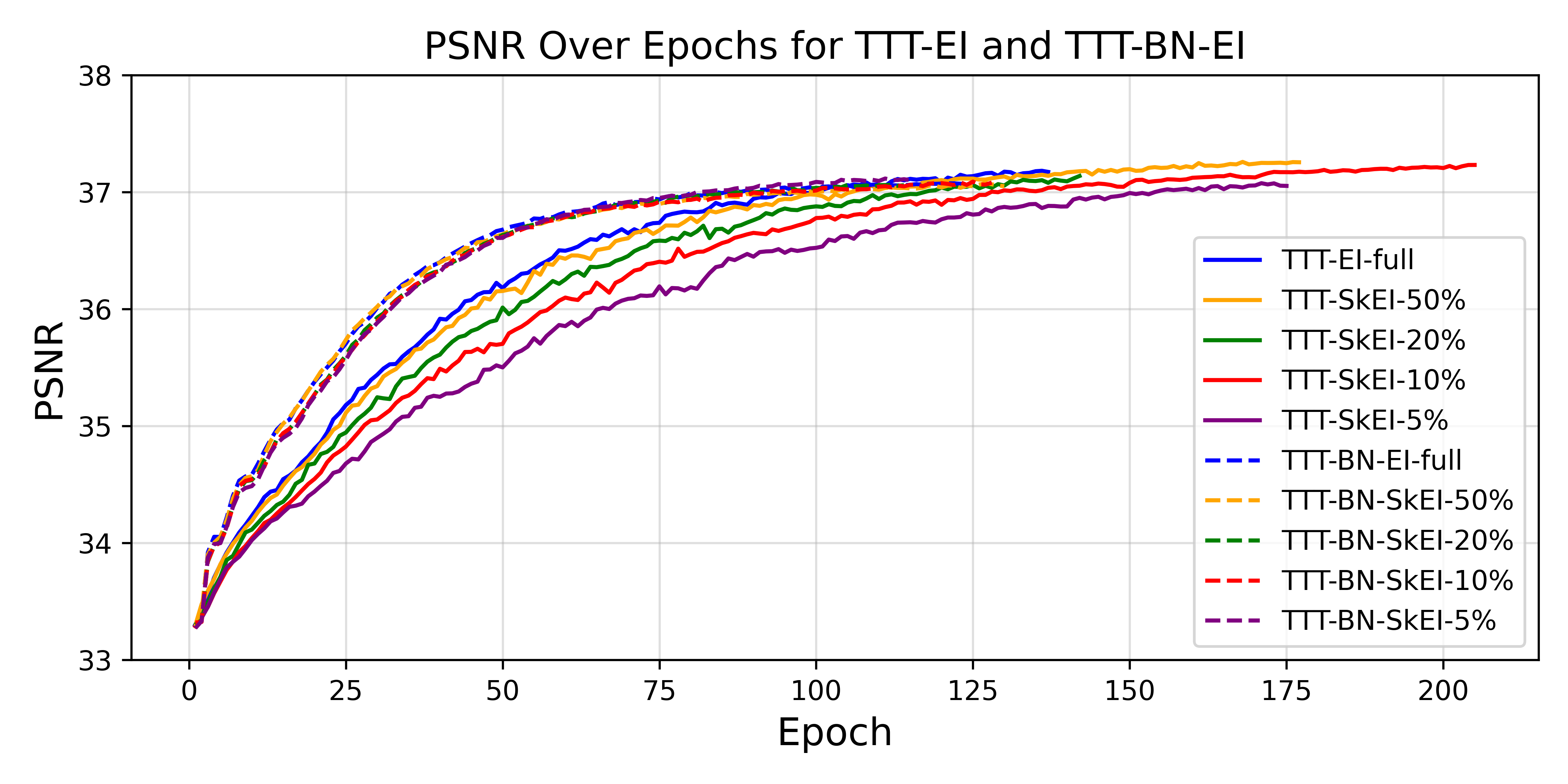} 
        \caption*{(a): Trend of PSNR with epoch in TTT-EI and TTT-BN-EI schemes.}
    \end{minipage}
    \hfill
    \begin{minipage}{0.48\textwidth}
        \centering
        \includegraphics[width=1.1\textwidth]
        {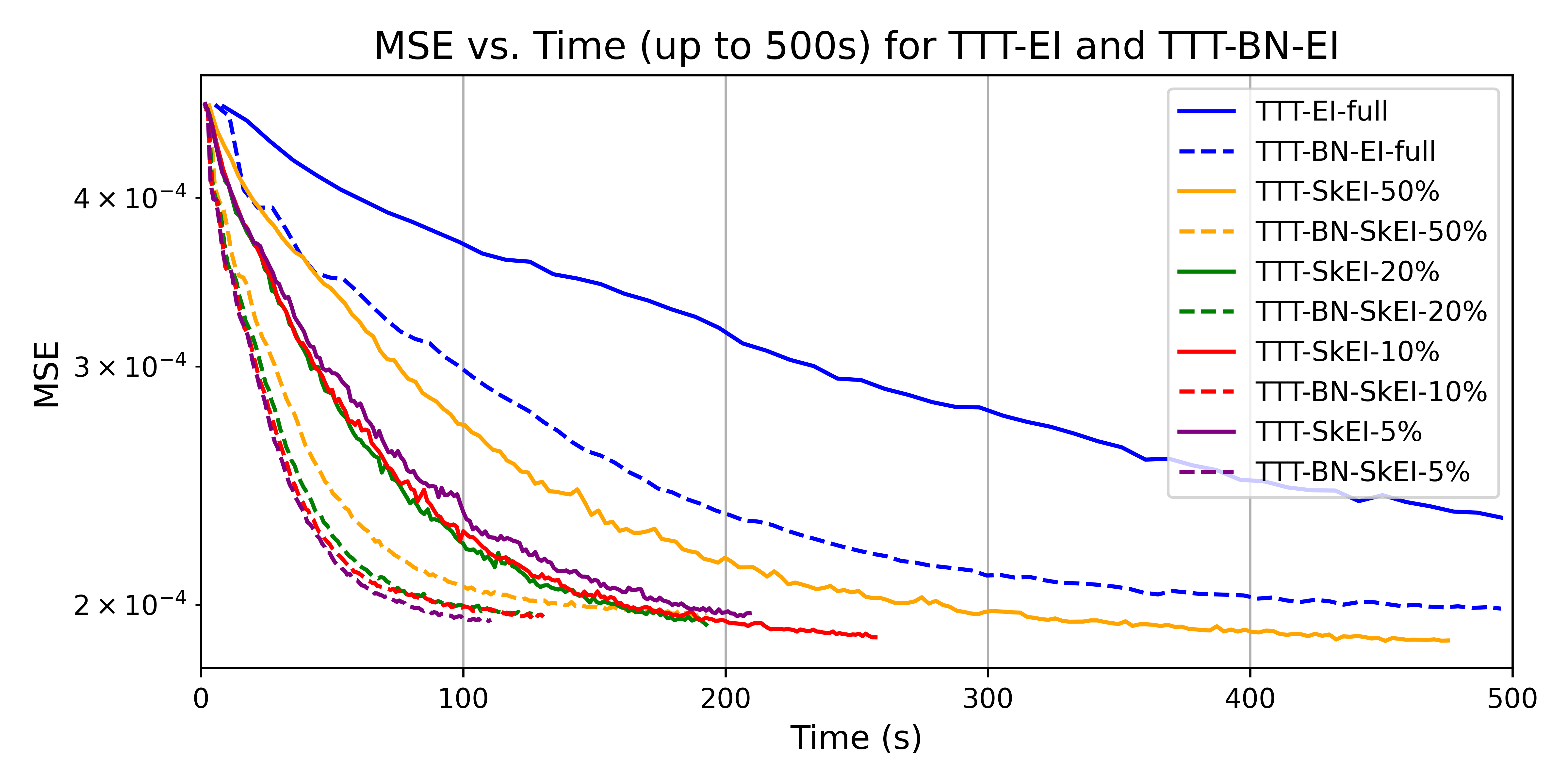}
        \caption*{(b): Trend of MSE with time (up to 500 seconds) in TTT-EI and TTT-BN-EI schemes.}
    \end{minipage}
    
\caption{Curves of PSNR and MSE of Test Time Training (TTT-EI) and Test Time Training with BathNorm only (TTT-BN-EI) schemes in CT reconstruction by Sketched EI. '-full' uses 100 CT scans, while 50\%, 20\%, 10\% and 5\% means only 50\%, 20\%, 10\% and 5\% of 100 CT scans used per iteration.}
\label{fig:NA-EI-DIP-time&psnr}
\end{figure}

In the next experiment, we fixed all other model parameters during {Test Time Training}, fine-tuning only the BatchNorm layer parameters (TTT-BN-EI). 
The results, as shown in Figure~\ref{fig:NA-EI-DIP}, indicate that the fine-tuning {only of the BatchNorm layers achieves a comparable performance compared to the fine-tuning of the entire model} for the {Test Time Training} task, while both fine-tuning schemes achieve superior performance than their counterparts through direct training. We continue by providing some insights of both two fine-tuning methods by comparing their convergence speeds. Figure~\ref{fig:NA-EI-DIP-time&psnr}(a) shows that the BatchNorm only scheme (TTT-BN-EI) converges faster than fine-tuning all network parameters (TTT-EI), while both methods required significantly less computation time to achieve superior performance compared to the directly trained counterparts. Moreover, the sketch operation further speeds up the convergence in both methods, as illustrated in Figure~\ref{fig:NA-EI-DIP-time&psnr}(b).

\begin{table}[htp]
\renewcommand{\arraystretch}{1.2}
\centering
\begin{tabular}{c|cccc}
\toprule
& PSNR & Time (s) / Epoch &  Trainable Param ($\times 10^7$) & Training Epochs \\
\hline
 \xgx{TTT}& & (noise level 0.1) &   \\
 \hdashline[1pt/1pt]
EI-full (*) & 37.17 & 9.56 &  3.45 & 137 \\
{\small SkEI-50\%} & 37.26 & 2.66 &  3.45 & 177 \\
{\small SkEI-20\%} & 37.14 & 1.34 &  3.45 & 142 \\
{\small SkEI-10\%} & 37.23 & 1.25 &  3.45 & 205 \\
{\small SkEI-5\%} & 37.05 & 1.18 &  3.45 & 175 \\
\midrule
\xgx{TTT-BN} & & & (noise level 0.1) &   \\
 \hdashline[1pt/1pt]
EI-full & 37.08 & 5.45 &  0.0014 & 125 \\
{\small SkEI-50\%} & 37.05 & 1.46 &  0.0014 & 130 \\
{\small SkEI-20\%} & 37.06 & 1.09 &  0.0014 & 114 \\
{\small SkEI-10\%} & 37.08 & 1.00 &  0.0014 & 128 \\
{\small SkEI-5\%} & 37.10 & 0.95 &  0.0014 & 115 \\
\bottomrule
\end{tabular}
\caption{Further comparisons between TTT-EI and TTT-BN-EI schemes in CT reconstruction. EI-full(*) uses 100 CT scans, while 50\%, 20\%, 10\% and 5\% means only 50\%, 20\%, 10\% and 5\% of 100 CT scans used per iteration. (*) denotes the baseline.}
\label{tab: NA v.s. BN}
\end{table}

\begin{figure}[htp!]
    \centering

    \begin{subfigure}{0.15\textwidth}
        \centering
        \includegraphics[width=\textwidth]{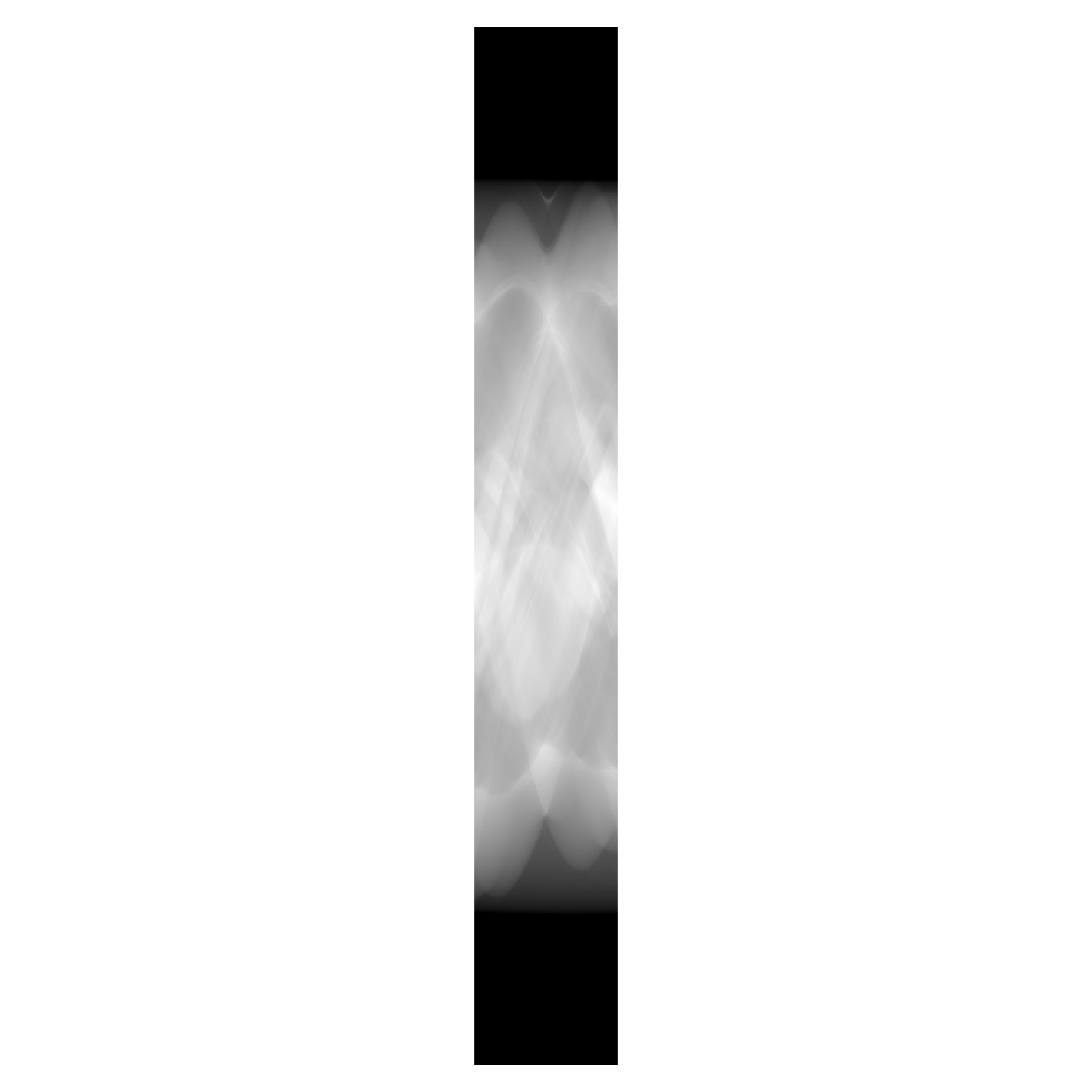} 
        \caption*{$y \sim \mathcal{N}(y^*, 0.05)$}
    \end{subfigure}
    \begin{subfigure}{0.15\textwidth}
        \centering
        \includegraphics[width=\textwidth]{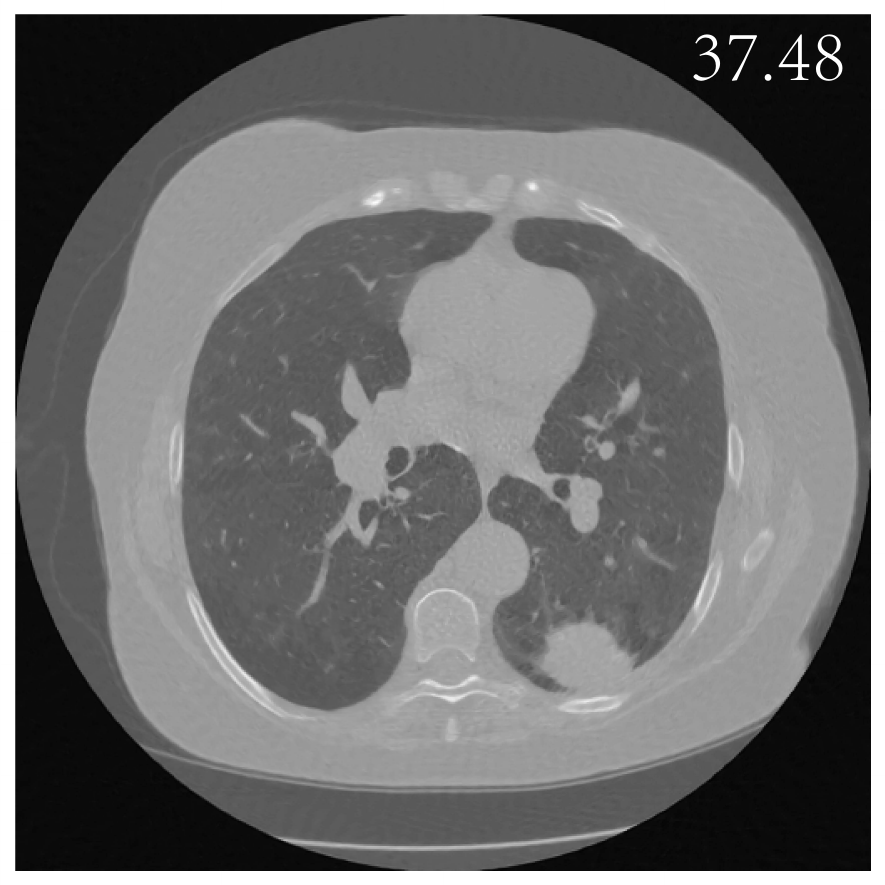} 
        \caption*{{TTT-EI-full(*)}}
    \end{subfigure}
    \begin{subfigure}{0.15\textwidth}
        \centering
        \includegraphics[width=\textwidth]{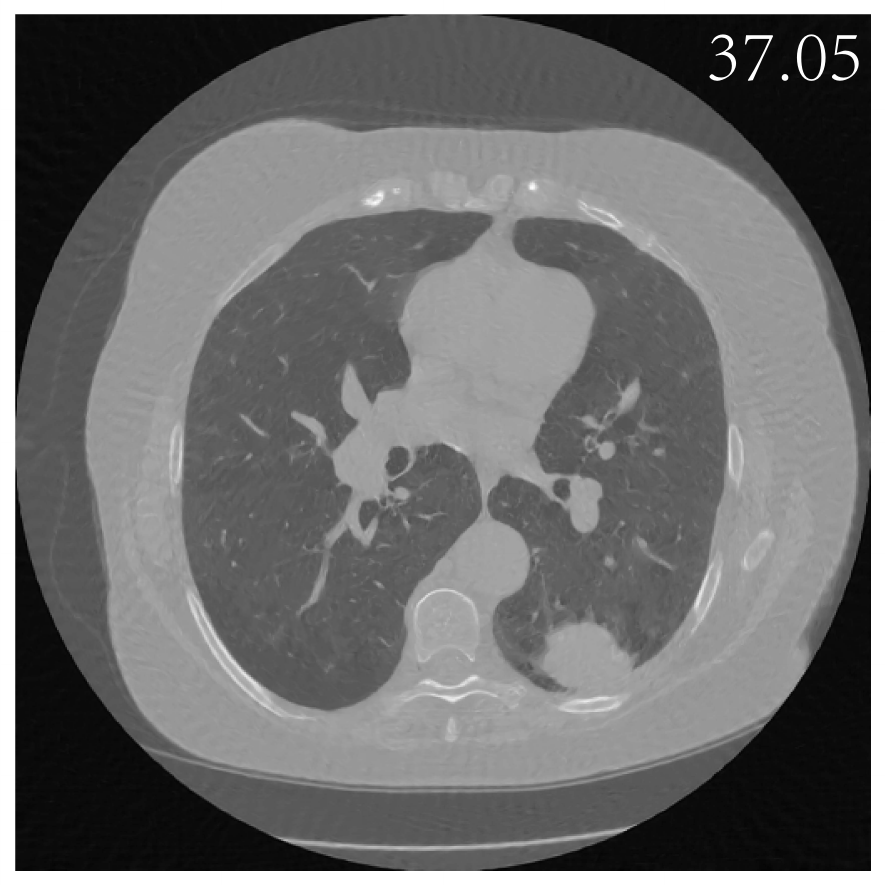} 
        \caption*{TTT-SkEI-5\%}
    \end{subfigure}
    \begin{subfigure}{0.15\textwidth}
        \centering
        \includegraphics[width=\textwidth]{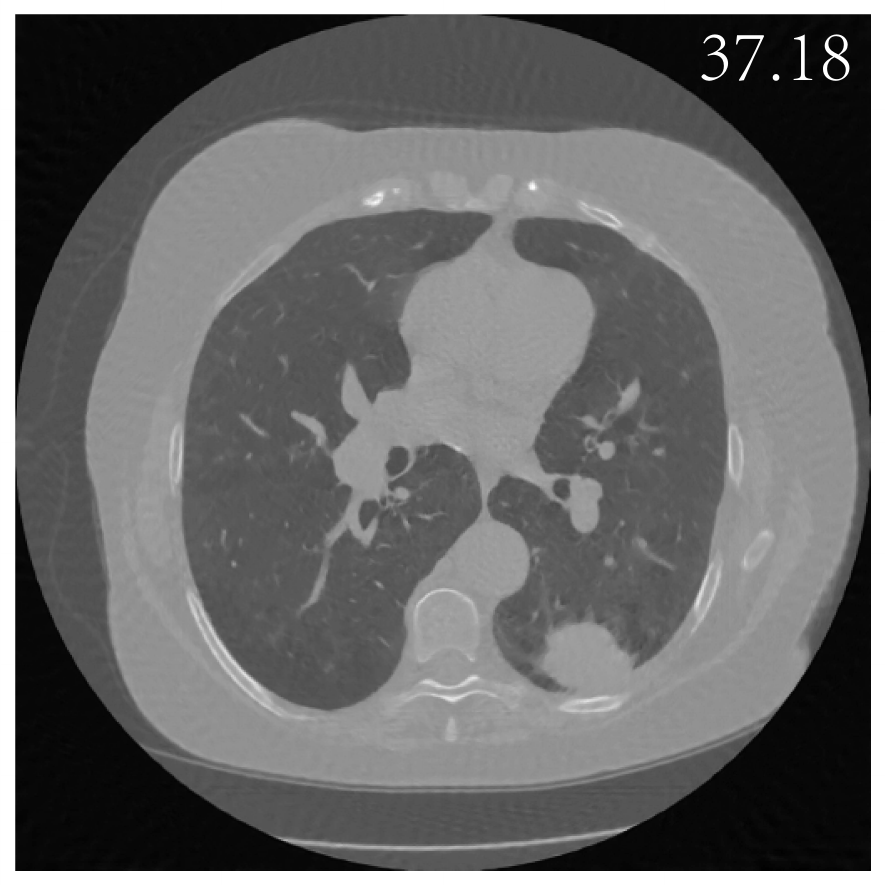} 
        \caption*{TTT-BN-EI-full}
    \end{subfigure}
    \begin{subfigure}{0.15\textwidth}
        \centering
        \includegraphics[width=\textwidth]{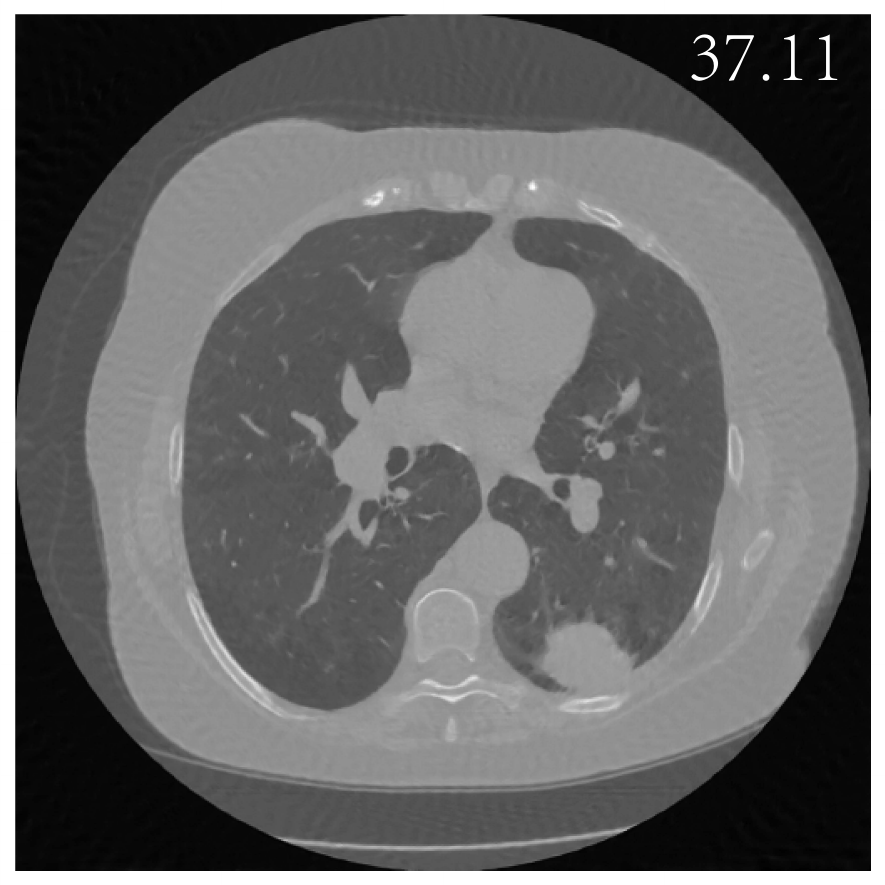} 
        \caption*{{\footnotesize TTT-BN-SkEI-5\%}}
    \end{subfigure}
    \begin{subfigure}{0.15\textwidth}
        \centering
        \includegraphics[width=\textwidth]{figures/3.2/512-CT/index-95/GTx-512.pdf} 
        \caption*{$x$ (GT)}
    \end{subfigure}

    \begin{subfigure}{0.15\textwidth}
        \centering
        \includegraphics[width=\textwidth]{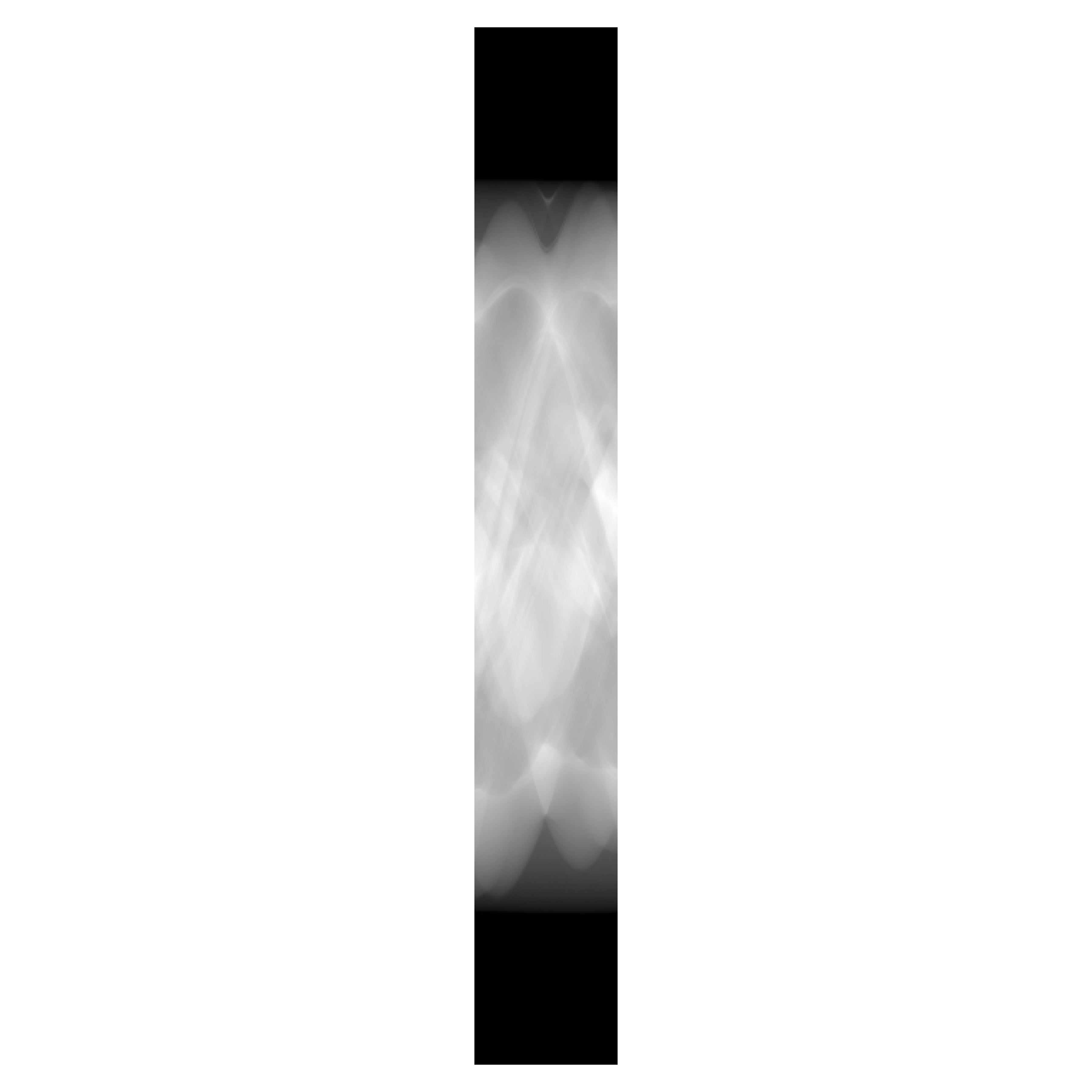} 
        \caption*{$y \sim \mathcal{N}(y^*, 0.2)$}
    \end{subfigure}
    \begin{subfigure}{0.15\textwidth}
        \centering
        \includegraphics[width=\textwidth]{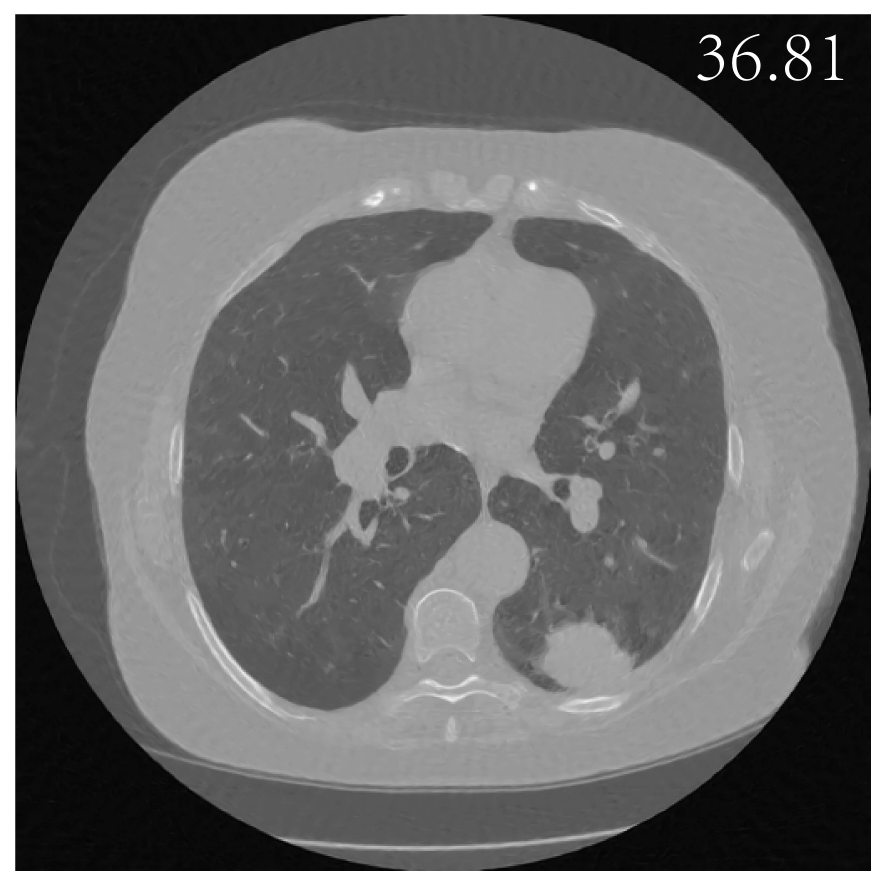} 
        \caption*{TTT-EI-full(*)}
    \end{subfigure}
    \begin{subfigure}{0.15\textwidth}
        \centering
        \includegraphics[width=\textwidth]{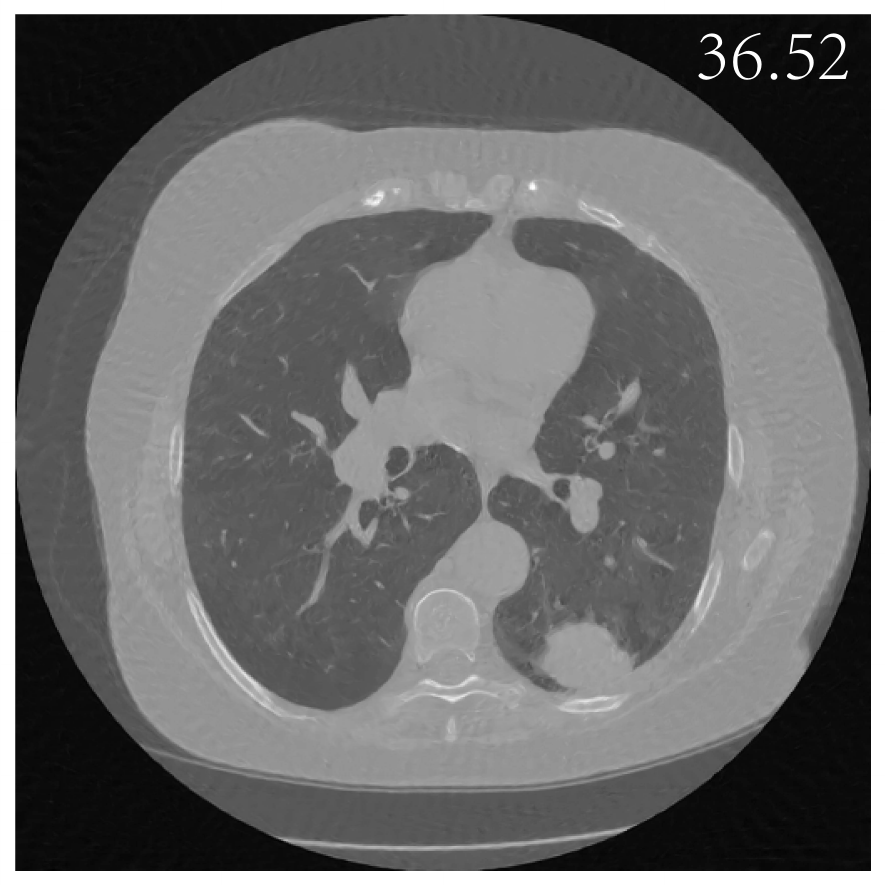} 
        \caption*{TTT-SkEI-5\%}
    \end{subfigure}
    \begin{subfigure}{0.15\textwidth}
        \centering
        \includegraphics[width=\textwidth]{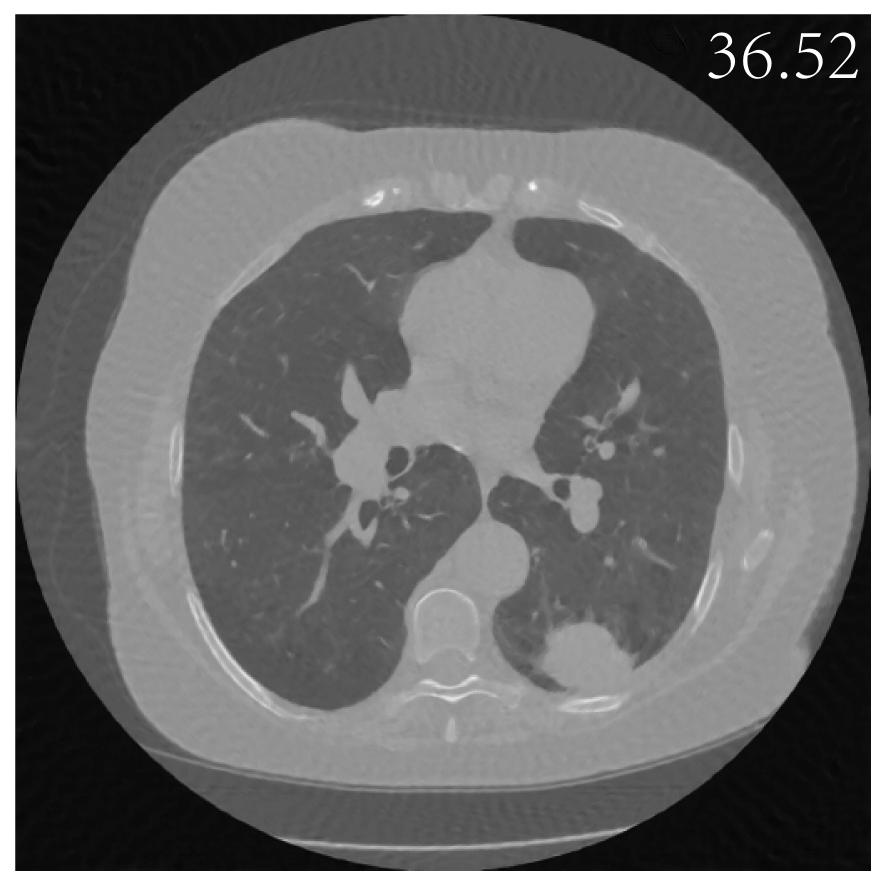} 
        \caption*{TTT-BN-EI-full}
    \end{subfigure}
    \begin{subfigure}{0.15\textwidth}
        \centering
        \includegraphics[width=\textwidth]{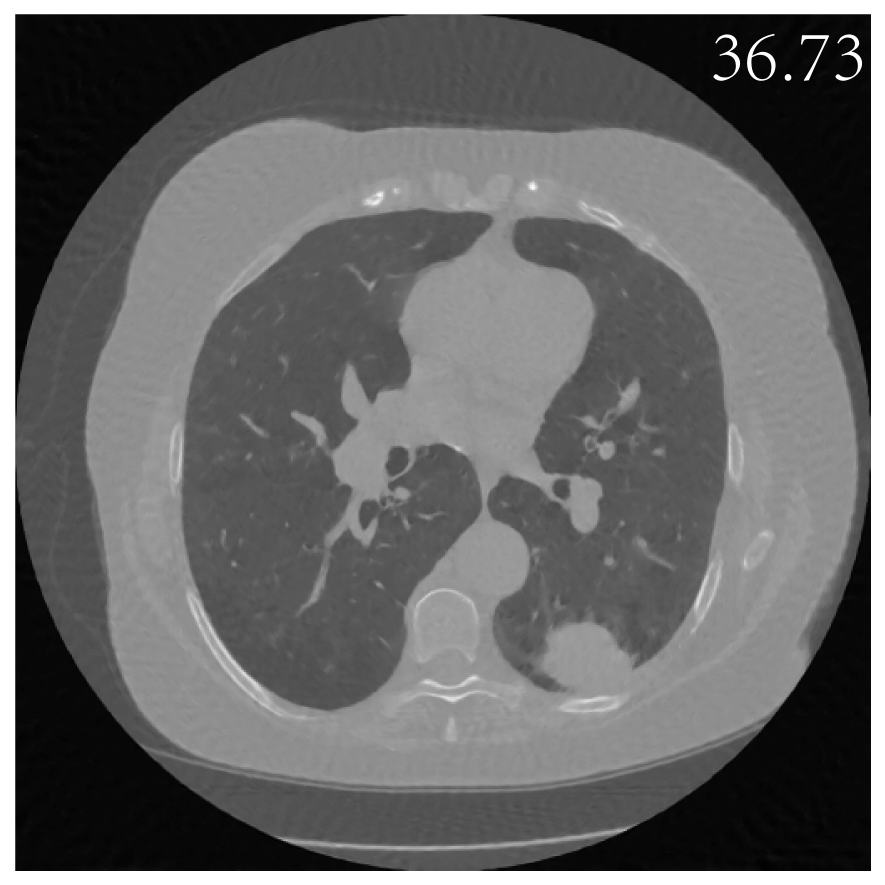} 
        \caption*{{\footnotesize TTT-BN-SkEI-5\%}}
    \end{subfigure}
    \begin{subfigure}{0.15\textwidth}
        \centering
        \includegraphics[width=\textwidth]{figures/3.2/512-CT/index-95/GTx-512.pdf} 
        \caption*{$x$ (GT)}
    \end{subfigure}

    \begin{subfigure}{0.15\textwidth}
        \centering
        \includegraphics[width=\textwidth]{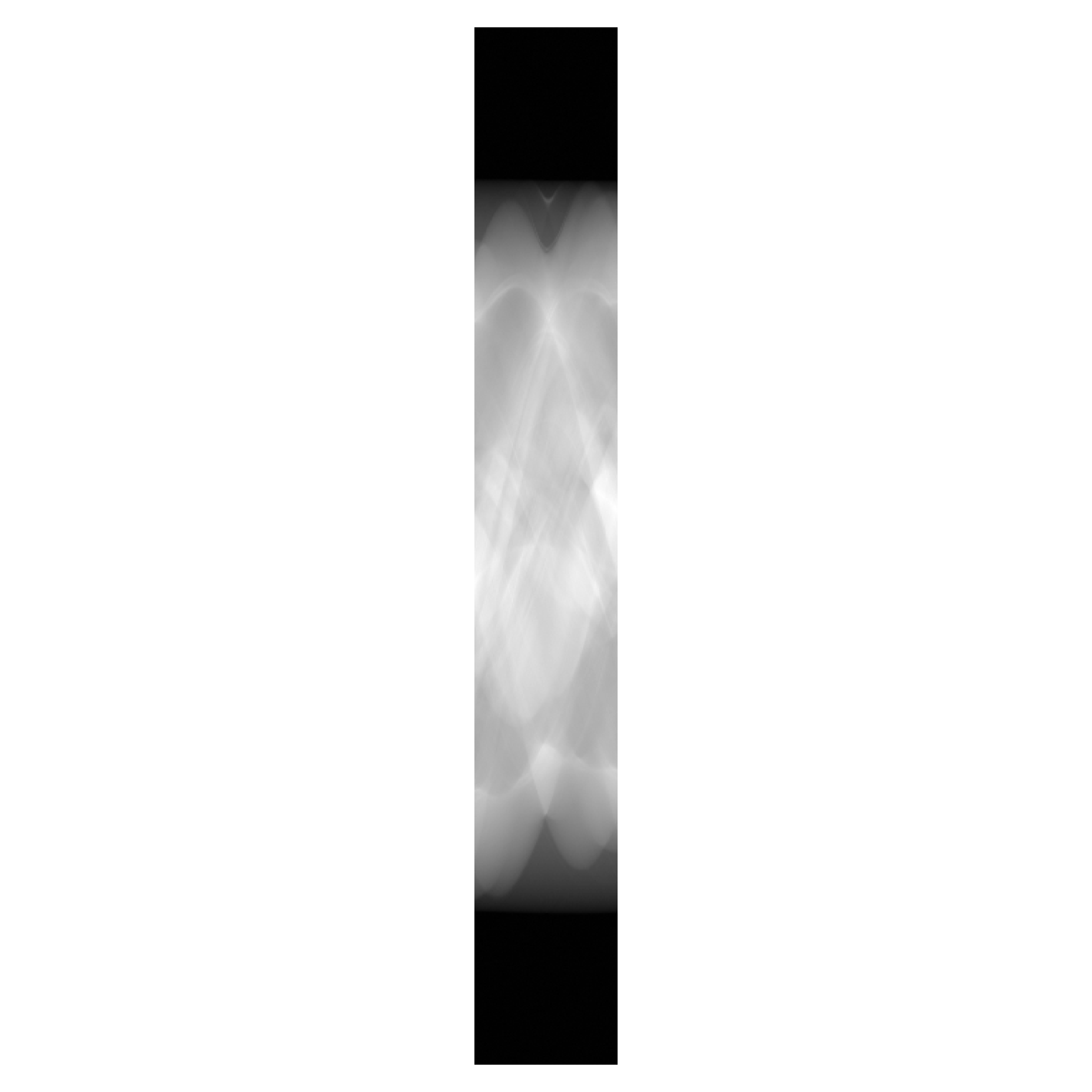} 
        \caption*{$y \sim \mathcal{N}(y^*, 0.5)$}
    \end{subfigure}
    \begin{subfigure}{0.15\textwidth}
        \centering
        \includegraphics[width=\textwidth]{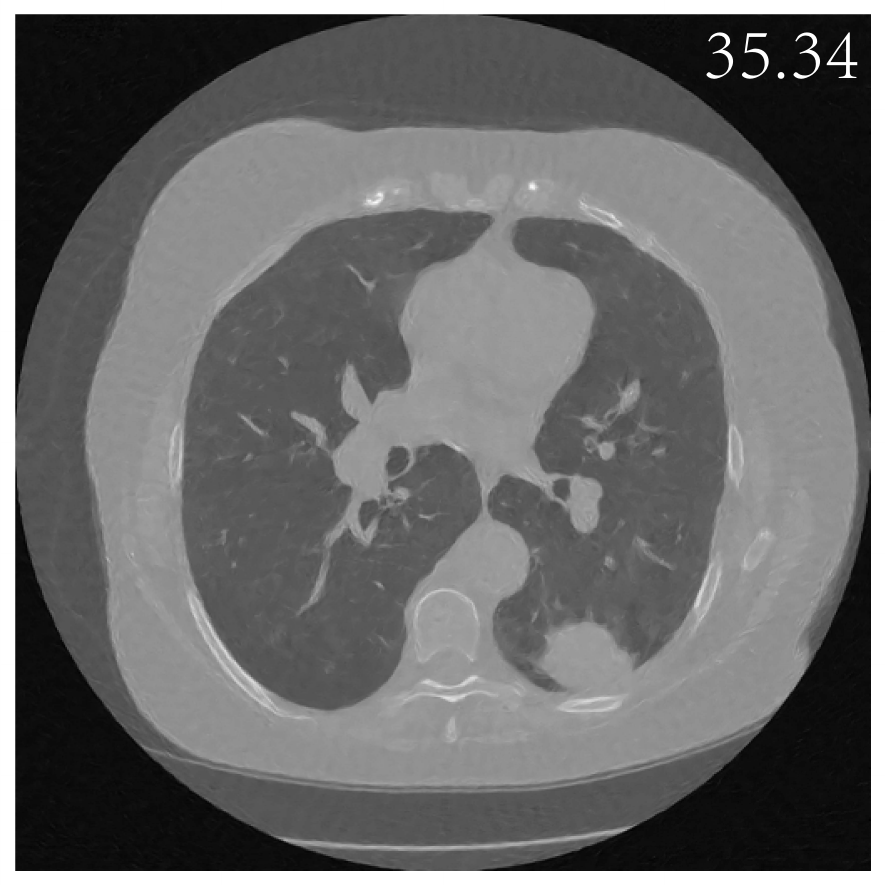} 
        \caption*{TTT-EI-full(*)}
    \end{subfigure}
    \begin{subfigure}{0.15\textwidth}
        \centering
        \includegraphics[width=\textwidth]{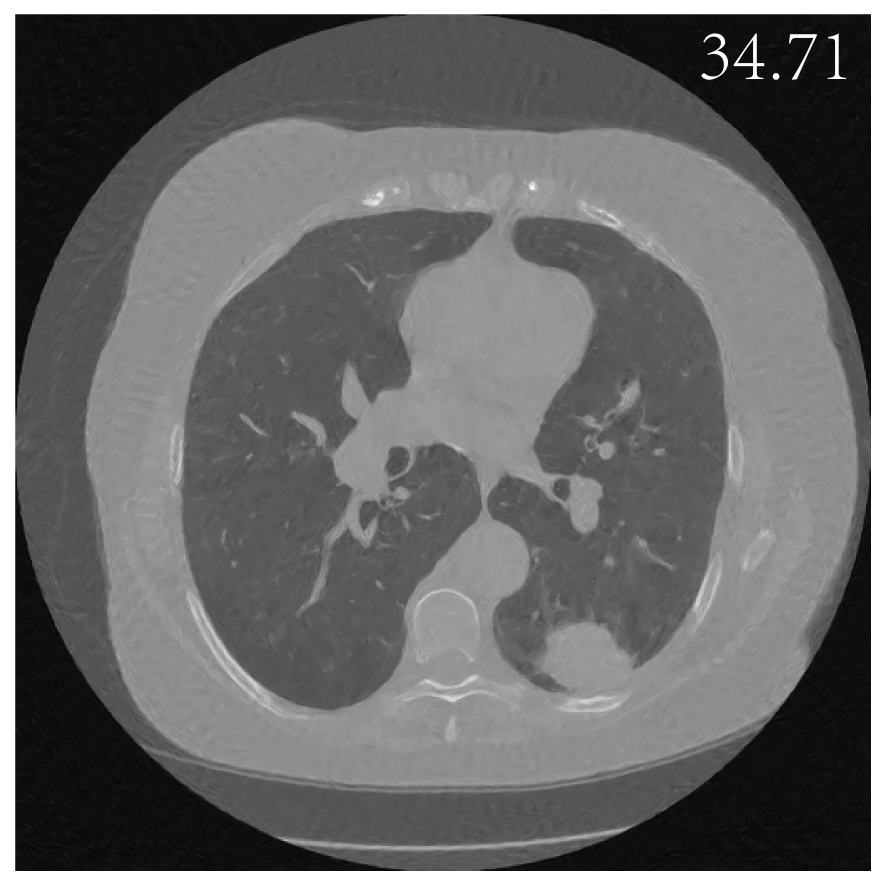} 
        \caption*{TTT-SkEI-5\%}
    \end{subfigure}
    \begin{subfigure}{0.15\textwidth}
        \centering
        \includegraphics[width=\textwidth]{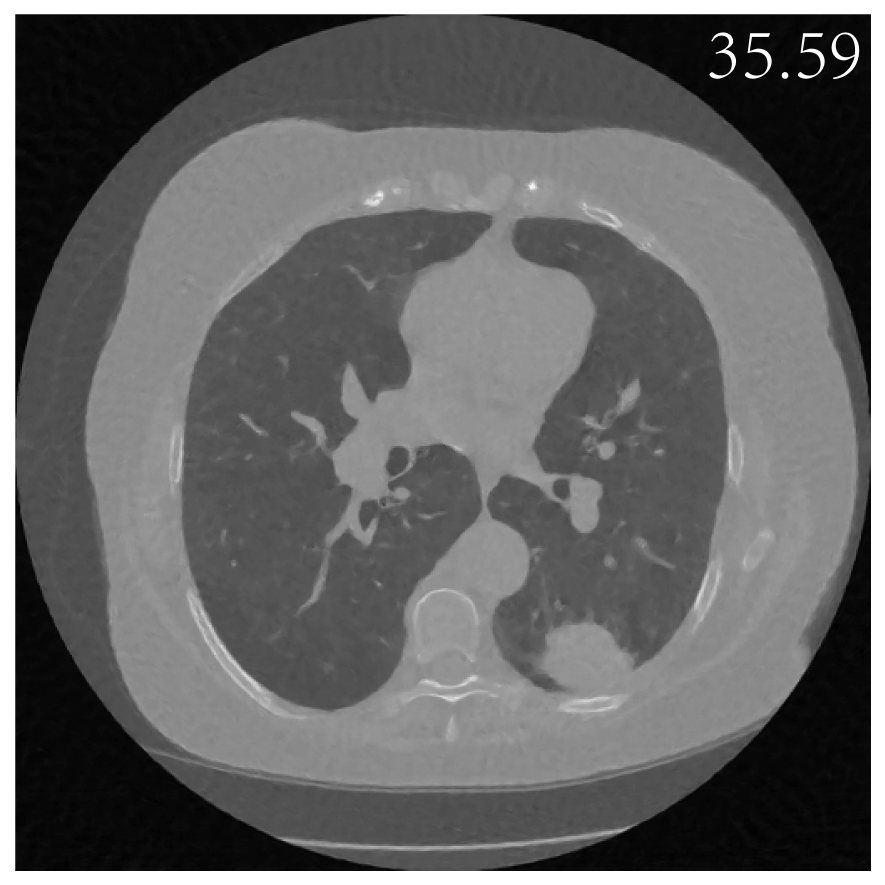} 
        \caption*{TTT-BN-EI-full}
    \end{subfigure}
    \begin{subfigure}{0.15\textwidth}
        \centering
        \includegraphics[width=\textwidth]{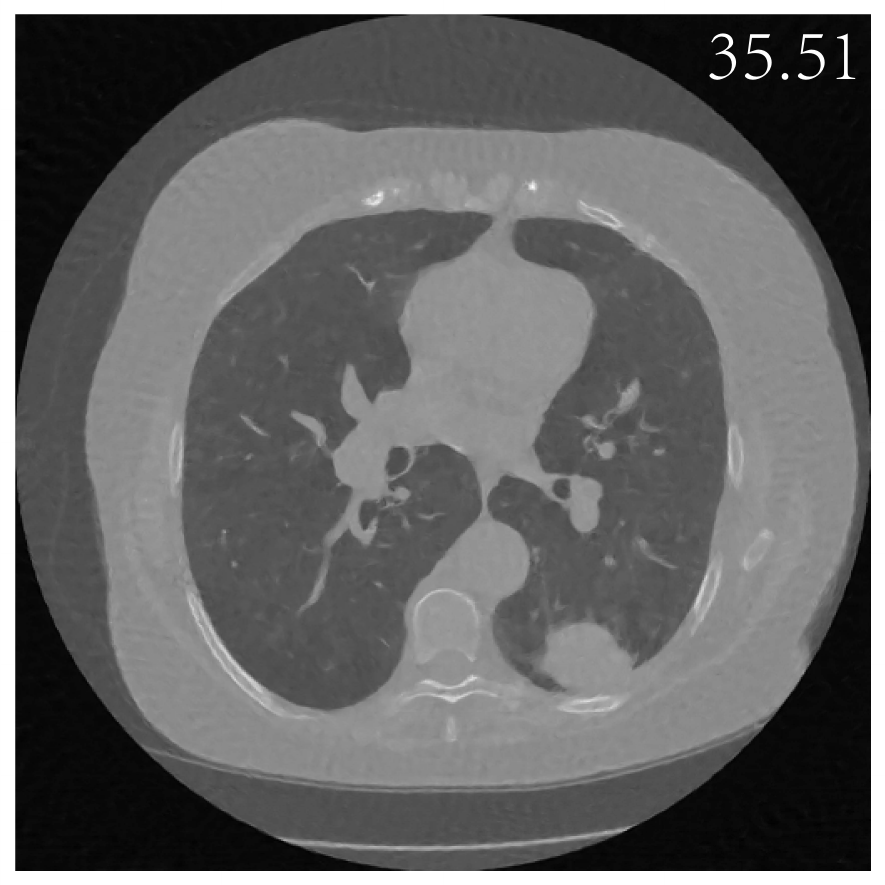} 
        \caption*{{\footnotesize TTT-BN-SkEI-5\%}}
    \end{subfigure}
    \begin{subfigure}{0.15\textwidth}
        \centering
        \includegraphics[width=\textwidth]{figures/3.2/512-CT/index-95/GTx-512.pdf} 
        \caption*{$x$ (GT)}
    \end{subfigure}

\caption{CT image reconstructions by Sketched EI of TTT-EI and TTT-BN-EI schemes with different noise level and different sketch size. TTT(TTT-BN)-full uses 100 CT scans, while Sketch-5\% means only 5\% of 100 CT scans used per iteration. (*) denotes the baseline.}
    \label{fig:NA-BN-EI-DIP}
\end{figure}

To better understand the benefits of the TTT-EI and TTT-BN-EI schemes, Table~\ref{tab: NA v.s. BN} summarizes the results of these two competing schemes in noisy measurement with a noise level of $\sigma=0.1$. As we can see in Table~\ref{tab: NA v.s. BN}, the time consumption per iteration decreased significantly with the sketch operation, while with increasing sketch size, the time consumption did not decrease significantly anymore. In addition, the training parameters used in the TTT-BN-EI scheme were significantly reduced, as expected.

\begin{table}[htp]
\renewcommand{\arraystretch}{1.0}
\setlength{\tabcolsep}{5.5pt}
\centering
\begin{tabular}{l|ccccc|ccccc}
\toprule
\diagbox{Noise}{PSNR}{Scheme} & & & TTT & & & & & TTT-BN & & \\
\hline
 & *full & 50\% & 20\% & 10\% & 5\% & full & 50\% & 20\% & 10\% & 5\% \\
\hline
0.05 & 37.48 & 37.46 & 37.30 & 37.23 & 37.05 & 37.18 & 37.17 & 37.08 & 37.15 & 37.11 \\
\hdashline[1pt/1pt]
0.2 & 36.81 & 36.51  & 36.70 & 36.19 & 36.52 & 36.52 & 36.72 & 36.54 & 36.76 & 36.73 \\
\hdashline[1pt/1pt]
0.5 & 35.34 & 35.25 & 34.89 & 34.45 & 34.71 & 35.59 & 35.48 & 35.59 & 35.52 & 35.51 \\
\bottomrule
\end{tabular}
\caption{PSNR of CT reconstructions by Sketched EI in TTT and TTT-BN schemes with increasing noise level from $\sigma = 0.05$ to $0.5$. 'full' means using 100 CT scans, while 50\%, 20\%, 10\% and 5\% means only 50\%, 20\%, 10\% and 5\% of 100 CT scans used per iteration. * denotes the baseline.}
\label{tab:2}
\end{table}

As the noise level increases, both fine-tuning schemes exhibit performance degradation to varying extents (Table~\ref{tab:2}, Figure~\ref{fig:NA-BN-EI-DIP}). The TTT-EI scheme demonstrates markedly higher sensitivity to noise than the TTT-BN-EI scheme. Moreover, larger sketch sizes exacerbate performance loss in the TTT-EI scheme, whereas the TTT-BN-EI scheme remains comparatively robust.

\subsubsection{Standard unsupervised training experiments}\label{unsup}
\begin{figure}[htp!]
    \centering
    \begin{minipage}{0.49\textwidth}
        \centering
        \begin{tabular}{c|ccc}
        \toprule
        Methods & $\text{PSNR}^1$ & $\text{PSNR}^2$ & Time (s) / Epoch  \\
        \hline
        EI-full (*) & 36.94 & 35.03  & 87.60 \\
        \hdashline[1pt/1pt]
        SkEI-50\% & \textbf{37.37} & \textbf{36.20} & 14.54 \\
        \hdashline[1pt/1pt]
        SkEI-20\% & 36.56 & 35.44 & 11.88 \\
        \hdashline[1pt/1pt]
        SkEI-10\% & 36.45 & 35.11 & 10.72 \\
        \hdashline[1pt/1pt]
        SkEI-5\% & 36.30 & 35.34 & 10.27 \\
        \bottomrule
        \end{tabular}
        \caption*{(a): Averaged PSNR and Time costs of the proposed EI with different sketched size. (*) means baseline. Superscript 1 means the averaged PSNR obtained from last epoch of train dataset; Superscript 2 means the averaged PSNR obtained from test dataset.}
    \end{minipage}
    \begin{minipage}{0.50\textwidth}
        \centering
        \includegraphics[width=1.0\textwidth]{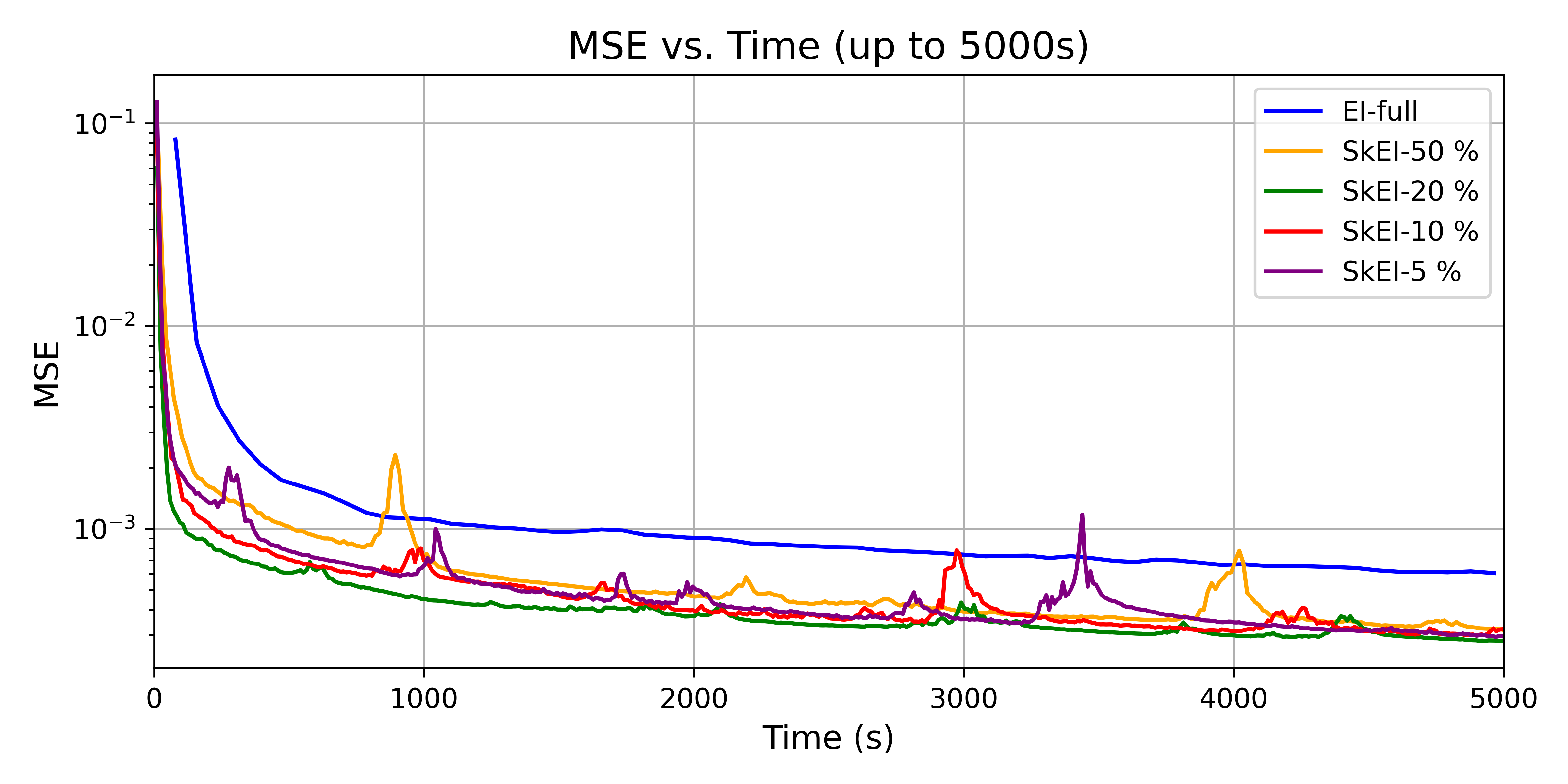} 
        \caption*{(b): MSE trend with time up to 5000 seconds of the proposed EI with different sketched size.}
    \end{minipage}
   
\caption{Averaged performance of the proposed Sketched EI with vanilla EI which trained on a small CT dataset of 10 samples. EI-full uses 100 CT scans, while 50\%, 20\%, 10\% and 5\% means only 50\%, 20\%, 10\% and 5\% of 100 CT scans used per iteration.}
\label{fig:ct-dataset}
\end{figure}

In addition, we evaluated the sketched EI on a small CT dataset consisting of only ten training examples. Specifically, we train our Sketched EI using \textit{CT100} dataset (index ranges from 1 to 10), then test on the same dataset with index ranges from 11 to 20. As the number of training samples increases, the quality of reconstruction improves steadily relative to the single input configuration, although this gain is accompanied by a higher computational cost (Figure \ref{fig:ct-dataset}(a)). Importantly, our sketched EI implementation achieves a reduction of approximately 86\% runtime while maintaining a reconstruction accuracy comparable to that of the full scan setting (Figure \ref{fig:ct-dataset}).

\subsubsection{Accelerated Multi-coil MRI}
In this subsection, we report the comparative performance of EI, SkEI and C-SkEI in multicoil MRI reconstruction task. We evaluate the performance of two distinct Sketched-EI methods proposed in this study for multicoil MRI, using a knee MRI image with 15 coils from~\cite{zbontar2018fastmri}. Subsequently, we performed a comparative analysis of the reconstruction results under two distinct sketch schemes and varying sketch sizes. Consistent with Section~\ref{sec: exp-ct}, we begin by comparing our proposed methods with the vanilla DIP method, considering the same DIP network architectures as outlined in Section~\ref{sec: exp-ct}. \par
For SkEI and C-SkEI schemes, $S$ is defined as the sketch operation that samples $N \leq 15$ coils from the original 15 coils, then forms a minibatch named $A_{S_N}$. However, these two sketch schemes differ in their minibatch partition strategies. Specifically, as described in Algorithm~\ref{alg:coil-sketch}, Coil-Sketch compresses the initial 15 coils and then selects the first $L$ compressed virtual coils as the new multi-coils. Subsequently, the first $R$ higher-energy virtual coils are retained, while the last $S$ lower-energy virtual coils undergo sketch operation. In contrast, Classical-Sketch directly acts on the original 15 coils, randomly sampling the $N$ coils into a minibatch. During each iteration, the minibatch is randomly chosen from the 15 coils for updates. For this study, we experiment with $N=10,5$ and
$2$, respectively.\par
\begin{figure}[htp!]
    \centering

    \begin{subfigure}{0.15\textwidth}
        \centering
        \includegraphics[width=\textwidth]{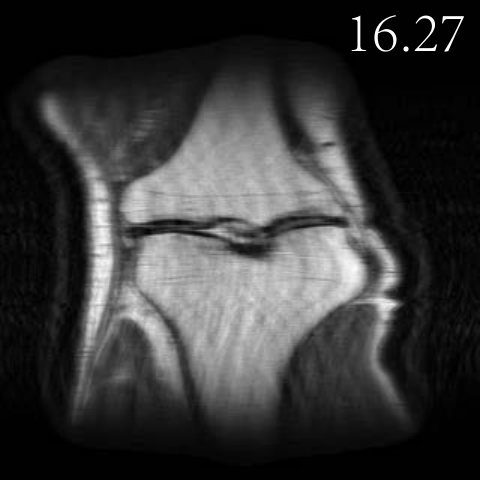} 
        \caption*{$A^{\dagger}y$}
    \end{subfigure}
    \begin{subfigure}{0.15\textwidth}
        \centering
        \includegraphics[width=\textwidth]{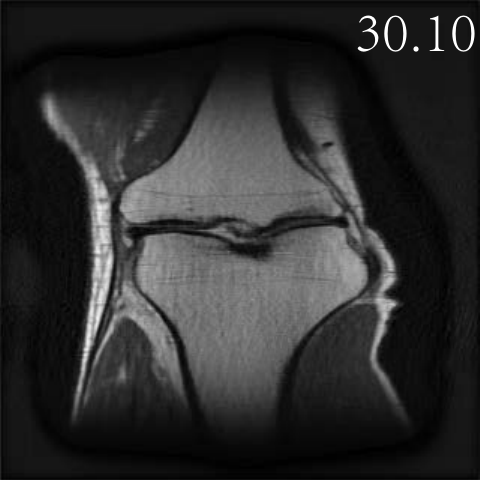} 
        \caption*{DIP}
    \end{subfigure}
    \begin{subfigure}{0.15\textwidth}
        \centering
        \includegraphics[width=\textwidth]{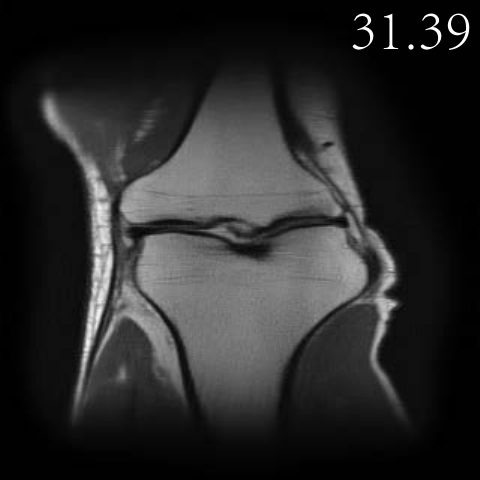} 
        \caption*{*EI{-full} (15coils)}
    \end{subfigure}
    \begin{subfigure}{0.15\textwidth}
        \centering
        \includegraphics[width=\textwidth]{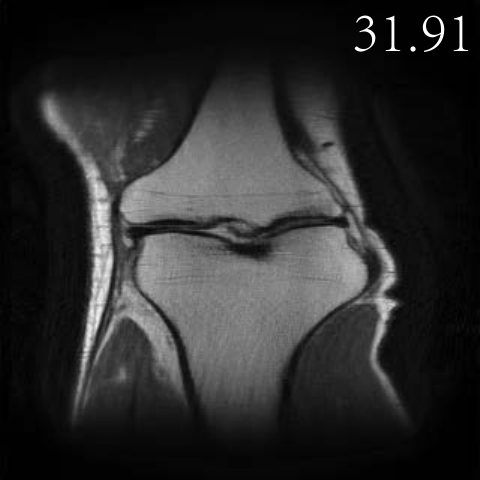} 
        \caption*{{SkEI-5coils}}
    \end{subfigure}
    \begin{subfigure}{0.15\textwidth}
        \centering
        \includegraphics[width=\textwidth]{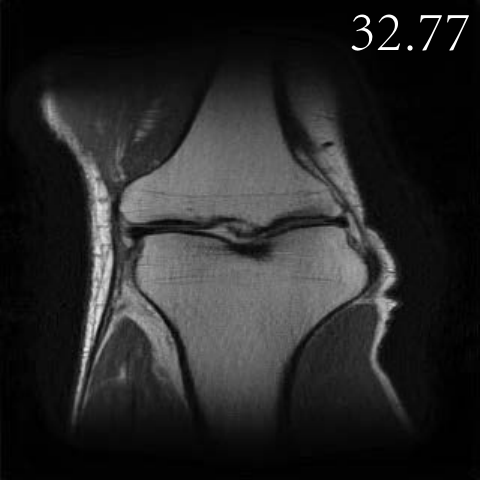} 
        \caption*{{C-SkEI-5coils}}
    \end{subfigure}
    \begin{subfigure}{0.15\textwidth}
        \centering
        \includegraphics[width=\textwidth]{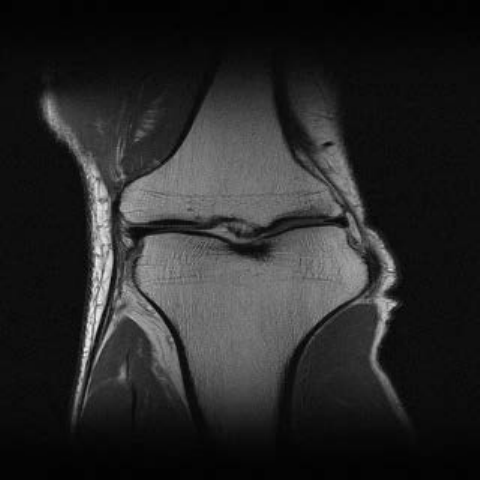} 
        \caption*{$x$ (GT)}
    \end{subfigure}

\caption{Multi-coil MRI Images (with corresponding PSNR) reconstructed by DIP, EI and our two Sketched EI methods, where `SkEI-' shorts for classical-sketch and `C-SkEI' for coil-sketch. `*' denotes the baseline.}
\label{fig:mri-comp}
\end{figure}
\begin{figure}[htp!]
    \centering
    \begin{minipage}{0.58\textwidth}
        \centering
        \includegraphics[width=1.0\textwidth]{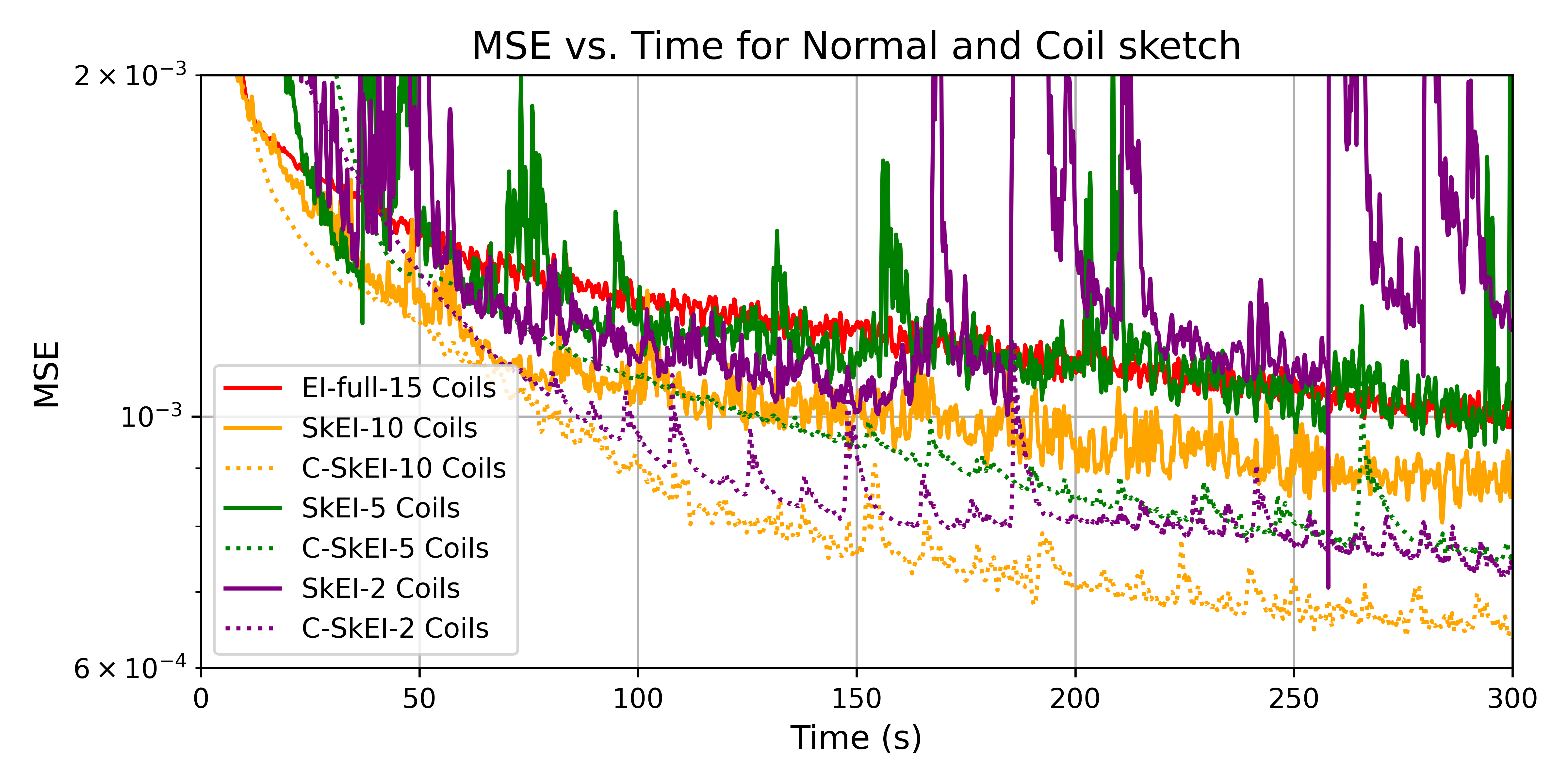} 
        \caption*{(a): MSE variation with time up to 300 seconds of the proposed EI with different sketched size for both classical sketch and coil sketch schemes.}
    \end{minipage}
    \hfill
    \begin{minipage}{0.34\textwidth}
        \centering
        \begin{tabular}{c|c}
            \hline
            Method & Time (s) / Epoch \\ 
            \hline
            {*EI-full} & 0.157  \\ 
            \hdashline
            C-SkEI-10 & 0.151 \\ 
            C-SkEI-5 & 0.134  \\ 
            C-SkEI-2 & 0.130  \\
            \hdashline
            SkEI-10 & 0.141  \\
            SkEI-5 & 0.129  \\
            SkEI-2 & 0.127  \\
            \hline
        \end{tabular}
        \caption*{(b): Time cost per iteration of the proposed SkEI / C-SkEI with different sketched size for both classical sketch and coil sketch schemes.}
    \end{minipage}
    
\caption{MSE and time cost comparisions of the proposed sketched EI and vanilla EI method in 15 coils knee MRI reconstruction.{`SkEI-' shorts for SkEI with subsampling sketch, `C-SkEI-' stands for C-SkEI with coil-sketch. `*' denotes the baseline.}}
\label{fig:mri-EI-DIP-time&psnr}
\end{figure}

The visualized comparisons are shown in Figure~\ref{fig:mri-comp}, as can be seen, the performance of the DIP method is still suboptimal, consistent with findings from the sparse view CT experiments. In comparison, the EI method achieves remarkable improvements in reconstruction quality due to the EI regularizer. Furthermore, the reconstruction results generated by these two proposed sketch methods exhibit almost no loss in performance and even surpass the vanilla EI method. Furthermore, as presented in Figure~\ref{fig:mri-EI-DIP-time&psnr} (b), the time cost per iteration of the sketched EI decreased compared to the vanilla EI method. Specifically, C-SkEI-2coils decreased 19\% per iteration compared to vanilla EI, while SkEI-2coils (with subsampling sketches) decreased 17\%. Moreover, we observe that C-SkEI demonstrates a much faster convergence rate compared to that of standard EI. \par

\begin{figure}[htp!]
    \centering

    \begin{subfigure}{0.23\textwidth}
        \centering
        \includegraphics[width=\textwidth]{figures/3.2/320-MRI/GT-x.pdf} 
        \caption*{$x$ (GT)}
    \end{subfigure}
    \begin{subfigure}{0.23\textwidth}
        \centering
        \includegraphics[width=\textwidth]{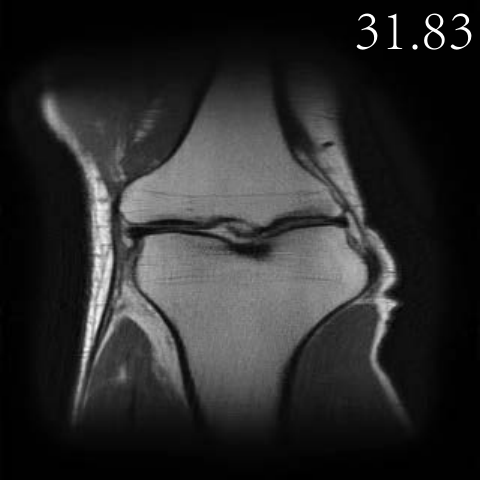} 
        \caption*{C-SkEI-10coils}
    \end{subfigure}
    \begin{subfigure}{0.23\textwidth}
        \centering
        \includegraphics[width=\textwidth]{figures/3.2/320-MRI/COIL-Coils5-32.77.pdf} 
        \caption*{C-SkEI-5coils}
    \end{subfigure}
    \begin{subfigure}{0.23\textwidth}
        \centering
        \includegraphics[width=\textwidth]{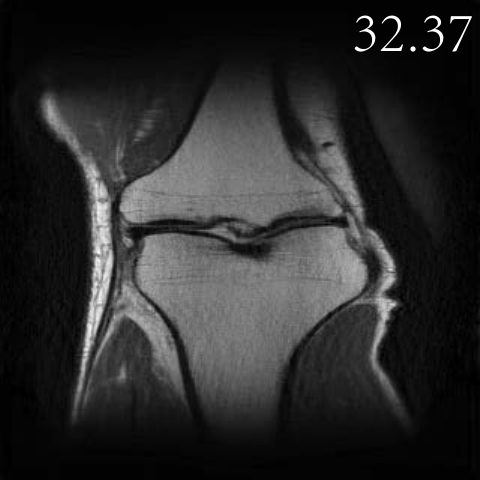} 
        \caption*{C-SkEI-2coils}
    \end{subfigure}

    \begin{subfigure}{0.23\textwidth}
        \centering
        \includegraphics[width=\textwidth]{figures/3.2/320-MRI/Coils15-31.39.pdf} 
        \caption*{*EI-full (15coils)}
    \end{subfigure}
    \begin{subfigure}{0.23\textwidth}
        \centering
        \includegraphics[width=\textwidth]{figures/3.2/320-MRI/Coils10-31.83.pdf} 
        \caption*{SkEI-10coils}
    \end{subfigure}
    \begin{subfigure}{0.23\textwidth}
        \centering
        \includegraphics[width=\textwidth]{figures/3.2/320-MRI/Coils5-31.91.pdf} 
        \caption*{SkEI-5coils}
    \end{subfigure}
    \begin{subfigure}{0.23\textwidth}
        \centering
        \includegraphics[width=\textwidth]{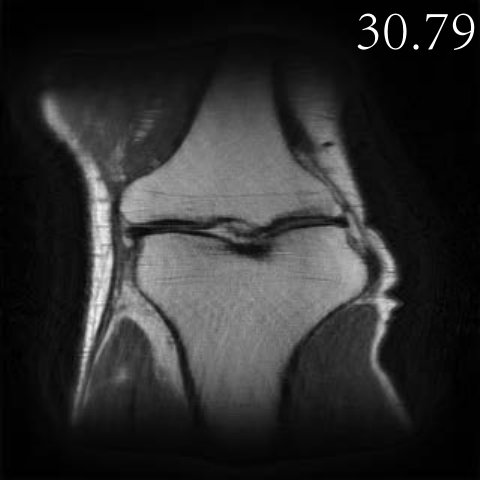} 
        \caption*{SkEI-2coils}
    \end{subfigure}

\caption{Multi-coil MRI Images (with corresponding PSNR) reconstructed by Sketched EI, with different sketch sizes, and `SkEI-' shorts for SkEI with subsampling sketch, `C-SkEI-' stands for C-SkEI with coil-sketch. `*' denotes the baseline.}
\label{fig:mri-sketch-comp}
\end{figure}

To further explore our schemes, we performed ablation studies with three different sketch sizes in two different sketch methods, as visualized in Figure~\ref{fig:mri-sketch-comp} and Figure~\ref{fig:mri-EI-DIP-time&psnr}(a). The results show that for C-SkEI, performance improves significantly when the sketch size increases to 5 but decreases slightly when the size reaches 2. Nevertheless, all three sketch sizes outperform the vanilla EI, attributed to the sketch algorithm optimized for multicoil MRI. In contrast, for plain SkEI with the classical subsampling sketch, performance initially increases with smaller sketch sizes, but significantly deteriorates once the size reaches 2, performing much worse than vanilla EI. Furthermore, as presented in Figure~\ref{fig:mri-EI-DIP-time&psnr}(a), compared to the vanilla EI method, both SkEI methods converge faster in the first 300 seconds of training. In particular, the C-SkEI method converges even faster than plain SkEI while maintaining performance levels that far exceed vanilla EI. Both schemes reach their own fastest convergence when the sketched size is 10.

\begin{figure}[htp!]
    \centering

    \begin{subfigure}{0.15\textwidth}
        \centering
        \includegraphics[width=\textwidth]{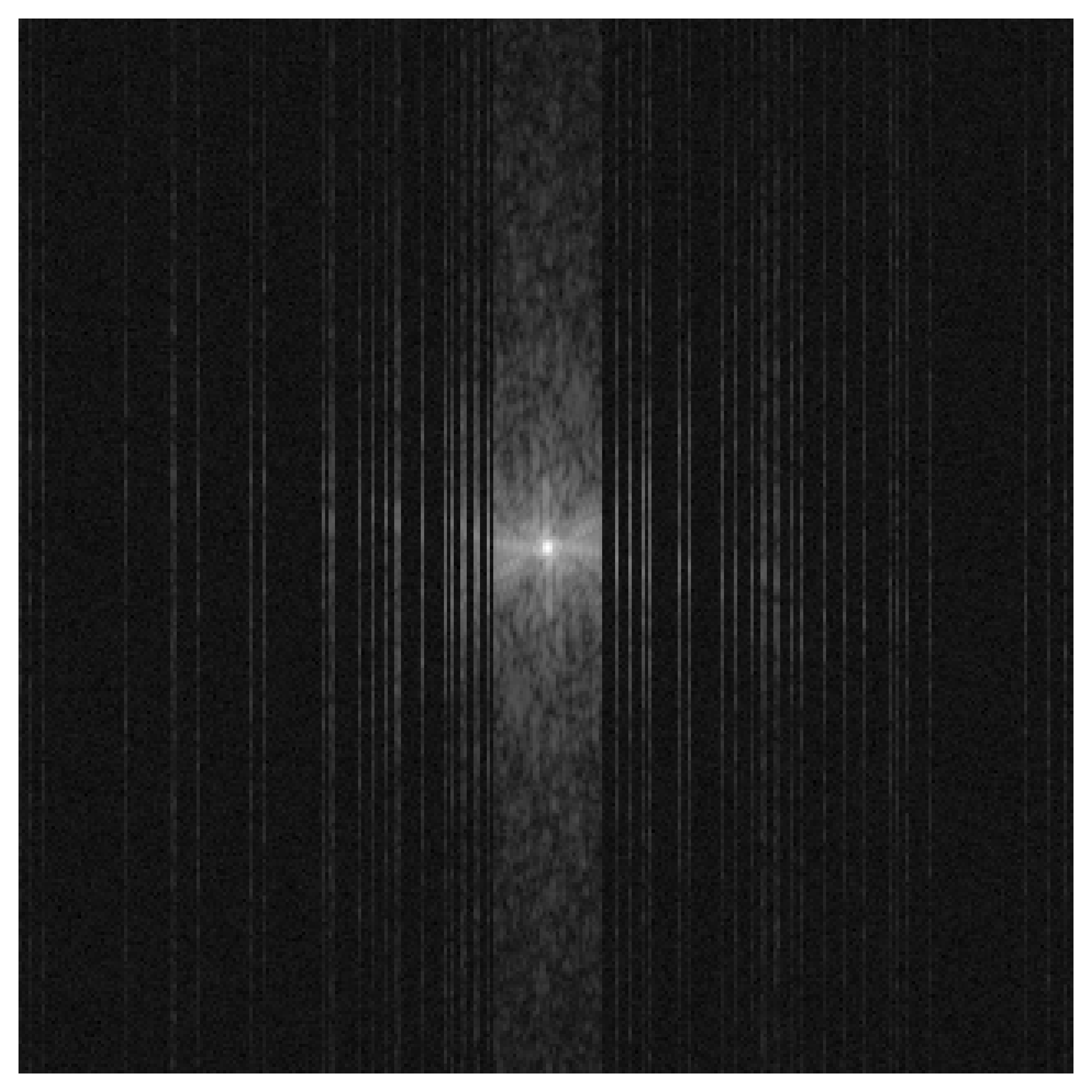} 
        \caption*{\centering $y \sim \mathcal{N}(y^*, 0.005)$}
    \end{subfigure}
    \begin{subfigure}{0.15\textwidth}
        \centering
        \includegraphics[width=\textwidth]{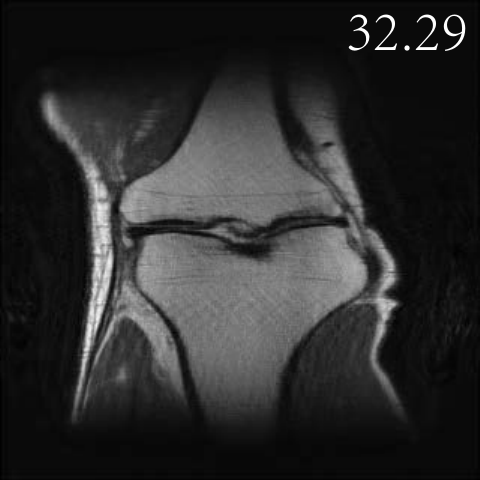} 
        \caption*{\centering {*TTT-EI- \\ full-15Coils}}
    \end{subfigure}
    \begin{subfigure}{0.15\textwidth}
        \centering
        \includegraphics[width=\textwidth]{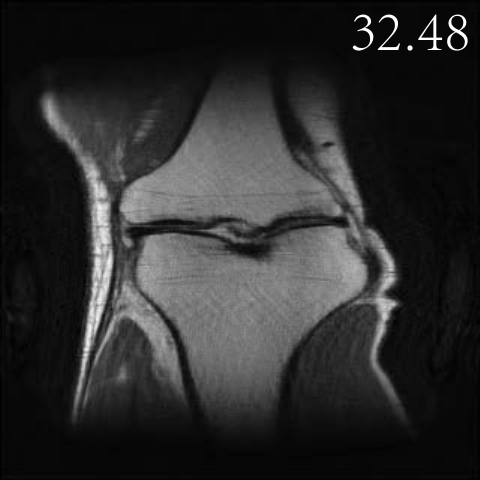} 
        \caption*{\centering TTT-\\C-SkEI-10Coils}
    \end{subfigure}
    \begin{subfigure}{0.15\textwidth}
        \centering
        \includegraphics[width=\textwidth]{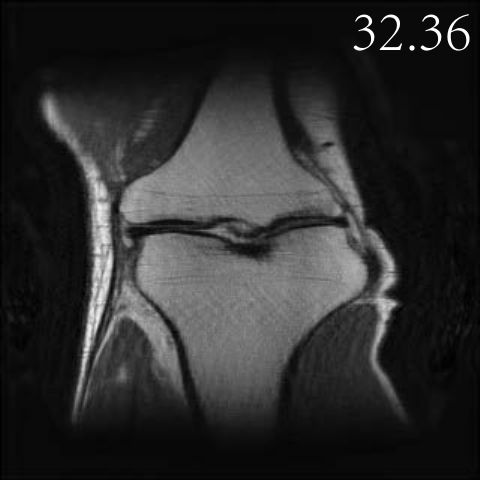} 
        \caption*{\centering TTT-\\C-SkEI-5Coils}
    \end{subfigure}
    \begin{subfigure}{0.15\textwidth}
        \centering
        \includegraphics[width=\textwidth]{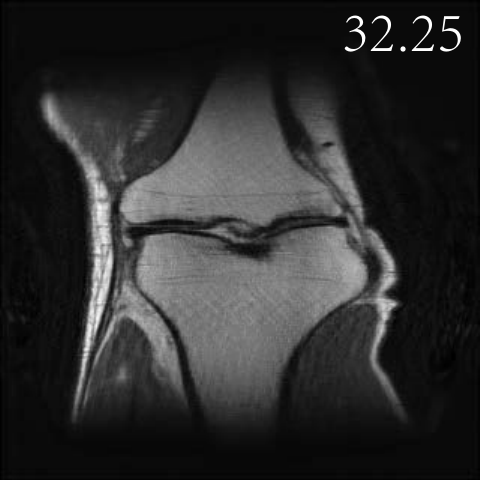} 
        \caption*{\centering TTT-\\C-SkEI-2Coils}
    \end{subfigure}
    \begin{subfigure}{0.15\textwidth}
        \centering
        \includegraphics[width=\textwidth]{figures/3.2/320-MRI/GT-x.pdf} 
        \caption*{$x$ (GT) \\ \quad}
    \end{subfigure}

    \begin{subfigure}{0.15\textwidth}
        \centering
        \includegraphics[width=\textwidth]{figures/3.2/320-MRI/GTy-320-0_005.png} 
        \caption*{\centering $y \sim \mathcal{N}(y^*, 0.005)$}
    \end{subfigure}
    \begin{subfigure}{0.15\textwidth}
        \centering
        \includegraphics[width=\textwidth]{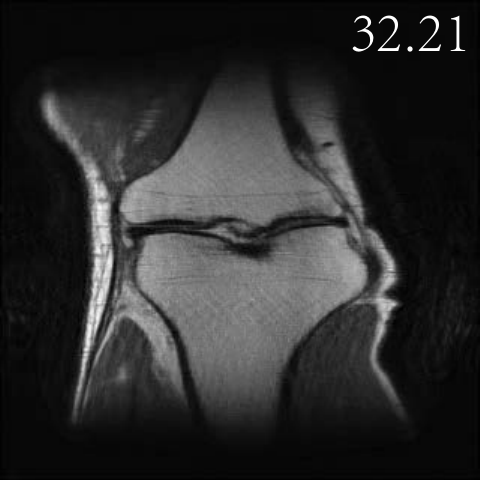} 
        \caption*{\centering TTT-BN-EI- \\ full-15Coils}
    \end{subfigure}
    \begin{subfigure}{0.15\textwidth}
        \centering
        \includegraphics[width=\textwidth]{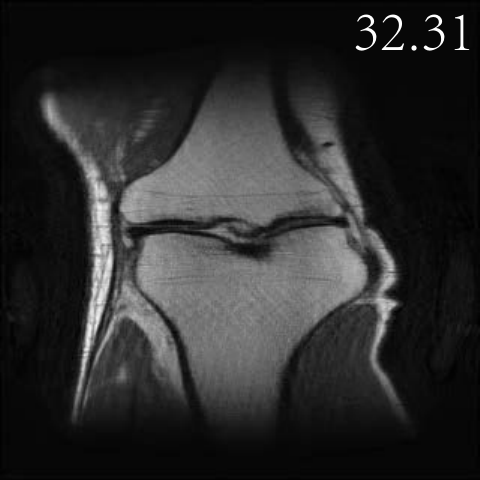} 
        \caption*{\centering TTT-BN-\\C-SkEI-10Coils}
    \end{subfigure}
    \begin{subfigure}{0.15\textwidth}
        \centering
        \includegraphics[width=\textwidth]{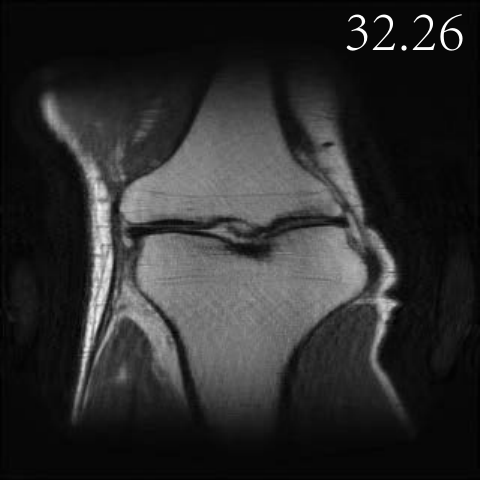} 
        \caption*{\centering TTT-BN-\\C-SkEI-5Coils}
    \end{subfigure}
    \begin{subfigure}{0.15\textwidth}
        \centering
        \includegraphics[width=\textwidth]{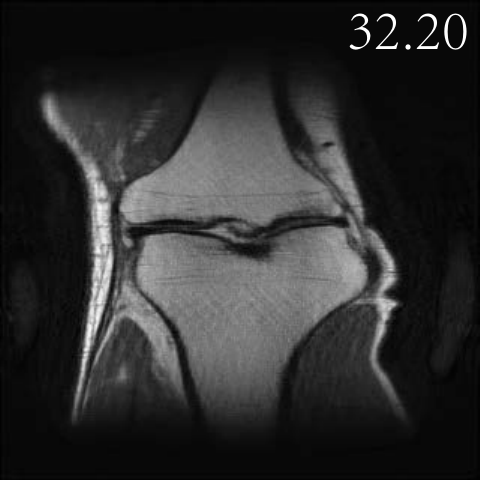} 
        \caption*{\centering TTT-BN-\\C-SkEI-2Coils}
    \end{subfigure}
    \begin{subfigure}{0.15\textwidth}
        \centering
        \includegraphics[width=\textwidth]{figures/3.2/320-MRI/GT-x.pdf} 
        \caption*{$x$ (GT) \\ \quad}
    \end{subfigure}

\caption{Multi-coil MRI image reconstructions in {Test Time Training} task with a single noisy measurement (noise level $\sigma=0.005$). Top row shows reconstructions of fine-tuning entire network (TTT) while bottom row fine-tuning only the BatchNorm layers (TTT-BN), both with various sketch size. `*' denotes the baseline.}
    \label{fig:mri-NA-EI-DIP}
\end{figure}

Next, we investigate the performance of our proposed SkEI and C-SkEI in {Test Time Training} task. Specifically, we fine-tune the model 
 trained on the fastMRI knee dataset in~\cite{darestani2022test}, with the experimental results illustrated in Figure~\ref{fig:mri-NA-EI-DIP}. As we can see, the single measurement was corrupted by Gaussian noise with noise level $\sigma = 0.005$, and the visualized images indicate that, in both the full-parameter {Test Time Training (TTT)} and {Test Time Training with BatchNorm only (TTT-BN)} scenarios, our proposed method yields reconstruction results that remain comparable to those obtained under noiseless conditions (as illustrated by comparing Figure~\ref{fig:mri-sketch-comp} with Figure~\ref{fig:mri-NA-EI-DIP}), even in the presence of noise. Additionally, coil-sketching operation appears to have minimal influence on reconstruction quality when noisy measurements are used, regardless of the fine-tuning strategy employed.

\begin{figure}[htp!]
    \centering
    \begin{minipage}{0.48\textwidth}
        \centering
        \includegraphics[width=1.0\textwidth]{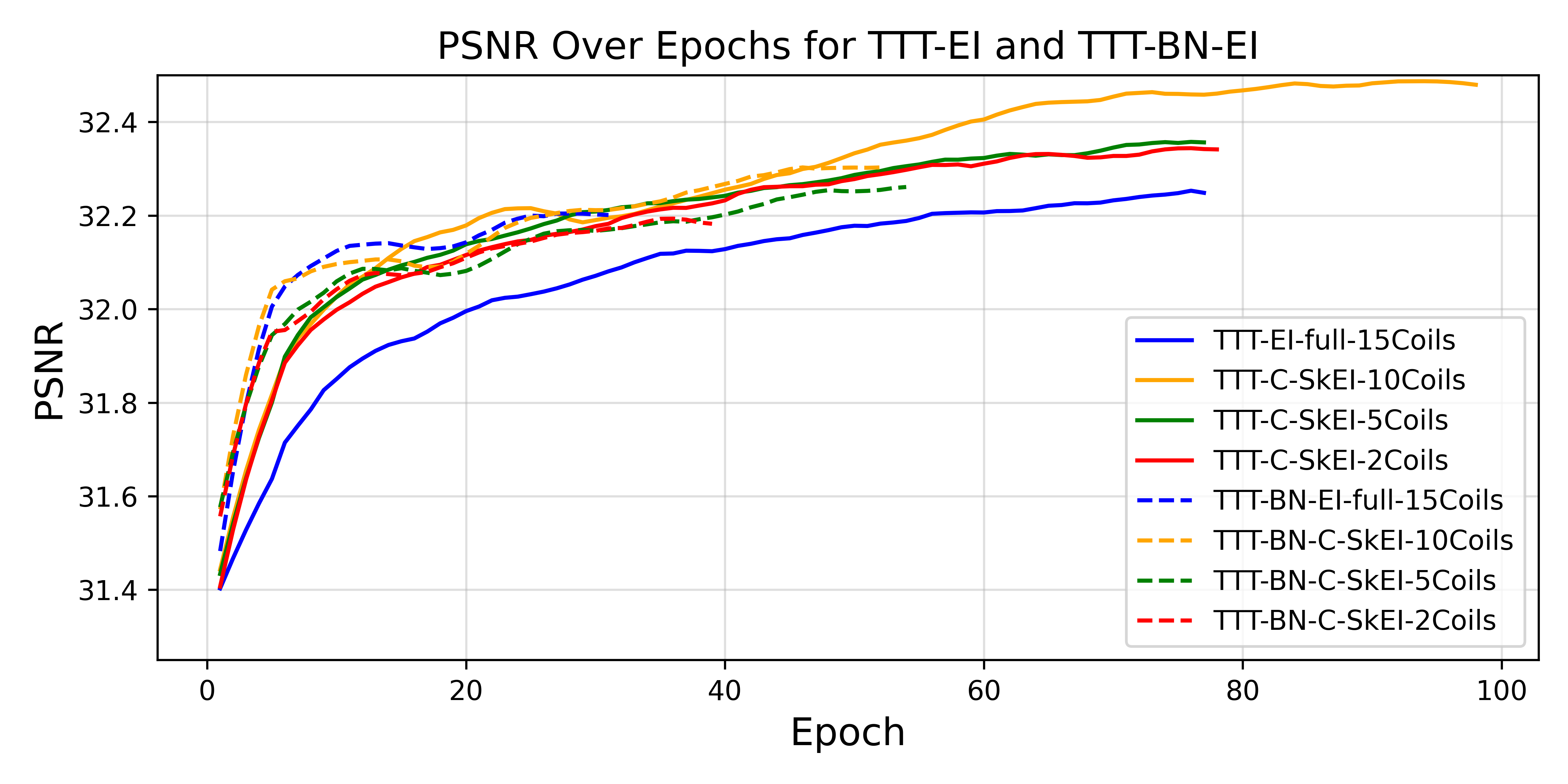} 
        \caption*{(a): Trend of PSNR with epoch in TTT-EI and TTT-BN-EI schemes.}
    \end{minipage}
    \hfill
    \begin{minipage}{0.48\textwidth}
        \centering
        \includegraphics[width=1.0\textwidth]
        {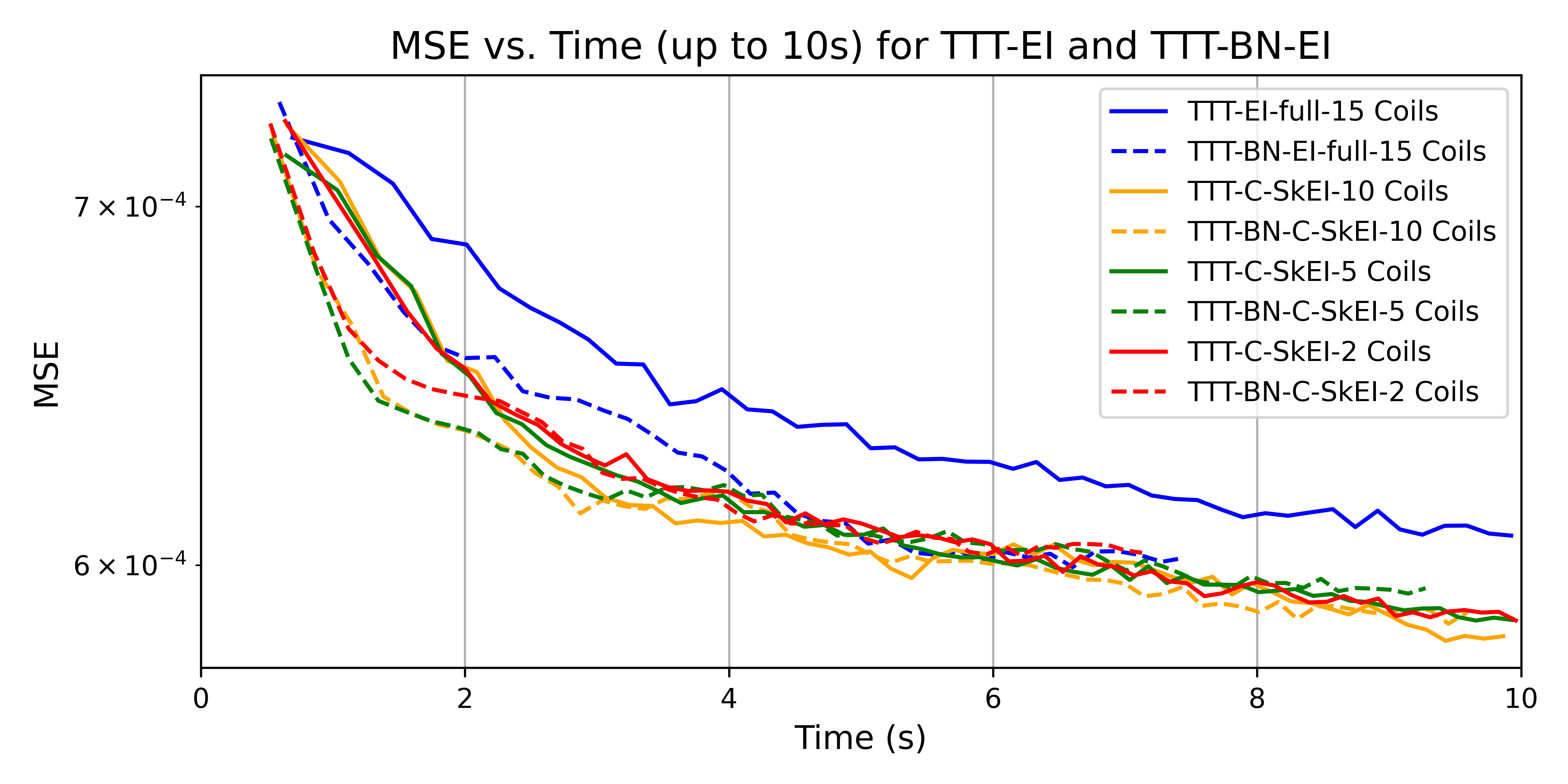}
        \caption*{(b): Trend of MSE with time (up to 10 seconds) in TTT-EI and TTT-BN-EI schemes.}
    \end{minipage}
    
\caption{Curves of PSNR and MSE of Test Time Training (TTT-EI) and Test Time Training with BatchNorm only (TTT-BN-EI) schemes in multi-coil MRI reconstruction. The solid blue curve (TTT-EI-full-15Coils) denotes the baseline.}
\label{fig:mri-bna}
\end{figure}

To gain a deeper understanding, we proceed by comparing the convergence speeds and efficiency of these two test-time training strategies. As shown in Figure~\ref{fig:mri-bna}, the {TTT-BN-EI} scheme converges more rapidly than the {TTT-EI} scheme, although TTT-EI ultimately yields better reconstruction results. Furthermore, both the TTT-EI and TTT-BN-EI schemes demonstrate robustness to the sketching operation, as evidenced in Figure~\ref{fig:mri-bna}. Table~\ref{tab: NA v.s. BN-mri} summarizes the comparative results of the two test-time training strategies under noisy conditions with $\sigma = 0.005$. 
Furthermore, the TTT-BN-EI scheme achieves this with a notable reduction in the number of trainable parameters.


\begin{table}[htp]
\renewcommand{\arraystretch}{1.2}
\centering
\begin{tabular}{c|ccc}
\toprule
& PSNR &  Trainable Param ($\times 10^7$) & Training Epochs \\
\hline
\xgx{TTT} & (noise level 0.005) &   \\
 \hdashline[1pt/1pt]
*EI-full (15 coils) & 32.29 &  3.45 & 82 \\
C-SkEI-10coils & 32.48  &  3.45 & 98 \\
C-SkEI-5coils & 32.36  &  3.45 & 82 \\
C-SkEI-2coils & 32.35  & 3.45 & 83 \\
\midrule
\xgx{TTT-BN} & & (noise level 0.005) &   \\
 \hdashline[1pt/1pt]
EI-full (15 coils) & 32.21 &  0.0014 & 36 \\
C-SkEI-10coils & 32.31  &  0.0014 & 57 \\
C-SkEI-5coils & 32.26  &  0.0014 & 59 \\
C-SkEI-2coils & 32.20  & 0.0014 & 44 \\
\bottomrule
\end{tabular}
\caption{Further comparisons between TTT-EI and TTT-BN-EI schemes in multi-coil MRI reconstruction. `*' denotes the baseline.}
\label{tab: NA v.s. BN-mri}
\end{table}

\begin{figure}[htp!]
    \centering

    \begin{subfigure}{0.15\textwidth}
        \centering
        \includegraphics[width=\textwidth]{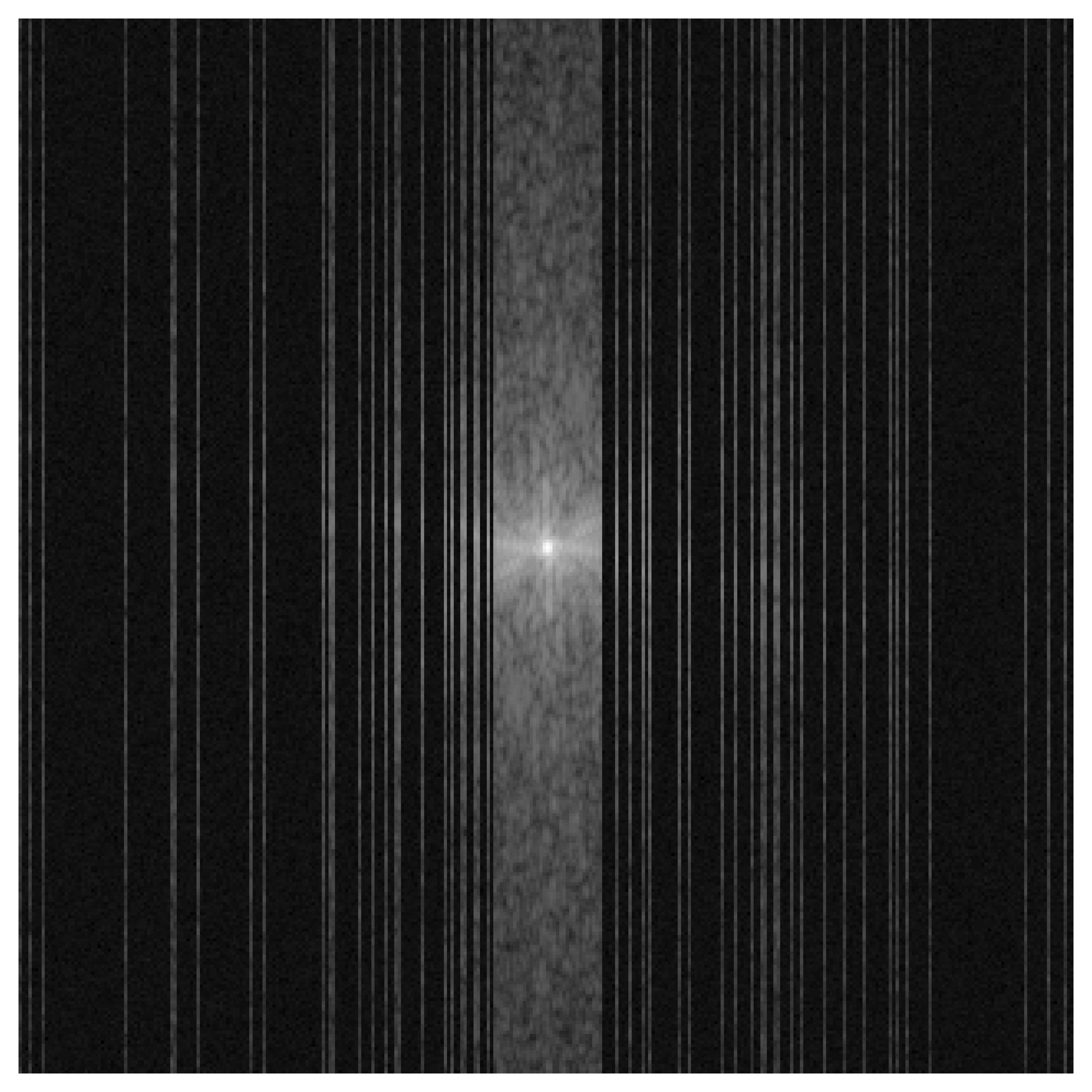} 
        \caption*{\centering$y \sim \mathcal{N}(y^*, 0.001)$}
    \end{subfigure}
    \begin{subfigure}{0.15\textwidth}
        \centering
        \includegraphics[width=\textwidth]{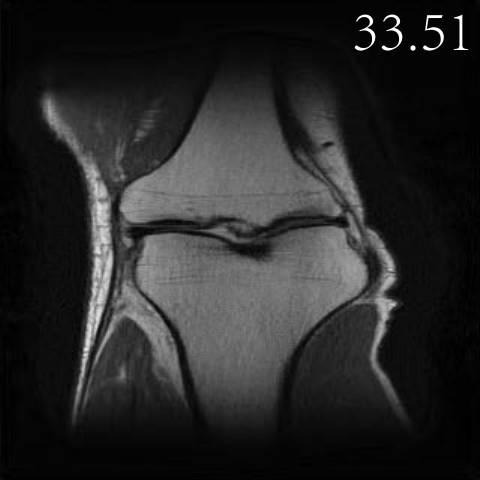} 
        \caption*{\centering *TTT-EI- \\ full-15Coils}
    \end{subfigure}
    \begin{subfigure}{0.15\textwidth}
        \centering
        \includegraphics[width=\textwidth]{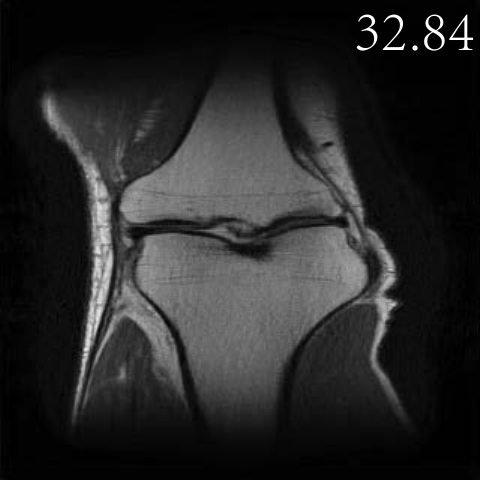} 
        \caption*{\centering TTT-\\C-SkEI-2Coils}
    \end{subfigure}
    \begin{subfigure}{0.15\textwidth}
        \centering
        \includegraphics[width=\textwidth]{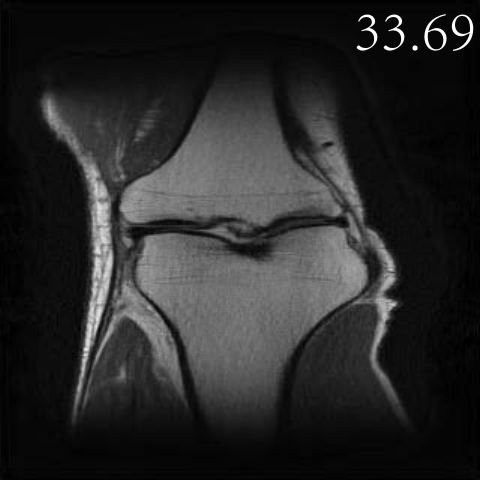} 
        \caption*{\centering TTT-BN-\\C-SkEI-15Coils}
    \end{subfigure}
    \begin{subfigure}{0.15\textwidth}
        \centering
        \includegraphics[width=\textwidth]{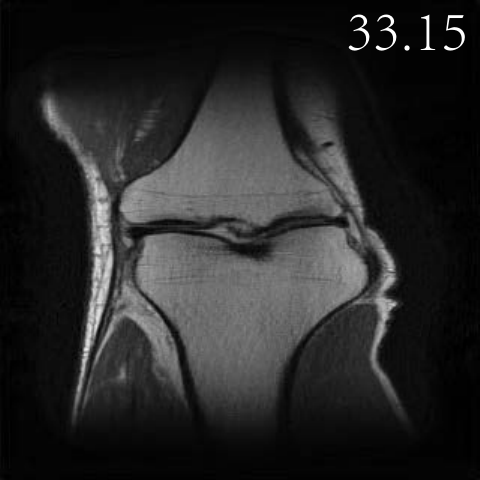} 
        \caption*{\centering TTT-BN-\\C-SkEI-2Coils}
    \end{subfigure}
    \begin{subfigure}{0.15\textwidth}
        \centering
        \includegraphics[width=\textwidth]{figures/3.2/320-MRI/GT-x.pdf} 
        \caption*{$x$ (GT) \\ \quad}
    \end{subfigure}

    \begin{subfigure}{0.15\textwidth}
        \centering
        \includegraphics[width=\textwidth]{figures/3.2/320-MRI/GTy-320-0_005.png} 
        \caption*{\centering$y \sim \mathcal{N}(y^*, 0.005)$}
    \end{subfigure}
    \begin{subfigure}{0.15\textwidth}
        \centering
        \includegraphics[width=\textwidth]{figures/3.2/320-MRI/NA-Coils15-0.005-32.29.pdf} 
        \caption*{\centering *TTT-EI- \\ full-15Coils}
    \end{subfigure}
    \begin{subfigure}{0.15\textwidth}
        \centering
        \includegraphics[width=\textwidth]{figures/3.2/320-MRI/NA-Coils2-0.005-32.25.pdf} 
        \caption*{\centering TTT-\\C-SkEI-2Coils}
    \end{subfigure}
    \begin{subfigure}{0.15\textwidth}
        \centering
        \includegraphics[width=\textwidth]{figures/3.2/320-MRI/BN-Coils15-0.005-32.21.pdf} 
        \caption*{\centering TTT-BN-\\C-SkEI-15Coils}
    \end{subfigure}
    \begin{subfigure}{0.15\textwidth}
        \centering
        \includegraphics[width=\textwidth]{figures/3.2/320-MRI/BN-Coils2-0.005-32.20.pdf} 
        \caption*{\centering TTT-BN-\\C-SkEI-2Coils}
    \end{subfigure}
    \begin{subfigure}{0.15\textwidth}
        \centering
        \includegraphics[width=\textwidth]{figures/3.2/320-MRI/GT-x.pdf} 
        \caption*{$x$ (GT) \\ \quad}
    \end{subfigure}

    \begin{subfigure}{0.15\textwidth}
        \centering
        \includegraphics[width=\textwidth]{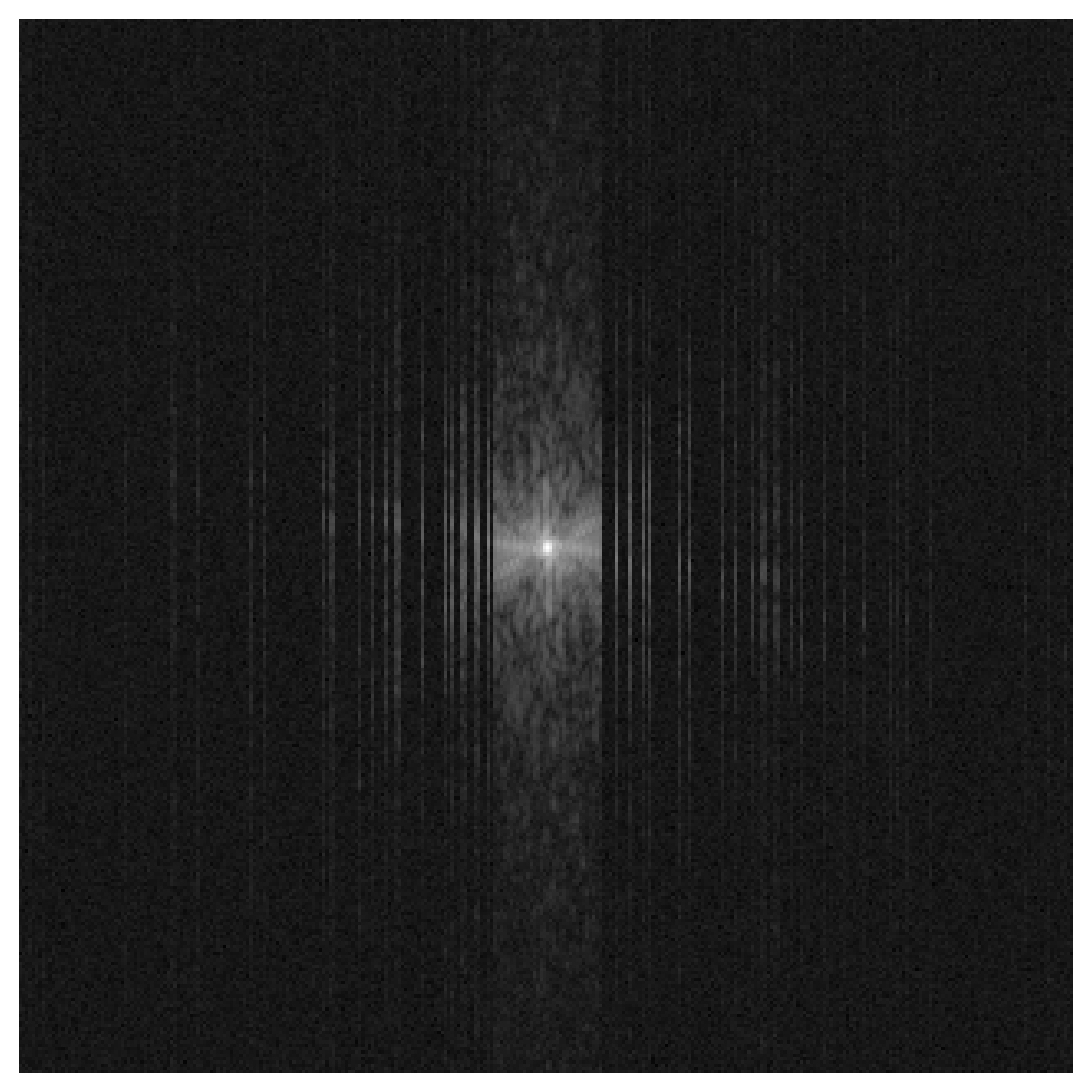} 
        \caption*{$y \sim \mathcal{N}(y^*, 0.01)$ \\ \quad}
    \end{subfigure}
    \begin{subfigure}{0.15\textwidth}
        \centering
        \includegraphics[width=\textwidth]{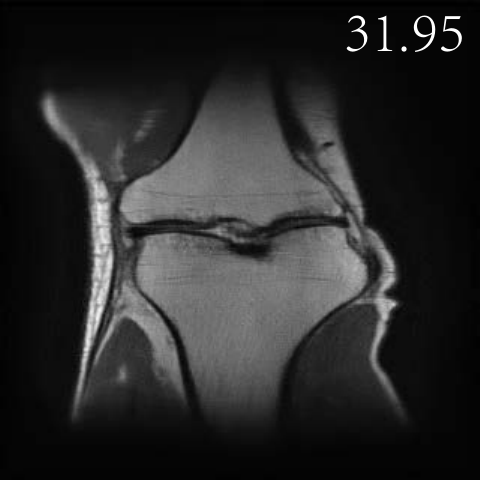} 
        \caption*{\centering *TTT-\\C-SkEI-15Coils}
    \end{subfigure}
    \begin{subfigure}{0.15\textwidth}
        \centering
        \includegraphics[width=\textwidth]{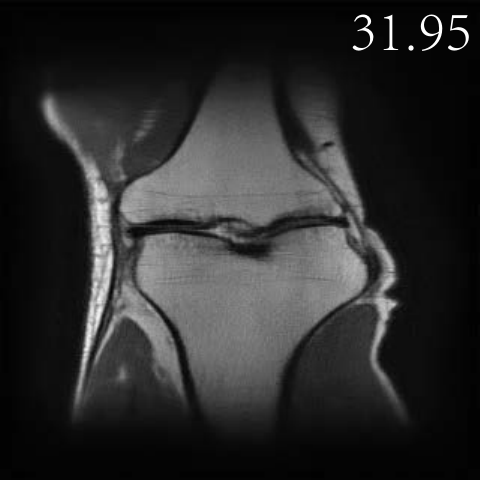} 
        \caption*{\centering TTT-\\C-SkEI-2Coils}
    \end{subfigure}
    \begin{subfigure}{0.15\textwidth}
        \centering
        \includegraphics[width=\textwidth]{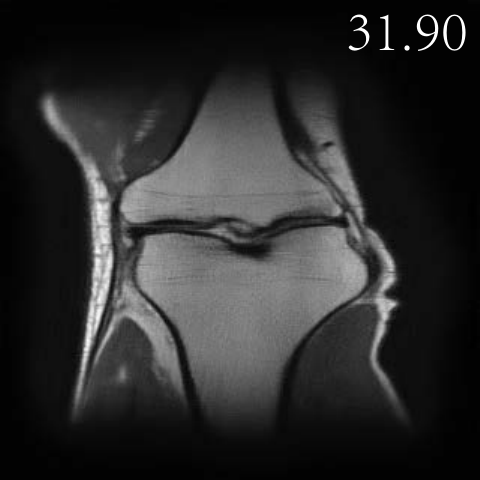} 
        \caption*{\centering TTT-BN-\\C-SkEI-15Coils}
    \end{subfigure}
    \begin{subfigure}{0.15\textwidth}
        \centering
        \includegraphics[width=\textwidth]{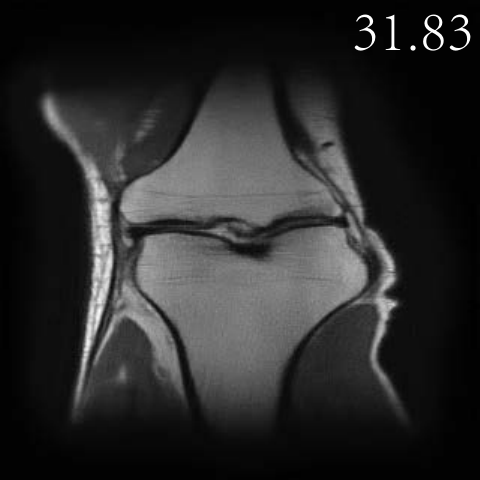} 
        \caption*{\centering TTT-BN-\\C-SkEI-2Coils}
    \end{subfigure}
    \begin{subfigure}{0.15\textwidth}
        \centering
        \includegraphics[width=\textwidth]{figures/3.2/320-MRI/GT-x.pdf} 
        \caption*{$x$ (GT) \\ \quad}
    \end{subfigure}

\caption{Multicoil MRI image reconstructions of Test Time Training (TTT) and Test Time Training with BatchNorm only (TTT-BN) schemes with different noise level and different sketch size. `*' denotes the baseline.}
    \label{fig:mri-NA-BN-EI-DIP}
\end{figure}

\begin{table}[htp]
\renewcommand{\arraystretch}{1.2}
\setlength{\tabcolsep}{6pt}
\centering
\begin{tabular}{l|cccc|cccc}
\toprule
\diagbox{noise level}{PSNR}{Scheme} & & TTT-EI & & & & TTT-BN-EI & & \\
\hline
Coil numbers & 15(*) & 10 & 5 & 2 & 15 & 10 & 5 & 2 \\
\hline
0.001 & 33.51 & 33.02 & 32.95 & 32.84 & 33.69 & 33.08 & 33.10 & 33.15 \\
\hdashline[1pt/1pt]
0.01 & 31.95  & 31.70 & 31.69 & 31.95 & 31.90 & 31.82 & 31.86 & 31.83 \\
\bottomrule
\end{tabular}
\caption{PSNR of multi-coil MRI reconstructions in TTT-EI and TTT-BN-EI schemes with increasing noise level from $\sigma = 0.001$ to $0.01$.  (*) denotes the baseline.}
\label{tab:noise-mri}
\end{table}

We further evaluated the performance of the two Test Time Training schemes, TTT-EI and TTT-BN-EI, on varying noise levels, as shown in Figure~\ref{fig:mri-NA-BN-EI-DIP} and Table~\ref{tab:noise-mri}. The results indicate that the performance of both methods gradually declines as the noise level increases. In particular, both approaches present robustness to the sketching operation at different noise levels.

\section{Conclusion}

In this work, we propose a sketched EI regularizer which can be efficiently applied for unsupervised training of deep imaging networks, especially in the real-time test-time training / network-adaptation setting. We provide a motivational theoretical analysis of the proposed sketching scheme demonstrating that it is an effective approximation of the original EI regularization proposed by \cite{chen2021equivariant}. In test-time training experiments for sparse-view X-ray CT, our Sketched-EI approach achieves a substantial speed-up compared with standard EI. For network adaptation of pre-trained models, we further identify a strong acceleration strategy that updates only the normalization layers, applicable to both EI and Sketched-EI. Finally, we introduce a coil-sketching extension tailored to multicoil MRI, which yields faster and higher-quality reconstructions than standard methods.

Looking ahead, the proposed framework opens several promising directions. A natural next step is a more comprehensive theoretical study of the sketching mechanism, including a more precise characterization of the approximation error and its impact on reconstruction quality. Another possibility is to develop adaptive or data-driven sketching strategies that optimize sketch dimensions or operators during training, potentially improving both efficiency and stability. Extending EI and Sketched-EI to nonlinear inverse problems also represents an important direction, as it would significantly broaden the applicability of EI-based regularization. We plan to pursue these directions in future work.
\section{Appendix}

\subsection{Proof of Theorem \ref{thm:sketch}}
In this section of appendix, we provide the proof for the motivational bound presented in Theorem 1 for the approximation of EI regularizer. We start by proving the upper bound:
\begin{equation}
\begin{aligned}
        \|v - \Ft(A_S^\dagger A_S v)\|_2 &= \|v - \Ft(A_S^\dagger A_S v) + \Ft(A^\dagger A v)- \Ft(A^\dagger A v)\|_2\\ &\leq \|v - \Ft(A^\dagger A v)\|_2 + \| \Ft(A_S^\dagger A_S v) - \Ft(A^\dagger A v)\|_2.
\end{aligned}
\end{equation}
Due to the assumption on the $L$-Lipschitz continuity of the network, we have the following.
\begin{equation}
\begin{aligned}
     & \| \Ft(A_S^\dagger A_S v) - \Ft(A^\dagger A v)\|_2 \\&\leq L\|A_S^\dagger A_S v - A^\dagger A v\|_2 \\ &\leq L\|A^\dagger S^\dagger SA - A^\dagger A\|_2\|v\|_2 \\
      & \leq L\|A^\dagger(S^TS - I)A\|_2\|v\|_2
\end{aligned}
\end{equation}
Denote $A$'s rank $k \leq d$, let's write the SVD as $A = U\Sigma V^T$ with semi-unitary matrices $U \in \R^{n \times k}$, $V^T \in \R^{k \times d}$ such that $U^TU = V^TV = I$, and diagonal matrix $\Sigma \in \R^{k \times k}$ whose diagonal contains the non-zero singular values $\Sigma = \mathrm{diag}[\sigma_1, ..., \sigma_{k} ]$, we have:
\begin{equation}
\begin{aligned}
     & \| \Ft(A_S^\dagger A_S v) - \Ft(A^\dagger A v)\|_2 \\
      & \leq L\|A^\dagger(S^TS - I)A\|_2\|v\|_2\\
      & \leq Lr\|A^T(AA^T)^{-1}(S^TS - I)A\|_2\\
      & = Lr\|V\Sigma U^TU\Sigma^{-2}U(S^TS - I)U\Sigma V^T\|_2\\
      & = Lr\|V\Sigma^{-1}U(S^TS - I)U\Sigma V^T\|_2\\
      &\leq \frac{Lr\sigma_1}{\sigma_{k}}\|U^T(S^TS - I)U\|_2
\end{aligned}
\end{equation}
The remaining challenge is to bound the term $\|U^T(S^TS - I)U\|_2 \leq \delta$ almost surely, whose actual value depends on the choice of the sketching operator $M$. Here we illustrate with the choice of sub-Gaussian sketches, while subsampling and randomized orthogonal sketches also satisfy the bound with different values of $\delta$ and probability. According to \cite[Proposition 1]{2015_Pilanci_Randomized}, if $M$ is a $\sigma$-sub-Gaussian sketch with sketch size $m$, then we only need $\delta = c_0\sqrt{\frac{k}{m}} + \delta_0$ for Theorem 1 to hold.

For the lower bound, we use the same reasoning:
\begin{equation}
\begin{aligned}
        \|v - \Ft(A^\dagger A v)\|_2 &= \|v - \Ft(A_S^\dagger A_S v) + \Ft(A_S^\dagger A_S v)- \Ft(A^\dagger A v)\|_2\\ &\leq \|v - \Ft(A_S^\dagger A_S v)\|_2 + \| \Ft(A_S^\dagger A_S v) - \Ft(A^\dagger A v)\|_2\\
        &\leq \|v - \Ft(A_S^\dagger A_S v)\|_2 + Lr\delta
\end{aligned}
\end{equation}
Then we immediately have $\|v - \Ft(A_S^\dagger A_S v)\|_2 \geq \|v - \Ft(A^\dagger A v)\|_2 - Lr\delta$. Thus, we finish the proof.


\paragraph{Regarding approximately low-rank operators.} In real-world applications, such as tomographic imaging (CT/MRI/PET), the measurement operator are approximately low-rank. For example, the spectrum of the $k$-approximate low-rank operator is $\Sigma = \mathrm{diag}[\sigma_1, ..., \sigma_{k}, ... , \sigma_\mathrm{min} ]$, where $\sigma_1 = O(1)$, $\sigma_{k} = O(1)$, $\sigma_{k-1} = o(1)$, $\sigma_\mathrm{min} = o(1)$, then the above bound would be:
\begin{equation}
\begin{aligned}
      \| \Ft(A_S^\dagger A_S v) - \Ft(A^\dagger A v)\|_2 \leq O(1)\|U^T(S^TS - I)U\|_2 + o(1).
\end{aligned}
\end{equation}
Considering the results for low-rank and approximate low-rank cases, we can observe that as long as the measurement operator $A$ has a fast decaying spectrum, where $k \ll d$, then the approximation of the sketched EI regularizer can be accurate.


\bibliographystyle{abbrv}
\bibliography{main.bib}

\begin{thebibliography}{10}

\bibitem{batson2019noise2self}
J.~Batson and L.~Royer.
\newblock Noise2self: Blind denoising by self-supervision.
\newblock In {\em International Conference on Machine Learning}, pages 524--533, 2019.

\bibitem{beck2009fast}
A.~Beck and M.~Teboulle.
\newblock Fast gradient-based algorithms for constrained total variation image denoising and deblurring problems.
\newblock {\em IEEE Transactions on Image Processing}, 18(11):2419--2434, 2009.

\bibitem{Buehrer2007Array}
M.~Buehrer, K.~P. Pruessmann, P.~Boesiger, and S.~Kozerke.
\newblock Array compression for mri with large coil arrays.
\newblock {\em Magnetic Resonance in Medicine}, 57(6):1131--1139, 2007.

\bibitem{cai2024nf}
Z.~Cai, J.~Tang, S.~Mukherjee, J.~Li, C.-B. Sch{\"o}nlieb, and X.~Zhang.
\newblock Nf-ula: Normalizing flow-based unadjusted langevin algorithm for imaging inverse problems.
\newblock {\em SIAM Journal on Imaging Sciences}, 17(2):820--860, 2024.

\bibitem{carioni2023unsupervised}
M.~Carioni, S.~Mukherjee, H.~Y. Tan, and J.~Tang.
\newblock Unsupervised approaches based on optimal transport and convex analysis for inverse problems in imaging.
\newblock {\em Radon Series on Computational and Applied Mathematics}, 2024.

\bibitem{chen2021equivariant}
D.~Chen, J.~Tachella, and M.~E. Davies.
\newblock Equivariant imaging: Learning beyond the range space.
\newblock In {\em Proceedings of the IEEE/CVF International Conference on Computer Vision}, pages 4379--4388, 2021.

\bibitem{chen2022robust}
D.~Chen, J.~Tachella, and M.~E. Davies.
\newblock Robust equivariant imaging: a fully unsupervised framework for learning to image from noisy and partial measurements.
\newblock In {\em Proceedings of the IEEE/CVF Conference on Computer Vision and Pattern Recognition}, pages 5647--5656, 2022.

\bibitem{clark2013cancer}
K.~Clark, B.~Vendt, K.~Smith, J.~Freymann, J.~Kirby, P.~Koppel, S.~Moore, S.~Phillips, D.~Maffitt, M.~Pringle, et~al.
\newblock The cancer imaging archive (tcia): maintaining and operating a public information repository.
\newblock {\em Journal of digital imaging}, 26:1045--1057, 2013.

\bibitem{darestani2022test}
M.~Z. Darestani, J.~Liu, and R.~Heckel.
\newblock Test-time training can close the natural distribution shift performance gap in deep learning based compressed sensing.
\newblock In {\em International conference on machine learning}, pages 4754--4776. PMLR, 2022.

\bibitem{ehrhardt2024guide}
M.~J. Ehrhardt, Z.~Kereta, J.~Liang, and J.~Tang.
\newblock A guide to stochastic optimisation for large-scale inverse problems.
\newblock {\em arXiv preprint arXiv:2406.06342}, 2024.

\bibitem{frankle2021training}
J.~Frankle, D.~J. Schwab, and A.~S. Morcos.
\newblock Training batchnorm and only batchnorm: On the expressive power of random features in {\{}cnn{\}}s.
\newblock In {\em International Conference on Learning Representations}, 2021.

\bibitem{gandelsman2022test}
Y.~Gandelsman, Y.~Sun, X.~Chen, and A.~Efros.
\newblock Test-time training with masked autoencoders.
\newblock {\em Advances in Neural Information Processing Systems}, 35:29374--29385, 2022.

\bibitem{HUANG2008133}
F.~Huang, S.~Vijayakumar, Y.~Li, S.~Hertel, and G.~R. Duensing.
\newblock A software channel compression technique for faster reconstruction with many channels.
\newblock {\em Magnetic Resonance Imaging}, 26(1):133--141, 2008.

\bibitem{johnson2013accelerating}
R.~Johnson and T.~Zhang.
\newblock Accelerating stochastic gradient descent using predictive variance reduction.
\newblock In {\em Advances in neural information processing systems}, pages 315--323, 2013.

\bibitem{kingma2014adam}
D.~P. Kingma and J.~Ba.
\newblock Adam: A method for stochastic optimization.
\newblock {\em Proceedings of 3rd International Conference on Learning Representations}, 2015.

\bibitem{lehtinen2018noise2noise}
J.~Lehtinen, J.~Munkberg, J.~Hasselgren, S.~Laine, T.~Karras, M.~Aittala, and T.~Aila.
\newblock Noise2noise: Learning image restoration without clean data.
\newblock {\em arXiv preprint arXiv:1803.04189}, 2018.

\bibitem{liu2020rare}
J.~Liu, Y.~Sun, C.~Eldeniz, W.~Gan, H.~An, and U.~S. Kamilov.
\newblock Rare: Image reconstruction using deep priors learned without groundtruth.
\newblock {\em IEEE Journal of Selected Topics in Signal Processing}, 14(6):1088--1099, 2020.

\bibitem{liu2019image}
J.~Liu, Y.~Sun, X.~Xu, and U.~S. Kamilov.
\newblock Image restoration using total variation regularized deep image prior.
\newblock In {\em ICASSP 2019-2019 IEEE International Conference on Acoustics, Speech and Signal Processing (ICASSP)}, pages 7715--7719. Ieee, 2019.

\bibitem{liu2021ttt++}
Y.~Liu, P.~Kothari, B.~Van~Delft, B.~Bellot-Gurlet, T.~Mordan, and A.~Alahi.
\newblock Ttt++: When does self-supervised test-time training fail or thrive?
\newblock {\em Advances in Neural Information Processing Systems}, 34:21808--21820, 2021.

\bibitem{mataev2019deepred}
G.~Mataev, P.~Milanfar, and M.~Elad.
\newblock Deepred: Deep image prior powered by red.
\newblock In {\em Proceedings of the IEEE/CVF International Conference on Computer Vision Workshops}, pages 0--0, 2019.

\bibitem{mueller2024normalization}
M.~Mueller, T.~Vlaar, D.~Rolnick, and M.~Hein.
\newblock Normalization layers are all that sharpness-aware minimization needs.
\newblock {\em Advances in Neural Information Processing Systems}, 36, 2024.

\bibitem{niu2022efficient}
S.~Niu, J.~Wu, Y.~Zhang, Y.~Chen, S.~Zheng, P.~Zhao, and M.~Tan.
\newblock Efficient test-time model adaptation without forgetting.
\newblock In {\em International conference on machine learning}, pages 16888--16905. PMLR, 2022.

\bibitem{Oscanoa2024Coil}
J.~A. Oscanoa, F.~Ong, S.~S. Iyer, Z.~Li, C.~M. Sandino, B.~Ozturkler, D.~B. Ennis, M.~Pilanci, and S.~S. Vasanawala.
\newblock Coil sketching for computationally efficient mr iterative reconstruction.
\newblock {\em Magnetic Resonance in Medicine}, 91(2):784--802, 2024.

\bibitem{2015_Pilanci_Randomized}
M.~Pilanci and M.~J. Wainwright.
\newblock Randomized sketches of convex programs with sharp guarantees.
\newblock {\em Information Theory, IEEE Transactions on}, 61(9):5096--5115, 2015.

\bibitem{pilanci2017newton}
M.~Pilanci and M.~J. Wainwright.
\newblock Newton sketch: A near linear-time optimization algorithm with linear-quadratic convergence.
\newblock {\em SIAM Journal on Optimization}, 27(1):205--245, 2017.

\bibitem{ronneberger2015u}
O.~Ronneberger, P.~Fischer, and T.~Brox.
\newblock U-net: Convolutional networks for biomedical image segmentation.
\newblock In N.~Navab, J.~Hornegger, W.~M. Wells, and A.~F. Frangi, editors, {\em Medical Image Computing and Computer-Assisted Intervention -- MICCAI 2015}, pages 234--241, Cham, 2015. Springer International Publishing.

\bibitem{rudi2015less}
A.~Rudi, R.~Camoriano, and L.~Rosasco.
\newblock Less is more: Nystr{\"o}m computational regularization.
\newblock {\em Advances in neural information processing systems}, 28, 2015.

\bibitem{ryu2019plug}
E.~Ryu, J.~Liu, S.~Wang, X.~Chen, Z.~Wang, and W.~Yin.
\newblock Plug-and-play methods provably converge with properly trained denoisers.
\newblock In {\em International Conference on Machine Learning}, pages 5546--5557, 2019.

\bibitem{scanvic2023self}
J.~Scanvic, M.~Davies, P.~Abry, and J.~Tachella.
\newblock Self-supervised learning for image super-resolution and deblurring.
\newblock {\em arXiv preprint arXiv:2312.11232}, 2023.

\bibitem{sun2020test}
Y.~Sun, X.~Wang, Z.~Liu, J.~Miller, A.~Efros, and M.~Hardt.
\newblock Test-time training with self-supervision for generalization under distribution shifts.
\newblock In {\em International conference on machine learning}, pages 9229--9248. PMLR, 2020.

\bibitem{sun2019online}
Y.~Sun, B.~Wohlberg, and U.~S. Kamilov.
\newblock An online plug-and-play algorithm for regularized image reconstruction.
\newblock {\em IEEE Transactions on Computational Imaging}, 2019.

\bibitem{tachella2023sensing}
J.~Tachella, D.~Chen, and M.~Davies.
\newblock Sensing theorems for unsupervised learning in linear inverse problems.
\newblock {\em Journal of Machine Learning Research}, 24(39):1--45, 2023.

\bibitem{tachella2024unsure}
J.~Tachella, M.~Davies, and L.~Jacques.
\newblock Unsure: Unknown noise level stein's unbiased risk estimator.
\newblock {\em arXiv preprint arXiv:2409.01985}, 2024.

\bibitem{tachella2020neural}
J.~Tachella, J.~Tang, and M.~Davies.
\newblock The neural tangent link between cnn denoisers and non-local filters.
\newblock {\em IEEE/CVF Conference on Computer Vision and Pattern Recognition}, 2021.

\bibitem{tan2024provably}
H.~Y. Tan, S.~Mukherjee, J.~Tang, and C.-B. Sch{\"o}nlieb.
\newblock Provably convergent plug-and-play quasi-newton methods.
\newblock {\em SIAM Journal on Imaging Sciences}, 17(2):785--819, 2024.

\bibitem{tang2020practicality}
J.~Tang, K.~Egiazarian, M.~Golbabaee, and M.~Davies.
\newblock The practicality of stochastic optimization in imaging inverse problems.
\newblock {\em IEEE Transactions on Computational Imaging}, 6:1471--1485, 2020.

\bibitem{tang2017gradient}
J.~Tang, M.~Golbabaee, and M.~E. Davies.
\newblock Gradient projection iterative sketch for large-scale constrained least-squares.
\newblock In {\em International Conference on Machine Learning}, pages 3377--3386. PMLR, 2017.

\bibitem{tirer2024deep}
T.~Tirer, R.~Giryes, S.~Y. Chun, and Y.~C. Eldar.
\newblock Deep internal learning: Deep learning from a single input.
\newblock {\em IEEE Signal Processing Magazine}, 41(4):40--57, 2024.

\bibitem{Uecker2013ESPIRiT}
M.~Uecker, P.~Lai, M.~J. Murphy, P.~Virtue, M.~Elad, J.~M. Pauly, S.~S. Vasanawala, and M.~Lustig.
\newblock Espirit—an eigenvalue approach to autocalibrating parallel mri: Where sense meets grappa.
\newblock {\em Magnetic Resonance in Medicine}, 71(3):990--1001, 2014.

\bibitem{ulyanov2018deep}
D.~Ulyanov, A.~Vedaldi, and V.~Lempitsky.
\newblock Deep image prior.
\newblock In {\em Proceedings of the IEEE Conference on Computer Vision and Pattern Recognition}, pages 9446--9454, 2018.

\bibitem{wang2024fully}
A.~Wang and M.~Davies.
\newblock Fully unsupervised dynamic mri reconstruction via diffeo-temporal equivariance.
\newblock {\em arXiv preprint arXiv:2410.08646}, 2024.

\bibitem{wang2024perspective}
A.~Wang and M.~Davies.
\newblock Perspective-equivariant imaging: an unsupervised framework for multispectral pansharpening.
\newblock {\em arXiv preprint arXiv:2403.09327}, 2024.

\bibitem{wang2020tent}
D.~Wang, E.~Shelhamer, S.~Liu, B.~Olshausen, and T.~Darrell.
\newblock Tent: Fully test-time adaptation by entropy minimization.
\newblock {\em arXiv preprint arXiv:2006.10726}, 2020.

\bibitem{woodruff2014sketching}
D.~P. Woodruff et~al.
\newblock Sketching as a tool for numerical linear algebra.
\newblock {\em Foundations and Trends{\textregistered} in Theoretical Computer Science}, 10(1--2):1--157, 2014.

\bibitem{yuan2023robust}
L.~Yuan, B.~Xie, and S.~Li.
\newblock Robust test-time adaptation in dynamic scenarios.
\newblock In {\em Proceedings of the IEEE/CVF Conference on Computer Vision and Pattern Recognition}, pages 15922--15932, 2023.

\bibitem{zbontar2018fastmri}
J.~Zbontar, F.~Knoll, A.~Sriram, T.~Murrell, Z.~Huang, M.~J. Muckley, A.~Defazio, R.~Stern, P.~Johnson, M.~Bruno, et~al.
\newblock fastmri: An open dataset and benchmarks for accelerated mri.
\newblock {\em arXiv preprint arXiv:1811.08839}, 2018.

\bibitem{Zhang2013Coil}
T.~Zhang, J.~M. Pauly, S.~S. Vasanawala, and M.~Lustig.
\newblock Coil compression for accelerated imaging with cartesian sampling.
\newblock {\em Magnetic Resonance in Medicine}, 69(2):571--582, 2013.

\end{thebibliography}

\end{document}